\documentclass[aps,prd,onecolumn,11pt,tightenlines,a4paper,superscriptaddress,preprintnumbers,floatfix,nofootinbib]{revtex4-2}
\usepackage{physics}
\usepackage{pifont}
\usepackage[utf8]{inputenc}
\usepackage[T1]{fontenc}
\usepackage{microtype}
\usepackage{amsmath,amssymb,amsfonts}
\usepackage{graphicx}
\usepackage{braket,booktabs}
\usepackage{xcolor}
\usepackage{bm}
\usepackage{tikz}
\usetikzlibrary{arrows.meta, positioning}
\usepackage[margin=1in]{geometry} 
\usepackage{multirow}
\usepackage{hyperref}
\hypersetup{colorlinks=true,linkcolor=blue,citecolor=blue,urlcolor=blue}

\definecolor{darkpurple}{RGB}{102,0,153}

\makeatletter
\renewcommand{\thesection}{\arabic{section}}
\renewcommand{\thesubsection}{\thesection.\arabic{subsection}}
\renewcommand{\thesubsubsection}{\thesubsection.\arabic{subsubsection}}
\def\p@subsection{}
\def\p@subsubsection{}
\newcommand{\jcaptocline}[6]{%
  \ifnum #1>\c@tocdepth\relax\else
    \addpenalty{\@secpenalty}%
    \vskip #6\relax
    \begingroup
      \parindent #2\relax
      \leftskip #2\relax
      \rightskip 2.4em plus 1fil\relax
      \parfillskip -2.4em\relax
      \interlinepenalty\@M
      \leavevmode
      \@tempdima #3\relax
      \advance\leftskip\@tempdima
      \null\nobreak\hskip-\leftskip
      #4\nobreak\hfill\nobreak
      \hbox to 2.4em{\hfil\textcolor{blue}{#5}}\par
    \endgroup
  \fi
}

\renewcommand*\l@section[2]{%
 \jcaptocline{1}{0em}{1.6em}{\bfseries #1}{#2}{0.22em}%
}
\renewcommand*\l@subsection[2]{%
 \jcaptocline{2}{1.7em}{1.85em}{#1}{#2}{0pt}%
}
\renewcommand*\l@subsubsection[2]{%
 \jcaptocline{3}{3.65em}{3.1em}{#1}{#2}{0pt}%
}
\renewcommand*\l@paragraph[2]{}
\renewcommand*\l@subparagraph[2]{}

\def\tocdepth@munge{}
\def\tocdepth@restore{}
\renewcommand{\tableofcontents}{%
  \begingroup
    \pagestyle{empty}%
    \thispagestyle{empty}%
    \vspace*{-0.67in}%
    \parskip=0pt\relax
    \baselineskip=14.1pt\relax
    \hypersetup{linkcolor=black}%
    \par\noindent\rule{\linewidth}{0.45pt}\par
    \vspace{0.65em}%
    \noindent\textbf{Contents}\par
    \vspace{1.52em}%
    \let\appendix\@empty
    \@starttoc{toc}%
    \vspace{0.65em}\par\noindent\rule{\linewidth}{0.45pt}\par
  \endgroup
}
\makeatother

\newcommand{\eq}[1]{Eq.~\eqref{#1}}
\newcommand{\Figref}[1]{Fig.~\ref{#1}}
\newcommand{\Tabref}[1]{Tab.~\ref{#1}}
\newcommand{\Secref}[1]{Sec.~\ref{#1}}
\newcommand{\Appref}[1]{App.~\ref{#1}}
\newcommand{\be}{\begin{equation}}
\newcommand{\ee}{\end{equation}}
\def\bea{\begin{eqnarray} }
\def\eea{ \end{eqnarray} }
\renewcommand{\log}{\operatorname{Log}}
\DeclareMathOperator{\logU}{LogU}
\DeclareMathOperator{\logN}{LogN}

\begin{document}

\title{LIGO-Virgo-KAGRA stochastic constraints on first-order phase transitions:
\texorpdfstring{\\}{ }
A tale of priors
\texorpdfstring{\\\hfill\normalfont\small\smash{\raisebox{2cm}{KIAS-P26088}}}{ }
}

\author{S.~AbdusSalam}
\email{abdussalam@sbu.ac.ir}
\affiliation{Department of Physics, Shahid Beheshti University, Tehran 1983969411, Iran}

\author{J.~Kersten}
\email{joern.kersten@uib.no}
\affiliation{Department of Physics and IPAP, Yonsei University, Seoul 03722, South Korea}
\affiliation{Department of Physics and Technology, University of Bergen, Postboks 7803, 5020 Bergen, Norway}
\affiliation{School of Physics, Korea Institute for Advanced Study, 85 Hoegi-ro, Seoul 02455, South Korea}

\author{S.~A.~Ojaghi}
\affiliation{Department of Physics, Shahid Beheshti University, Tehran 1983969411, Iran}

\author{L.~Velasco-Sevilla}
\email{liliana.velascosevilla@gmail.com}
\affiliation{Department of Physics and IPAP, Yonsei University, Seoul 03722, South Korea}

\author{A.~Zafari}
\affiliation{Department of Physics, Shahid Beheshti University, Tehran 1983969411, Iran}

\date{\today}

\begin{abstract}
\noindent We re-examine the methodology for constraining
cosmological first-order phase transitions from gravitational-wave
searches using Bayesian statistical analysis. Specifically, we use
results from LIGO-Virgo-KAGRA searches for a stochastic gravitational
wave background during observing runs O1--O4a.
We explore multiple prior choices across three categories of fit parameters: geometrical parameters, thermodynamical parameters of phase transitions, and the parameters of an Abelian Higgs particle physics model.
Our key finding is that the constraints on most parameters describing
gravitational waves from a phase transition depend strongly on the prior.
Consequently, the data do not yet allow to derive robust constraints on
these parameters.
Besides, even if a parameter is meaningfully constrained, the upper limit can vary significantly with the prior range.
We provide practical recommendations for assessing the robustness of parameter estimates and guidelines for rigorously testing prior independence in future analyses.
\end{abstract}

\maketitle

\clearpage
\tableofcontents
\thispagestyle{empty}

\section{Motivation and overview}
Gravitational waves (GW) provide a probe of the early Universe that complements what photons and neutrinos can access: once produced, GW propagate essentially freely, and therefore carry information about epochs and energy scales that are otherwise out of reach.  A stochastic GW background (SGWB) of cosmological origin, a diffuse signal made up of
many independent, unresolved contributions from across cosmological history, which is potentially accessible to space and ground-based interferometers, and can thus reveal physics at energy scales inaccessible to current colliders \cite{Caprini:2018mtu}.

First-order phase transitions (FOPT) associated with spontaneous symmetry breaking in the early Universe are a particularly well-motivated source of a SGWB\@.  As the temperature decreases, the temperature-dependent effective potential of a scalar field $\phi$ develops a minimum at a non-zero field value $v$.  The minimum at $\phi=0$ becomes a local minimum
(false vacuum with unbroken symmetry), and the global minimum is located at $\phi=v$ (true vacuum with spontaneously broken symmetry).  In a \emph{first-order} phase transition, the transition from the false to the true vacuum proceeds via the nucleation of bubbles, inside of which $\phi=v$.  Subsequently, these bubbles expand and combine until the
symmetry is broken in the entire Universe \cite{Hindmarsh:2020hop}.  The resulting signal is characterised by a small set of thermodynamical parameters: the transition temperature, strength, duration and bubble wall velocity, all of which we introduce in \Secref{sec:ThermoPars}.

Several mechanisms contribute to the GW signal.  Collisions of expanding bubble walls in the scalar field sector produce a characteristic broken-power-law spectrum determined by the typical bubble size at collision.  Early studies based on the envelope approximation have been refined by fully relativistic lattice simulations, which confirm the qualitative features of these spectra while highlighting deviations due to nonlinear field dynamics and overlapping bubble walls \cite{Cutting:2018tjt,Jinno:2016vai}.  Although this contribution is typically short-lived, it can dominate in strongly supercooled
transitions, where the Universe is vacuum-dominated during the transition and the GW amplitude is boosted \cite{Cutting:2018tjt}.  In thermal phase transitions, in contrast, a significant fraction of the released vacuum energy is transferred to the surrounding plasma.  After bubble collisions, this energy sources long-lived sound waves in the form of expanding acoustic shells; numerical simulations with relativistic hydrodynamics show that these coherent bulk flows persist for a Hubble time or longer and source gravitational radiation efficiently \cite{Hindmarsh:2015qta}.  The acoustic contribution is therefore often the dominant GW source in realistic thermal FOPT, as supported by both numerical and semi-analytical
studies \cite{Hindmarsh:2016lnk,Caprini:2019egz}.  At later times, the breakdown of coherent sound waves may lead to magnetohydrodynamic (MHD) turbulence, an additional, typically subdominant source that generates a
broad spectrum whose detailed shape depends on the turbulence model and its decay properties \cite{Caprini:2009yp,Kahniashvili:2008er,Brandenburg:2016jwe}. Although subject to larger theoretical uncertainties, it is relevant for a complete characterisation of the signal.

The LIGO-Virgo-KAGRA (LVK) collaboration searches for an isotropic SGWB by cross-correlating the data of its detectors.  No statistically significant detection has been reported so far, and the resulting upper limits on the SGWB amplitude in the frequency band from roughly $20$ to $100$\,Hz are starting to constrain parts of the parameter space of strongly supercooled FOPT \cite{Abbott:2021xyo,Romero:2021kby,Badger:2022jzg,LVK:2025O4a}.
Interpreting these limits in terms of FOPT parameters requires templates for the GW spectrum \cite{Caprini:2024hue,Caprini:2024gyk}, a model of the astrophysical foreground from unresolved compact binary coalescences, and a
choice of priors for the fit parameters.  The earlier recasts of the O1--O3 results \cite{Romero:2021kby,Badger:2022jzg} relied on the scale-invariant Jeffreys prior only, and the recent release of data from
the first part of the fourth observing run (O4a)
\cite{LVK:2025O4a,LVK:2025O4aData} makes it timely to revisit this
procedure.

A central aim of this work is to address a critical gap in the Bayesian inference of constraints on SGWB, the systematic assessment of robustness under prior change.
Robust Bayesian inference should be approximately prior-independent to allow for meaningful conclusions, making a comparison of multiple priors essential in order to investigate to which extent the limits inferred from GW data are determined by the likelihood and to which extent they are influenced by assumptions about the prior probability distribution. This question is especially relevant for searches for a SGWB from a FOPT, where the available data do not yet provide a significant detection and where the parameter space spans several orders of magnitude. In such a situation, apparently quantitative upper limits or preferred regions potentially reflect the prior volume as much as the information contained in the data.

We examine three complementary approaches to constraining GW parameters from LVK data.  First, a geometric approach extracts limits on the parameters appearing directly in the templates for the GW spectrum, most
importantly the peak amplitude and frequency as well as the amplitude of the astrophysical foreground from unresolved mergers. The second approach constrains the characteristic parameters of a FOPT such as the nucleation temperature and the strength of the phase transition.
Third, a model-based approach places bounds directly on the Lagrangian
parameters of a generic Abelian Higgs model in a dark sector, from which we compute the FOPT parameters (using dimensional reduction to accurately determine the
effective scalar potential at finite temperature) and the resulting GW spectrum.

For each approach, we systematically test at least two different priors, including the scale-invariant Jeffreys prior, a prior informed by the detector sensitivity, and priors motivated by the expected distribution of
FOPT signals.  We then compare the resulting posterior distributions and quantify their degree of agreement. 
This thorough prior sensitivity analysis demonstrates the profound impact of prior choice and establishes a methodology we advocate for broader adoption across the community.

On the technical side, we also improve the description of the peak of
the GW spectrum stemming from sound waves and MHD turbulence, which is
needed to relate the geometrical parameters (peak amplitude and
frequency) to the break frequencies and amplitudes of the
double-broken-power-law (DBPL) template for the spectrum. The details
are given in \Secref{sec:templates}.

The remainder of this paper is organised as follows.
Section~\ref{sec:FOPT} reviews the phenomenology of FOPT and collects the spectral templates used throughout the analysis: the broken-power-law (BPL)
description of bubble collisions, the DBPL description of sound waves and MHD turbulence including the refined peak relations discussed above, and the astrophysical foreground from compact
binary coalescence, from which the cosmological signal must be separated.
Section~\ref{sec:Bayes} sets up the Bayesian inference framework by defining the likelihood for the LVK cross-correlation data and introducing as well as
motivating the prior choices compared in this work.
Section~\ref{sec:U1_direct_sampling} introduces a minimal dark Abelian Higgs model as a concrete particle-physics realisation of a supercooled FOPT, whose parameters are sampled directly.
In \Secref{sec:results}, we present our results for the three complementary approaches outlined above: constraints on the geometrical parameters of the templates, constraints on the underlying thermodynamical parameters of the
FOPT, and constraints on the fundamental parameters of the Abelian Higgs
model. Throughout, the robustness of the constraints under a change of prior is quantified using the diagnostics collected in
\Appref{app:robustness}\@.
We conclude in \Secref{sec:Conclusions}. Technical details are collected in Appendices~\ref{app:CalcLimit}--\ref{app-abelian-higgs-procedure}.

\section{Gravitational wave spectra} \label{sec:FOPT}
\subsection{Characteristic parameters of first-order phase transitions}
\label{sec:ThermoPars}

A cosmological FOPT is characterised by the following
\emph{thermodynamical} parameters.
\begin{itemize}
  \item The temperature of the thermal bath at the end of the phase transition, $T_*$.%
  \footnote{We follow the convention that quantities at the time of GW production are denoted by an asterisk subscript.}
  \item The nucleation (or percolation) temperature $T_n$, at which one
bubble is formed per Hubble volume on average (approximately equal to
$T_*$ if no significant reheating occurs).
  \item The strength parameter $\alpha$, defined as the ratio of latent heat to radiation energy density,
	$\alpha \equiv \frac{1}{\rho_\text{rad}} \left.\left[
	\Delta\rho_\text{vac} -
	\frac{T}{4}\frac{\partial\Delta\rho_\text{vac}}{\partial T}
	\right]\right|_{T=T_n}$,
  where $\Delta\rho_\text{vac}$ is the released vacuum energy density.
  \item The inverse transition duration $\beta$ (commonly expressed as $\beta/H_*$ with $H_*$ being the Hubble rate at $T_*$); rapid transitions have large $\beta/H_*$.
  \item The bubble wall velocity $v_w$, which influences the partition of energy between scalar field gradients and bulk fluid motion.
  \item In addition, the SGWB sourced by turbulence depends on the fraction $\varepsilon$ of the released energy that is converted into turbulent motion during the transition.
\end{itemize}

For a \emph{supercooled} FOPT, the nucleation temperature is much lower
than the intrinsic symmetry breaking scale, $T_n \ll v$.  In the extreme
limit $\alpha\gg1$, the false vacuum energy is much larger than the
radiation density.  Thus, the Universe is vacuum-dominated during the
phase transition, which can significantly boost GW amplitudes
\cite{Cutting:2018tjt}.  
For the non-experts, we summarise in \Tabref{tab:supercooling_comparison}
the differences between supercooled and non-supercooled transitions.
\begin{table}
\centering
\begin{tabular}{lcc}
\toprule
Feature &
Non-supercooled transition &
Supercooled transition \\ \midrule
Completion temperature &
$T_* \sim v$ & $T_* \ll v$
\\
Strength parameter &
$\alpha \ll 1$ &
$\alpha \gtrsim 1$
\\
Bubble wall dynamics &
Subsonic or &
Ultra-relativistic,\\
& mildly supersonic & possibly runaway
\\
Dominant GW source &
Sound waves &
Bubble wall collisions\\
& in plasma & (often)
\\
Typical GW peak frequency &
Higher &
Lower
\\
Plasma effects &
Strong coupling &
Plasma dilution,\\
& to plasma & weak friction
\\
Cosmological impact &
Radiation domination & Vacuum domination \\
\bottomrule
\end{tabular}
\caption{Qualitative comparison between non-supercooled and supercooled cosmological first-order phase transitions. Supercooling delays bubble nucleation to temperatures well below the 
symmetry breaking scale $v$, enhancing the vacuum energy release and modifying both bubble dynamics and the resulting stochastic gravitational wave background~\cite{Caprini:2015zlo,Caprini:2019egz,Espinosa:2010hh}.
}
\label{tab:supercooling_comparison}
\end{table}

\subsection{Spectrum templates for first-order phase transitions} \label{sec:templates}
The GW spectrum is represented as
\begin{equation}
    \Omega_\text{GW}(f) =
    \frac{1}{\rho_c}
    \frac{d\rho_\text{GW}}{d\ln f} \,,
\end{equation}
where $f$ is the frequency, $\rho_\text{GW}$ is the GW energy density,
and $\rho_c$ is the critical density of the Universe.
It is difficult to accurately calculate the relations between
$\Omega_\text{GW}$ and the above thermodynamical parameters.  Therefore,
bounds from experiments on these parameters cannot be obtained directly.  

In practice, the GW spectrum is hence modelled in terms of a
\emph{template} $\mathcal{F}(f,f_p,q)$ as
\begin{equation}
\label{eq:OmegaGW_gen_temp}
\Omega_\text{GW}(f) = \Omega_p \, \mathcal{F}(f,f_p,q) \,,
\end{equation}
where $\Omega_p$ and $f_p$ are the peak amplitude and the peak frequency,
respectively.  The template is obtained from simulations of a given GW
source and also depends on a set of additional parameters $q$
appropriate for that source.  Together with $\Omega_p$ and $f_p$, they
are referred to as \emph{geometrical} parameters.  The data directly
constrain these parameters.  With the help of relations
between geometrical and thermodynamical parameters obtained by analytical
calculations and simulations, these constraints on the former parameters
can then be matched onto the latter.
The SGWB resulting from bubble collisions is commonly modelled by a
broken-power-law (BPL) spectrum, whereas spectra originating from sound
waves and turbulence have been described by double broken power laws
(DBPL) in recent years \cite{Caprini:2024hue,Caprini:2024gyk}.

In order to derive constraints from current LVK data, we adopt the approach
proposed in Refs.~\cite{Romero:2021kby,Badger:2022jzg}, which is basically
testing for the hypothesis that the data contain a signal from a phase
transition,
using the templates from Ref.~\cite{Caprini:2024hue} for the GW spectra,
which we will describe in the following.

\subsubsection{Bubble collisions}
For GW production from bubble collisions and highly relativistic
fluid shells, which dominate for FOPT with vacuum domination, i.e.,
$\alpha \gg 1$, we use a BPL,
\begin{equation} \label{eq:BPL}
\Omega_{\rm GW}^{\rm BPL}(f) =
\Omega_b
\left(\frac{f}{f_b}\right)^{n_1}
\left[
\frac{1}{2}
+
\frac{1}{2}
\left(\frac{f}{f_b}\right)^{a_1}
\right]^{\frac{n_2-n_1}{a_1}},
\end{equation}
where $f_b$ is the break frequency.
From the asymptotic behaviour
\begin{equation}
\Omega_\text{GW}^\text{BPL} \propto
\begin{cases}
f^{n_1}, & f \ll f_b,\\
f^{n_2}, & f \gg f_b ,
\end{cases}
\label{eq:BPL_asymptotics}
\end{equation}
we see that the spectral slopes $n_1>0$ and $n_2<0$ determine the rise
at low frequencies and the fall-off at large frequencies, respectively. The
smoothness parameter $a_1$ governs the width of the transition region.
In our analysis, we fix these parameters to the values given in the first
line of \Tabref{tab:spectral_indices}, which were determined by simulations
\cite{Lewicki:2022pdb}.

The amplitude and frequency of the peak of the spectrum, denoted by
$\Omega_p$ and $f_p$, respectively, are different from $\Omega_b$ and
$f_b$ in general and given by Eqs.~(\ref{eq:BPL_fp},\,\ref{eq:BPL_Omegap}).
They are related to the parameters characterising the FOPT by
Eqs.~(\ref{eq:Ktilde}--\ref{eq:Astrong}), which can be summarised as
\begin{eqnarray}
	h^2 \Omega_p &\simeq&
	8.2 \times 10^{-7}
	\left( \frac{100}{g_*(T_*)} \right)^{1/3}
	\left( \frac{\alpha}{1+\alpha} \right)^2
	\left( \frac{H_*}{\beta} \right)^2 ,
\label{eq:BPL_peakamplitude}
\\
	f_p &\simeq&
	1.8 \times 10^{-6} \,\text{Hz}
	\left( \frac{g_*(T_*)}{100} \right)^{1/6}
	\left( \frac{T_*}{100\,\text{GeV}} \right)
	\left( \frac{\beta}{H_*} \right) ,
\label{eq:BPL_peakfrequency}
\end{eqnarray}
where $g_*$ is the effective number of degrees of freedom in the thermal
bath.

\subsubsection{Sound waves}
Sound waves and MHD turbulence dominate GW production for $\alpha < 1$.
They produce a spectrum that we model as a DBPL,
\begin{equation}
\Omega_\text{GW}^\text{DBPL}(f) = \Omega_2\,\frac{S(f)}{S(f_2)}
\label{eq:DBPL}
\end{equation}
with the shape function
\begin{equation} \label{eq:Sf}
S(f) = \mathcal{N} \left(\frac{f}{f_1}\right)^{n_1}
\left[1+\left(\frac{f}{f_1}\right)^{a_1}\right]^{-\frac{n_1-n_2}{a_1}}
\left[1+\left(\frac{f}{f_2}\right)^{a_2}\right]^{-\frac{n_2-n_3}{a_2}} ,
\end{equation}
where the normalisation factor $\mathcal{N}$ ensures that
$\int S(f) \, d\log f = 1$.
The break frequencies are $f_1 < f_2$,
and $\Omega_2$ is the GW density at the second break.
Again, we consider the asymptotic behaviour,
\begin{equation}
\Omega_\text{GW}^\text{DBPL} \propto
\begin{cases}
f^{n_1}, & f \ll f_1,\\
f^{n_2}, & f_1 \ll f \ll f_2,\\
f^{n_3}, & f \gg f_2 .
\end{cases}
\label{eq:DBPL_asymptotics}
\end{equation}
Thus, the increase at frequencies below the peak is governed by the
positive parameters $n_1$ and $n_2$, while the decrease at higher
frequencies is determined by $n_3<0$.

For sound waves,
we use the values given in the second line of \Tabref{tab:spectral_indices} for these
parameters as well as the smoothness parameters $a_1$ and $a_2$ \cite{Jinno:2022mie}.
The other parameters of the DBPL spectrum 
are related to the thermodynamical parameters via Eqs.~(\ref{eq:SW_f1}--\ref{eq:AuxOmegas}),
which yield
\begin{equation} \label{eq:f2byf1}
	\frac{f_2}{f_1} \simeq 2.5 \, \Delta_w^{-1} \,.
\end{equation}
Compared to the BPL case, the bubble wall velocity $v_w$ becomes relevant,
entering via the parameter
\begin{equation} \label{eq:DefDeltaw}
\Delta_w = \frac{v_\text{shell}}{\max(v_w,c_s)}
\end{equation}
that is related to the sound shell thickness,
where $v_\text{shell}=|v_w-c_s|$ is the shell velocity
and $c_s=1/\sqrt{3}$ is the sound speed.

The peak frequency $f_p$ is close but not equal to the second break
frequency $f_2$.  For $f_p, f_2 \gg f_1$, the peak frequency and
amplitude can be approximated by relations analogous to 
Eqs.~(\ref{eq:BPL_fp},\,\ref{eq:BPL_Omegap}) \cite{Caprini:2024hue}.
For the general case without a strong hierarchy between the break
frequencies, we will now derive precise relations that are used in our
numerical analysis.

The peak frequency is obtained by maximizing $S(f)$, or equivalently $\ln S(f)$, leading to
\begin{equation}
n_1
+(n_2-n_1)\frac{\left(\frac{f_p}{f_1}\right)^{a_1}}{1+\left(\frac{f_p}{f_1}\right)^{a_1}}
+(n_3-n_2)\frac{\left(\frac{f_p}{f_2}\right)^{a_2}}{1+\left(\frac{f_p}{f_2}\right)^{a_2}}=0
\,.
\label{eq:peak_exact}
\end{equation}
Introducing
\begin{equation}
z=\frac{f_p}{f_2} \,,
\end{equation}
the peak equation \eqref{eq:peak_exact} becomes
\begin{equation}
n_1
+(n_2-n_1)
\frac{\left(\frac{f_2}{f_1}\right)^{a_1}z^{a_1}}
{1+\left(\frac{f_2}{f_1}\right)^{a_1}z^{a_1}}
+(n_3-n_2)
\frac{z^{a_2}}{1+z^{a_2}}
=0 \,.
\label{eq:peak_z}
\end{equation}
This equation is transcendental and, in general, must be solved numerically.

However, in the limit $f_2/f_1\rightarrow\infty$, the first transition has already been completed before the second one is reached,
and thus \eq{eq:peak_z} reduces to
\begin{equation}
n_2+(n_3-n_2)\frac{z^{a_2}}{1+z^{a_2}}=0 \,,
\end{equation}
whose solution is
\begin{equation}
z_0=
\left(-\frac{n_2}{n_3}\right)^{1/a_2}
\end{equation}
or equivalently
\begin{equation} \label{eq:fp0}
f_p^{(0)}
=f_2
\left(-\frac{n_2}{n_3}\right)^{1/a_2} ,
\end{equation}
which is the approximation used in Ref.~\cite{Caprini:2024hue}.

For finite values of $f_2/f_1$, the first transition is not fully
saturated. Rather than replacing the first transition factor in
\eq{eq:peak_z} by unity, one may evaluate it at the zeroth-order solution,
\begin{equation} \label{eq:R}
R\equiv
\frac{\left(\frac{f_2}{f_1}\right)^{a_1}
z_0^{a_1}}
{1+
\left(\frac{f_2}{f_1}\right)^{a_1}
z_0^{a_1}}
=
\frac{
\left(\frac{f_2}{f_1}\right)^{a_1}
\left(-\frac{n_2}{n_3}\right)^{a_1/a_2}}
{
1+
\left(\frac{f_2}{f_1}\right)^{a_1}
\left(-\frac{n_2}{n_3}\right)^{a_1/a_2}
}
\end{equation}
and plug it into \eq{eq:peak_z} to obtain the more accurate approximation
\begin{equation}
n_1
+(n_2-n_1) R
+(n_3-n_2)\frac{z^{a_2}}{1+z^{a_2}}
\simeq0 \,.
\end{equation}
This is solved by
\begin{equation} \label{eq:zapprox}
z^{a_2} \simeq
\frac{-n_1+(n_1-n_2)R}{n_3+n_1-n_2+(n_2-n_1)R} \,,
\end{equation}
and therefore
\begin{equation} \label{eq:fpapprox}
f_p
\simeq
f_2
\left[
\frac{-n_1+(n_1-n_2)R}{n_3+n_1-n_2+(n_2-n_1)R}
\right]^{1/a_2}.
\end{equation}
This expression interpolates smoothly between the asymptotic solution
\eqref{eq:fp0} for $f_2/f_1\rightarrow\infty$ ($R\to1$)
and the regime of moderate hierarchy, where the finite value of $R-1$ accounts for the incomplete saturation of the first spectral transition.

Having determined the peak frequency, we can find the peak amplitude
$\Omega_p$ and its relation to the break amplitude $\Omega_2$
by considering \eq{eq:DBPL} at $f=f_p$,
\begin{equation} \label{eq:OmegaP2}
	\Omega_p =
	\Omega_\text{GW}^\text{DBPL}(f_p) =
	\Omega_2 \, \frac{S(f_p)}{S(f_2)} \,.
\end{equation}
Conveniently, this relation is independent of the normalisation factor
$\mathcal{N}$.

\subsubsection{Magnetohydrodynamic turbulence}
As mentioned, the shape of the GW spectrum produced by MHD turbulence is 
modelled by a DBPL as well, but now the shape parameters are given by the
third line of \Tabref{tab:spectral_indices} \cite{Caprini:2024hue}, and
the relations with the thermodynamical parameters are given by
Eqs.~(\ref{eq:Omegas}--\ref{eq:AMHD}).  They determine the ratio of the
break frequencies as
\begin{equation}
	\frac{f_2}{f_1} \simeq 5.1 \, \Omega_s^{-1/2} \,,
\end{equation}
where the energy in the GW source stemming from turbulence, $\Omega_s$,
appears as a new parameter.  It is proportional to the parameter
$\varepsilon$ introduced in \Secref{sec:ThermoPars}, see \eq{eq:Omegas}.
As MHD turbulence yields a subdominant contribution to the GW signal, we
include it only in the analysis in terms of thermodynamical parameters.

\subsection{Astrophysical foregrounds}
Unresolved compact binary coalescences (CBC) produce a well-known
astrophysical foreground that must be modelled or subtracted when
searching for cosmological SGWB\@.  Following previous analyses
\cite{Abbott:2021xyo,Romero:2021kby,Badger:2022jzg,LVK:2025O4a}, we
assume a power-law spectrum at frequencies relevant for LVK observations,
\begin{equation} \label{eq:CBC}
	\Omega_\text{GW}^\text{CBC}(f) =
	\Omega_\text{ref} \left( \frac{f}{f_\text{ref}} \right)^\frac{2}{3} ,
\end{equation}
where $f_\text{ref}$ is a reference frequency typically chosen within
the frequency range probed by the experiment, conventionally
$f_\text{ref}=25$\,Hz, and serves to normalise the spectrum.
The amplitude of the SGWB from CBC at this frequency is denoted by
$\Omega_\text{ref}$.

Other astrophysical contributions like rotating neutron stars,
magnetars, and core collapse supernovae can be modelled with specific
energy spectrum integrals over cosmic source rates (see, e.g.,
\cite{Schneider:2010ks,Kuroyanagi:2018csn,Christensen:2018iqi,Abbott:2021xyo}
and references therein)
but are not included here, as they are assumed to be less important than
CBC.

\section{Bayesian inference and LVK stochastic searches} \label{sec:Bayes}
\subsection{Current limits from LVK stochastic searches} \label{sec:Searches}
The LVK detectors searched for isotropic and anisotropic stochastic
backgrounds using data from observing runs O1--O4a.  Employing the
cross-correlation between the different detectors to distinguish an SGWB
from noise, they reported upper limits on $\Omega_{\rm GW}(f)$ for a
power-law spectrum with a range of spectral indices~\cite{Abbott:2021xyo}.
The combined O3 result (including O1 and O2) has been recast into
constraints on geometrical and thermodynamical parameters associated
with FOPT by assuming templates for $\Omega_\text{GW}(f)$
\cite{Romero:2021kby,Badger:2022jzg}.

The LVK collaborations have since extended this search to the first part
of the fourth observing run (O4a), combining the new data with the full
O1--O3 dataset \cite{LVK:2025O4a,LVK:2025O4aData}. As in the earlier
runs, the O4a cross-correlation analysis finds no statistically
significant isotropic stochastic background. For a CBC-like spectrum
with spectral index $2/3$, the search constrains $\Omega_\text{GW}^\text{CBC}(25\,{\rm Hz})\leq2.0\times10^{-9}$ at $95\%$ credibility, a factor of $1.7$ improvement over the corresponding O3 limit, while a scale-invariant background is constrained to $\Omega_{\rm GW}(25\,{\rm Hz})\leq2.8\times10^{-9}$, a factor of $2.1$ improvement.

In the following, we recast both the older dataset and the latest one
into constraints on FOPT parameters by assuming the templates for
$\Omega_{\rm GW}(f)$ described above.  We will refer to the O1--O3 data
as ``O3'' and to the most comprehensive available dataset from runs O1--O4a
as ``O4a''.

\subsection{Likelihood model}
We adopt a simplified Gaussian likelihood for cross-correlation
estimators $C_{IJ}(f)$ between detectors $I$ and $J$ \cite{Mandic:2012pj},
\begin{equation}
\label{eq:lkdef}
  \mathcal{L}(\theta) = p(\{C_{IJ}(f)\}|\theta)\propto \exp\!\bigg[-\frac{1}{2}\sum_{I<J}\sum_f \frac{\big[C_{IJ}(f)-\Gamma_{IJ}(f)\,\Omega_{\rm GW}(f;\theta)\big]^2}{\sigma^2_{IJ}(f)}\bigg],
\end{equation}
where $\Gamma_{IJ}(f)$ is the overlap reduction function%
\footnote{To the best of our knowledge, the effect of the overlap
function is taken into account in the publicly released LVK data.
Consequently, we use $\Gamma_{IJ}=\delta_{IJ}$.}
and $\sigma^2_{IJ}(f)$ the variance for the estimator at frequency bin $f$.
This form of the likelihood is commonly used when interpreting LVK cross-correlation results (see, e.g., \cite{Abbott:2021xyo,Badger:2022jzg}).

Based on this likelihood, we perform a Bayesian analysis to determine
the allowed regions for the fit parameters $\theta$, for which we
consider three different sets that will be described in the following sections.
To sample the parameter space, we use nested
sampling~\cite{Skilling:2004pqw,Skilling:2006gxv} as implemented in the
\texttt{dynesty} package~\cite{Speagle:2019ivv,koposov2024}. Each run
starts with a static nested sampling stage, which maps out the overall structure of the likelihood, and then switches to dynamic nested sampling~\cite{Higson:2018cwj}, which reallocates the sampling effort to the regions that contribute most to the posterior. We chose this combination because it gives us a well-sampled posterior and the Bayesian evidence from a single run, without having to fix the sampling resolution in advance.

\subsection{Nested-sampling convergence and evidence stability}
\label{sec:nested_logz_convergence}

To assess the numerical stability of the nested-sampling inference, we compare independent \texttt{dynesty} runs performed with different sampling resolutions. For a fixed likelihood $\mathcal{L}(d\mid\theta)$ and prior
$\pi(\theta)$, the Bayesian evidence is
\begin{equation}
\label{eq:logz_evidence_definition}
Z = \int \mathcal{L}(d\mid\theta)\,\pi(\theta)\,d\theta .
\end{equation}
Each run provides an estimate $\log Z_i$ and an associated numerical uncertainty $\sigma_i\equiv\sigma_{\log Z_i}$. For two independently initialised runs $a$ and $b$, we define the evidence-consistency statistic
\begin{equation}
\label{eq:logz_convergence_statistic}
R_Z^{(a,b)} \equiv
\frac{\left|\log Z_a-\log Z_b\right|}
{\sqrt{\sigma_a^2+\sigma_b^2}}.
\end{equation}
We adopt $R_Z^{(a,b)}\leq 1$ as a practical one-standard-deviation consistency check. This criterion tests whether the difference between the reported evidence estimates is compatible with their combined numerical uncertainties; it is not, by itself, a proof of sampler convergence. In particular, a small $R_Z$ is insufficient when both estimates have large
uncertainties. We therefore also inspect the stability of the marginalised posterior distributions and their credible regions as the sampling effort is increased.

The comparison is carried out only between runs with the same dataset, likelihood normalisation, physical model, and prior family and bounds; random seeds and sampling resolution may differ. For dynamic nested sampling, the number of initial live points is one measure of resolution, while the subsequent allocation of samples and stopping settings also affect
precision. Evidence comparisons between different prior choices are not interpreted as convergence tests, since the evidence itself depends on the prior. We use this diagnostic to assess numerical stability, separately from the prior-sensitivity measures considered below.

\subsection{Priors and parameter estimation for geometrical parameters}
\label{sec:pestimation_geometrical}

Bayesian analyses rely on priors to infer posterior probabilities for
the parameters that are to be constrained by data.  Consequently, in
order to obtain statistically robust results from such an analysis it is
indispensable to understand and justify the choice of priors and to
assess how it affects the output \cite{gelman:2013bayesian,Romero-Rodriguez:2024ldc}.
In physics applications, physical intuition or measurements should guide
the choice of appropriate models and parameter priors.
Sensible priors are chosen to span physically motivated ranges
(e.g., the region of detector sensitivity) and to follow a physically
motivated distribution.
However, in many cases, it is precisely the lack of well-justified
priors that is a challenging part of the analysis (see, e.g.,
Ref.~\cite{Efstathiou:2008ed} for a discussion in the context of
cosmology).

Previous analyses constraining the parameters of FOPT from LVK data
\cite{Romero:2021kby,Badger:2022jzg} used only the Jeffreys prior \cite{jeffreys1961}%
\footnote{That is, the log-uniform (scale-invariant) prior, defined for strictly positive scale parameters, which we review in the next subsection.}
without considering alternatives.
Although this is a conservative prior, Bayesian analysis relies on
testing different priors, and for this reason, we look for additional
well-motivated priors tailored to the LVK data. In the remainder of this section, we
justify the priors that we use for the analysis of the density of GW
fitted in terms of \eq{eq:OmegaGW_gen_temp} for the different templates
described in \Secref{sec:templates}.

For the analysis in terms of geometrical parameters, we assume the
GW spectrum potentially contained in the data to be the sum of an
astrophysical contribution of the form \eqref{eq:CBC} and a contribution
from a FOPT, described either by a BPL (dominant bubble collisions,
Eq.~\ref{eq:BPL}) or a DBPL (dominant sound waves, Eq.~\ref{eq:DBPL}).
Consequently, the geometrical fit parameters characterising the GW spectrum are
\begin{itemize}
\item $\Omega_\text{ref}$, the amplitude of the astrophysical foreground
at the reference frequency $f_\text{ref}=25\,$Hz,
\item $f_p$, the peak frequency of the SGWB produced by the FOPT,
\item $\Omega_p$, the corresponding peak amplitude
 $\Omega_\text{GW}^\text{BPL}(f_p)$ or
 $\Omega_\text{GW}^\text{DBPL}(f_p)$, and
\item $\Delta_w$ from \eq{eq:DefDeltaw} in the case of dominant sound
waves; this parameter depends on the bubble wall velocity and determines
the position of the first frequency break in the DBPL template.
\end{itemize}

\subsubsection{Jeffreys prior} 
For strictly positive scale parameters (e.g., cross sections, masses, or
amplitudes), the log-uniform (improper Jeffreys) prior takes the form
\begin{equation}
\label{eq:jeffreys} 
\pi(\theta) \propto \frac{1}{\theta\left[\log\theta_{\rm max}-\log\theta_{\rm min}\right]}
\equiv \logU(\theta_\text{min},\theta_\text{max})
\,, \quad \theta\in(\theta_{\min},\theta_{\max}) \,.
\end{equation} 
This prior expresses ignorance about the order of magnitude of $\theta$ rather than its absolute value. It is invariant under rescaling of units, but it is improper without bounds, so one must choose finite limits $\theta_{\min}$ and $\theta_{\max}$; the choice of these limits affects evidence calculations.%
\footnote{In general, the Jeffreys prior is defined by
$
\pi_J(\theta)
\;\propto\;
\sqrt{
-\left.\left\langle
\frac{\partial^2}{\partial \theta^2}
\log p(X \mid \theta)
\right\rangle\right|_{p(X \mid \theta)}}
$, that is, the average over the data $X$ assuming the model with likelihood $p(X \mid \theta)$ \cite{Jeffreys,Jaynes2003}. In practice, the prior can be written as in \eq{eq:jeffreys}.}
We can use this prior for all fit parameters.
For some parameters, such as $\Delta_w$, we use its linear counterpart
(uniform prior), which is obtained by replacing $\log\theta\to\theta$
and denoted as U$(\theta_\text{min},\theta_\text{max})$.

\subsubsection{Detector-sensitivity-based prior}
When external measurements, theoretical expectations, or calibration
data provide information about plausible values of a parameter, a
Gaussian (normal) prior is a convenient choice:
\begin{equation}\label{eq:gauss} \pi(\theta) = \frac{1}{\sqrt{2\pi}\sigma}\exp\left[-\frac{(\theta-\mu)^2}{2\sigma^2}\right]
\equiv \text{N}(\mu,\sigma) \,.
\end{equation}
This prior is useful for encoding existing constraints (e.g., a
previously measured particle mass or calibration offset). Choosing a
large standard deviation $\sigma$ yields a weakly informative prior;
conversely, a small $\sigma$ encodes strong prior belief. When physical
parameters must remain in a limited range, one often uses a truncated
normal.

The Jeffreys prior gives an implicit equal treatment (on a logarithmic
scale) over the entire parameter range.  Taking into consideration the
detector sensitivity, this is not well-justified, since it gives equal
probability to parameter regions where sensitivity is poor
and to regions of peak sensitivity.
 For example, in the context of LVK data analyses, using a Jeffreys prior
$\pi(f_p/\text{Hz})=\logU(10^{-2},10^3)$ for the peak frequency
\cite{Badger:2022jzg} can be considered naive, given that the
detectors have negligible sensitivity below about $10$\,Hz, where, on a
logarithmic scale, three of the five decades spanned by this prior (that
is, $60\%$ of its probability) are located.
This prior also ignores the detector noise, which varies by orders of
magnitude across the frequency band. Last but not least, the
Hanford-Livingston correlation coefficient 
varies significantly with frequency; this is not reflected in the
Jeffreys prior, but is taken into account in the released public data and
encoded in \eq{eq:lkdef}.

Therefore, we consider instead a detector-sensitivity-based prior, which
is a log-normal prior for $f_p$ centered at the LVK peak sensitivity
$f_\text{sens} \simeq 40$\,Hz,
\begin{equation}
\pi_\text{sens}(f_p) \propto \frac{1}{f_p}
\exp\left[-\frac{(\log_{10}(f_p/f_\text{sens}))^2}{2\sigma_{\log}^2}\right]
\equiv \logN\bigl(\log_{10}(f_\text{sens}/\text{Hz}),\sigma_{\log}\bigr)
\end{equation}
with $\sigma_{\log} = 0.8$ (corresponding to the $1$-$\sigma$ bandwidth
in $\log_{10}$ units).
Equivalently, the prior of the logarithm 
of the peak frequency is given by a Gaussian distribution,
\begin{equation}
\pi_\text{sens}\bigl(\log_{10}(f_p/\text{Hz})\bigr) = \text{N}(\log_{10} 40,\,0.8) \,,
\end{equation}
truncated to the frequency interval
$10\,\mathrm{Hz} \leq f_p \leq 1000\,\mathrm{Hz}$ where
LVK is sensitive.

We will use this prior for the peak frequency, and also in the two prior
sets we describe in the following.  In those sets, we vary the priors on
the SGWB amplitudes in addition.

\subsubsection{Model-based prior}
Although analyses typically adopt log-uniform or uniform priors for the
peak energy density $\Omega_p$ of a FOPT signal~\cite{Abbott:2021xyo,Romero:2021kby,Badger:2022jzg,LVK:2025O4a},
the distribution of GW amplitudes in realistic particle physics models
need not be log-uniform.
For example, Ref.~\cite{Alanne:2019bsm} maps the parameters of the phase
transition in a real-singlet extension of the Standard Model onto the peak
amplitude and frequency of the stochastic signal, and shows which region of
this observable space the model can populate.

Investigations of extended scalar sectors such as that of two-Higgs-doublet
models showed that viable parameter space points produce a wide range of
$\Omega_p$ values but not an extremely steep distribution~\cite{Zhou:2020irf}.
Therefore, we adopt the prior
\begin{equation}
\pi(\Omega_p)\propto \Omega_p^{-0.7}
\end{equation}
as a model-inspired prior reflecting the distribution of GW amplitudes
that emerge in realistic FOPT scenarios with signals strong enough to be
tested, in contrast to purely scale-invariant choices.
This prior biases the signal amplitude toward smaller values, consistent
with both theoretical expectations and the absence of a detection, but
includes a fairly heavy tail for strong signals.

In addition, we consider improved estimates of the CBC population based
on detections of CBC events and LVK O3 constraints \cite{Abbott:2021xyo}.
As a consequence, in the model-based prior set we narrow the range of
the log-uniform prior for the amplitude of the astrophysical foreground.

\subsubsection{Mixed prior}
In order to further illustrate the potential consequences of a prior
change, we use a mixed prior where
we retain the broad Jeffreys prior for the astrophysical contribution
$\Omega_\text{ref}$ but change the prior for the amplitude of the FOPT
contribution to a steep power law and a very small lower limit,
\begin{equation}
\pi(\Omega_p) \propto \Omega_p^{-3} \,,
\quad
\Omega_p \in (10^{-14}, 10^{-6}) \,.
\end{equation}
This steep distribution is meant as a deliberately extreme counterpart to
the model-based prior rather than as the result of a scan of a specific
model. It represents a population dominated by weak, sound-wave-dominated
transitions, i.e., many weak and few strong ones.  For such transitions
the amplitude is strongly suppressed; for instance, \eq{eq:SW_Omegint}
gives $h^2\Omega_p$ of order $10^{-13}$ for $K\sim H_*R_*\sim10^{-2}$,
which explains the very small lower limit of the range.  This prior thus
tests the sensitivity of the posteriors in the direction opposite to that
of the model-based prior.

In \Tabref{tab:PriorsGeom}, we present an overview of the combinations
of priors that we use for constraining the geometrical parameters of the
templates presented in \Secref{sec:templates} by the O3 data.  For the
parameter $\Delta_w$, we do not have a preferred range from theory and
do not expect tight constraints from the data either; therefore, we
simply adopt a uniform prior in all cases.

\begin{table}
\small
\centering
\setlength{\tabcolsep}{4pt}
\renewcommand{\arraystretch}{1.15}
\begin{tabular}{p{1.8cm}p{2.8cm}p{3.0cm}p{3.5cm}p{3.0cm}}
\hline
\textbf{Parameter} & \textbf{Jeffreys} &
\shortstack{\textbf{Detector-}\\\textbf{Sensitivity-}\\\textbf{Based}} &
\textbf{Model-Based} & \textbf{Mixed} \\
\hline
$\boldsymbol{\Omega_\textbf{ref}}$
& LogU$[10^{-10},10^{-7}]$
& LogU$[10^{-10},10^{-7}]$
& LogU$[5{\times}10^{-10},\,2{\times}10^{-8}]$
& LogU$[10^{-10},10^{-7}]$ \\
\hline
$\boldsymbol{\Omega_p}$
& LogU$[10^{-9},10^{-4}]$
& LogU$[10^{-9},10^{-4}]$
& \shortstack{$\propto \Omega_p^{-0.7}$\\ on $[10^{-11},10^{-4}]$}
& \shortstack{$\propto \Omega_p^{-3}$\\ on $[10^{-14},10^{-6}]$} 
\\
$\boldsymbol{f_p/\textbf{Hz}}$
& LogU$(10^{-2},10^{3})$
& LogN$(\log_{10}40,\,0.8)$
& LogN$(\log_{10}40,\,0.8)$
& LogN$(\log_{10}40,\,0.8)$ \\
& & \\
\hline
$\boldsymbol{\Delta_w}$ & U$(0,1)$ & U$(0,1)$ & U$(0,1)$ & U$(0,1)$ \\
\hline
\hline
\end{tabular}
\caption{Prior specification for the O3 data analysis in terms of
geometrical parameters. $\Omega_\text{ref}$ is associated with the astrophysical foreground, while $\Omega_p$ and $f_p$ parameterise the signal from a cosmological phase transition. 
 Log-uniform (Jeffreys), log-normal, and uniform priors are denoted by LogU, LogN, and U, respectively.
Here ``U'', corresponds to a uniform prior in the corresponding range.
\label{tab:PriorsGeom}
}
\end{table}

\subsubsection{Prior modifications for the O4a analysis}
The LVK O4a isotropic SGWB analysis searches for an SGWB using power-law
and other spectral models, different from the BPL or DBPL in our analysis.
Consequently, there is no unique set of ``official O4a priors'' for the
parameters $\Omega_p$ and $f_p$.
The LVK analysis finds no statistically significant evidence for an
isotropic SGWB and provides substantially improved constraints on the
amplitude compared with previous observing runs \cite{LVK:2025O4a}.

\paragraph{Jeffreys prior}
To employ these data to constrain the geometrical BPL or DBPL
parameters, we adopt a reasonable broad choice for the Jeffreys prior
that extends to smaller amplitudes compared to the O3 analysis and is
given in the second column of \Tabref{tab:PriorGeom_O4a}.
For the amplitude $\Omega_\text{ref}$ of the astrophysical foreground,
we use the same prior as the LVK analysis.
The lower limit of $10^{-13}$, also adopted for $\Omega_p$, should be understood as a deliberately broad prior rather than as an
experimental sensitivity threshold. It allows the posterior to extend
well below the region where the LVK detectors are sensitive, thereby
avoiding an artificial truncation.  
Conversely, the conservative upper limits of $10^{-6}$ and $10^{-5}$,
respectively, avoid an artificial upper cutoff close to the expected
astrophysical contribution.

For the peak frequency, we retain the same broad range as in the O3 analysis.
In the spirit of the Jeffreys prior, it is not restricted to the most
sensitive part of the LVK band.

\begin{table}
\small
\centering
\setlength{\tabcolsep}{4pt}
\renewcommand{\arraystretch}{1.15}
\begin{tabular}{p{1.8cm}p{2.8cm}p{3.0cm}p{3.5cm}p{3.0cm}}
\hline
\textbf{Parameter} & \textbf{Jeffreys} &
\shortstack{\textbf{Detector-}\\\textbf{Sensitivity-}\\\textbf{Based}} &
\textbf{Model-Based} & \textbf{Mixed} \\
\hline
$\boldsymbol{\Omega_\textbf{ref}}$
& LogU$[10^{-13},10^{-6}]$
& LogU$[10^{-13},10^{-6}]$
& LogU$[2{\times}10^{-10},\,5{\times}10^{-9}]$
& LogU$[10^{-13},10^{-6}]$ \\
\hline
$\boldsymbol{\Omega_p}$
& LogU$[10^{-13},10^{-5}]$
& LogU$[10^{-10},10^{-4}]$
& \shortstack{$\propto\Omega_p^{-0.7}$\\ on $[10^{-12},10^{-4}]$}
& \shortstack{$\propto\Omega_p^{-3}$\\ on $[10^{-14},10^{-6}]$} \\
\hline
$\boldsymbol{f_p/\textbf{Hz}}$
& LogU$[10^{-2},10^{3}]$
& LogN$(\log_{10}25,\,0.8)$
& LogN$(\log_{10}25,\,0.8)$
& LogN$(\log_{10}25,\,0.8)$ \\
\hline
$\boldsymbol{\Delta_w}$ & U$(0,1)$ & U$(0,1)$ & U$(0,1)$ & U$(0,1)$ \\
\hline
\end{tabular}
\caption{Prior specification for the O4a analysis in terms of
geometrical parameters.  The notation is the same as in
\Tabref{tab:PriorsGeom}.
\label{tab:PriorGeom_O4a}
}
\end{table}

\paragraph{Detector-sensitivity-based prior}
For the O4a analysis, we center the sensitivity-based prior for $f_p$ on
the frequency $f_\text{sens}=25\,$Hz, following the LVK stochastic
background analysis of the first part of O4.  This differs from
the $40\,$Hz frequency used in our prior choice for O3.  The new choice
is particularly
appropriate for the O4a analysis because the LVK results are
reported in terms of $\Omega_{\rm GW}(25\,{\rm Hz})$.

\paragraph{Model-based prior}
The O4a results provide an updated astrophysical reference scale:
combining the stochastic-background analysis with the GWTC-4
population inference, the expected total CBC amplitude is
\begin{equation}
    \Omega_{\rm CBC}(25\,{\rm Hz})
    =
    0.9^{+1.1}_{-0.5}\times10^{-9}
\end{equation}
at 90\% credibility, with binary black holes providing the dominant
contribution \cite{LVK:2025O4a}.

We use this astrophysical prediction as motivation for the model-based
prior, which centers the range for $\Omega_\text{ref}$ on the central
value of $9\times10^{-10}$, rather than interpreting it as an
LVK-reported prior.

The shallow power-law prior for $\Omega_p$ is almost the same as for the O3
analysis, with only the lower limit reduced by an order of magnitude.

\paragraph{Mixed prior}
Here we retain the steep power law for $\Omega_p$ with the same range as
in the O3 analysis.  The priors for $\Omega_\text{ref}$ and $f_p$ are
adjusted as for the Jeffreys and the detector-sensitivity-based prior,
respectively.

\subsection{Priors and parameter estimation for thermodynamical parameters \label{sec:pestimation_thermo}}
For the analysis in terms of thermodynamical parameters, we follow the same
approach as for the geometrical parameters in principle.  In this case we
define two prior sets, where the first choice is again the Jeffreys prior
and the other one takes into account the experimental sensitivity.
For this second choice, which we call ``conditional prior'', we start
from the conditions
\begin{align}
	20\,\text{Hz} &< f_p < 1000\,\text{Hz} \,,
	\label{eq:FrequencyConstraint}
	\\
	\Omega_p h^2 &> 1\times10^{-9}
	\label{eq:AmplitudeRequirement}
\end{align}
for the O4a analysis, in order to restrict the sampled region to signals
that the data can actually test.  The first constraint keeps only points
whose peak frequency lies in the band in which the LVK cross-correlation
data are sensitive; for peaks outside this band the likelihood is to a
good approximation independent of the phase transition parameters, so
such points would only fill the posterior with prior volume.  The
requirement on $\Omega_p$ ensures that we avoid sampling points whose
peak amplitude is far below the sensitivity of the respective data set.
The threshold in \eq{eq:AmplitudeRequirement} is of the order of the
upper limit on the amplitude of an SGWB at $25$\,Hz reported by LVK,
cf.\ \Secref{sec:Searches}.  For the O3 data, this threshold is raised
to $6 \times 10^{-9}$.

\paragraph{BPL spectrum}
Assuming a cosmological SGWB dominantly produced by bubble collisions,
whose spectrum is thus described by the BPL of \eq{eq:BPL}, we use the
relevant thermodynamical parameters $T_*$, $\alpha$, and $\beta/H_*$
defined in \Secref{sec:ThermoPars} as fit parameters.  For simplicity,
we set $g_*(T_*)=100$.  In addition, the amplitude $\Omega_\text{ref}$
of the astrophysical foreground is used as before.
We present the two prior choices for these parameters in \Tabref{tab:analysis_nucl_O4a}.

\begin{table}
\centering
\begin{tabular}{ccc}
\hline
\textbf{Parameter} & \textbf{Jeffreys} & \textbf{Conditional} (O4a Detector-Based)\\
\hline
$\Omega_\text{ref}$ &
$\logU(10^{-13},10^{-6})$ & $\logU(10^{-13},10^{-6})$ \\
\hline
$\alpha$ &
LogU$(10^{-3},5)$ &
LogU$(0.036,5)$ \\
\hline
$\beta/H_*$ &
LogU$(1,10^4)$ &
LogU$\left(1,\, 28.6\,\frac{\alpha}{1+\alpha}\right)$ \\
\hline
$T_*/\text{GeV}$ &
LogU$(10^2,10^8)$ &
LogU$\left(1.1\times10^9\,(\beta/H_*)^{-1},\, 5.5\times10^{10}\,(\beta/H_*)^{-1}\right)$ \\
\hline
\end{tabular}
\caption{Prior specification for the O4a data analysis in terms of the
thermodynamical parameters $\alpha$, $\beta/H_*$, and $T_*$ described in
\Secref{sec:ThermoPars} for a BPL spectrum produced by bubble
collisions.  As before, $\Omega_\text{ref}$ parameterises the
astrophysical foreground.
\label{tab:analysis_nucl_O4a}}
\end{table}

The lower edge of the conditional prior on $\alpha$ follows from
\eq{eq:BPL_peakamplitude} in combination with the amplitude requirement
\eqref{eq:AmplitudeRequirement} and the condition for the phase
transition to complete, $\beta/H_*>1$.
For a given value of $\alpha$, \eq{eq:BPL_peakamplitude} also allows to
translate the amplitude requirement into an upper bound on $\beta/H_*$,
which is the upper prior edge for this parameter.
Finally, for each value of $\beta/H_*$ within its prior range, the
frequency constraint \eqref{eq:FrequencyConstraint} yields the prior
range for $T_*$ via \eq{eq:BPL_peakfrequency}.

\paragraph{DBPL spectrum}
In the analysis constraining an SGWB from weaker FOPT that is
characterised by the DBPL spectrum \eqref{eq:DBPL}, we take into account
both production mechanisms, i.e., the dominant sound waves and the
subdominant MHD turbulence.  Hence, we have two additional parameters,
the bubble wall velocity $v_w$ and the turbulent energy fraction
$\varepsilon$.
However, in this case, the relations \eqref{eq:DefDeltaw} and
(\ref{eq:H_star_R_star}--\ref{eq:AuxOmegas}) between the thermodynamical
parameters and the parameters of the DBPL spectrum are fairly
complicated.  In particular, Eqs.~\eqref{eq:DefDeltaw} and
\eqref{eq:H_star_R_star} as well as the discussion about the efficiency
factor $\kappa(\alpha,v_w)$ in App.~\ref{sec:soundwaves} show that
sampling over $\alpha$, $\beta/H_*$, and $v_w$ would require
distinguishing several cases depending on the size of the wall velocity
relative to the sound and Jouguet velocities (or adding yet another
parameter to sample).  Because of that, it is more convenient to trade
these fit parameters for suitable combinations.  This leads to the
following fit parameters for the thermodynamical DBPL analysis in
addition to $\Omega_\text{ref}$:
\begin{itemize}
\item the kinetic energy fraction $K$ defined in \eq{eq:eff-factor_K},
which is related to the strength of the phase transition,
\item the characteristic bubble separation $H_* R_*$, see \eq{eq:H_star_R_star},
\item the nucleation temperature $T_*$ in GeV,
\item $\Delta_w$ from \eq{eq:DefDeltaw} parameterising the width of the
sound-shell feature, and
\item the energy fraction of turbulence, $\varepsilon$.
\end{itemize}
This set of thermodynamical fit parameters is very similar to the one in
Ref.~\cite{Caprini:2024hue}, except that we replaced $v_w$ by $\Delta_w$.

This sampling is particularly important for the combination of sound waves
and MHD turbulence, because both sources are controlled by the same two
quantities, $K$ and $H_*R_*$: the sound-wave break frequencies and
amplitude depend on $K$, $H_*R_*$, and $\Delta_w$, see
Eqs.~(\ref{eq:SW_f1}--\ref{eq:SW_Omegint}), whereas the turbulent source
is characterised by $\Omega_s=\varepsilon K$ and the same
$H_*R_*$, see Eqs.~(\ref{eq:Omegas}--\ref{eq:MHD_Omega2}).  The combined
spectrum is thus a function of $(K,H_*R_*,\Delta_w,\varepsilon)$ and,
through $H_{*,0}$, of $T_*$ only, whereas a parametrisation in terms of
$(\alpha,\beta/H_*,v_w,\varepsilon)$ would require the case distinctions
discussed above.

The corresponding priors are specified in \Tabref{tab:O4_KR}.
Here we apply the frequency and amplitude requirements
(\ref{eq:FrequencyConstraint},\,\ref{eq:AmplitudeRequirement}) on the
sound-wave contribution, since the MHD contribution is subdominant and
hence does not affect the peak of the overall spectrum appreciably.
In order to determine the edges of the conditional priors, we start by
noting that \eq{eq:H_star_R_star} yields the conservative upper limit
$H_*R_* < (8\pi)^{1/3}$, since $v_w<1$ and $\beta/H_*>1$.  Next, we use
Eqs.~\eqref{eq:OmegaP2} and (\ref{eq:SW_Omegint}--\ref{eq:AuxOmegas})
to express the preak amplitude in terms of thermodynamical parameters.
Introducing the abbreviation
\begin{equation}
	x \equiv \frac{f_2}{f_1} \simeq \frac{2.5}{\Delta_w} \,,
\end{equation}
see \eq{eq:f2byf1}, and plugging in \eq{eq:FGW0} for $g_*(T_*)=100$,
we find
\begin{align}
	\Omega_p h^2 &= 1.8 \times 10^{-6} \,
	\frac{1}{\pi}\left(\sqrt{2}+\frac{2x}{1+x^2}\right)
	\frac{S(f_p)}{S(f_2)} \,
	K^2 \, H_*R_* \,
	\min\!\Big(1\,,\,\frac{2H_*R_*}{\sqrt{3K}}\Big)
\nonumber\\
	{}&\equiv
	c_p(\Delta_w) \, 
	\min\!\Big(K^2\,H_*R_*\,,\,\frac{2\,(H_*R_*)^2}{\sqrt{3}}\,K^{3/2}\Big)
	\,,
\label{eq:Omegap_sw}
\end{align}
where the ratio $S(f_p)/S(f_2)$ depends only on $x$ and $z=f_p/f_2$
determined by \eq{eq:zapprox}, which implies that the quantity $c_p$ is
a function of only $\Delta_w$.
As $\Omega_p h^2$ is the minimum of two expressions that both increase
with $H_*R_*$, the amplitude requirement \eqref{eq:AmplitudeRequirement}
is guaranteed to be satisfied if $H_*R_*$ is sufficiently large for the
$\Omega_p h^2$ values resulting from \emph{both} expressions to exceed
the threshold.  A conservative lower limit is therefore
\begin{equation} \label{eq:sw_hr}
	H_*R_* > H_*R_*{}^\text{min} \equiv \max\!\Bigg[
	\frac{10^{-9}}{c_p(\Delta_w)\,K^2} \,,\,
	\left(\frac{\sqrt3\times10^{-9}}{2c_p(\Delta_w)}\right)^{1/2} K^{-3/4}
	\Bigg].
\end{equation}
This condition can only be satisfied simultaneously with
$H_*R_* < (8\pi)^{1/3}$ if $K$ is not too small, resulting in the
lower prior edge for $K$,
\begin{equation}
	K > \left( \frac{10^{-9}}{(8\pi)^{1/3} \,c_p(\Delta_w)} \right)^{1/2} .
\end{equation}
The upper edge is $K < 0.6$ according to \eq{eq:eff-factor_K}, since the
efficiency factor $\kappa$ in that equation cannot exceed $1$.

Finally, we employ the frequency constraint to derive the prior range
for $T_*$ for given values of $H_*R_*$ and $\Delta_w$.  Using
Eqs.~\eqref{eq:Hstar0} and \eqref{eq:SW_f2} yields
\begin{equation}
	f_p = z \, f_2 =
	1.65\times10^{-5}\,\text{Hz} \, \frac{z}{2\,\Delta_w\,H_*R_*} \,
	\frac{T_*}{100\,\text{GeV}} \,.
\end{equation}
Solving for $T_*$ and applying the constraint
\eqref{eq:FrequencyConstraint} on $f_p$, we find
\begin{equation}
	2.4\times10^8 \, \frac{\Delta_w}{z} \, H_*R_* < T_*/\text{GeV} <
	1.2\times10^{10} \, \frac{\Delta_w}{z} \, H_*R_* \,.
\end{equation}

\begin{table}
\centering
\begin{tabular}{ccc}
\hline
\textbf{Parameter} & \textbf{Jeffreys} & \textbf{Conditional} (O4a Detector-Based) \\
\hline
$\Omega_\text{ref}$ &
$\logU(10^{-13},10^{-6})$ & $\logU(10^{-13},10^{-6})$ \\
\hline
$\Delta_w$ & U$(0.01,1)$ & U$(0.01,1)$\\
\hline
$K$ &
LogU$(10^{-4},1)$ &
LogU$(1.8\times10^{-5}\,c_p(\Delta_w)^{-1/2},\,0.6)$ \\
\hline
$H_*R_*$ &
LogU$(3\times10^{-4},3)$ &
LogU$\left(H_*R_*{}^\text{min},\, (8\pi)^{1/3} \right)$ \\
\hline
$T_*/\text{GeV}$ &
LogU$(10^2,10^8)$ &
LogU$\left(2.4\times10^8 \, \frac{\Delta_w}{z} \, H_*R_*,\,
 1.2\times10^{10} \, \frac{\Delta_w}{z} \, H_*R_*\right)$ \\
\hline
$\epsilon$ & U$(0,0.1)$ & U$(0,0.1)$ \\
\hline
\end{tabular}
\caption{Prior specification for the O4a data analysis in terms of
thermodynamical parameters for a DBPL spectrum.  Instead of the strength and duration parameters $\alpha$ and $\beta$ as well as the wall velocity, we use the combinations $K$,
$H_*R_*$, and $\Delta_w$ that appear in the relations determining the spectrum. As before, $\Omega_\text{ref}$ parameterises the astrophysical foreground.
The quantities $c_p(\Delta_w)$, $H_*R_*{}^\text{min}$, and $z$ are given in
\eq{eq:Omegap_sw}, \eq{eq:sw_hr}, and \eq{eq:zapprox}, respectively.
}
\label{tab:O4_KR}
\end{table}

\subsection{Direct sampling for $U(1)$ extensions of the Standard Model}
\label{sec:U1_direct_sampling}

Minimal $U(1)$ gauge extensions of the Standard Model (SM) are among the
simplest and best-motivated classes of physics beyond the SM: adding a
single new Abelian factor, spontaneously broken by a complex scalar that
gives mass to the associated gauge boson $Z'$, is renormalizable and
calculable, and its phenomenology is fixed almost entirely by the choice of
which SM (or hidden-sector) charges the new symmetry couples to. This
single construction therefore covers a remarkably broad range of
well-studied scenarios. Gauged $B-L$ (baryon minus lepton number) is one of
the simplest anomaly-free extensions of the SM once right-handed neutrinos
are included, and is directly tied to the origin of (Majorana) neutrino
masses~\cite{Davidson:1979px,Marshak:1979fm,Mohapatra:1980yp}; its breaking
has been studied as an explicit SGWB source in its own
right~\cite{Hasegawa:2019amx}, at a scale set by the same right-handed
neutrino Majorana mass that also controls the light-neutrino spectrum, so
that a GW detection would probe neutrino-mass physics directly.
Likewise, gauged $L_\alpha-L_\beta$ models ($\alpha,\beta \in
\{e,\mu,\tau\}$) are anomaly-free with the SM fermion content alone.
At the opposite
extreme, secluded ``dark-sector'' $U(1)$ groups, under which no SM field is
directly charged, communicating with the visible sector only through
kinetic mixing with hypercharge, underlie the broad class of dark-photon
models~\cite{Holdom:1985ag,Fabbrichesi:2020wbt}. In each of these
realizations, the spontaneous breaking of the new $U(1)$ symmetry can
proceed via a FOPT that sources an SGWB whose parameters
$(\alpha,\beta/H_\ast,T_\ast,v_w,\ldots)$ of \Secref{sec:thermodynamical_results}
can be computed directly from the Lagrangian parameters of the model
(charges, gauge coupling, quartic coupling, VEV) rather than treated as
free phenomenological quantities to be scanned independently. This is the
central appeal of the direct-sampling approach pursued in this section: it
maps an LVK bound directly onto the mass and coupling of a hypothetical new
gauge boson, in a form that can be confronted, model by model, with
collider, precision-electroweak, and fifth-force constraints on the same
particle  a level of interpretability that the geometrical and
template-based analyses of \Secref{sec:geometrical_results} and
\Secref{sec:thermodynamical_results} do not by themselves provide. Badger
et al.~\cite{Badger:2022jzg} used exactly this logic for the gauged $B-L$
case, converting LVK O1--O3 posterior/upper-limit results on a
phenomenological $\Omega_{\rm GW}(f)$ model into excluded regions of
$(m_{Z'},g_{B-L})$ parameter space; the pipeline we apply below follows the
same underlying logic, with the FOPT parameters computed from first
principles as described in \Appref{app-abelian-higgs-procedure} rather
than mapped through the original envelope/sound-wave fits of that analysis.

As a concrete worked example of this generic pipeline, we specialize in
this paper to the case of a $U(1)_X$ gauge symmetry living entirely within
a secluded dark sector, i.e.\ one under which no SM field is charged and
which communicates with the SM only thermally rather than through any
direct portal coupling. We choose this realization for two reasons. First,
it is arguably the simplest UV-complete setting in which the direct-sampling
pipeline can be demonstrated cleanly: the dark sector's own dynamics the
phase transition and the resulting GW signal decouple entirely from the
details of any portal coupling to the SM, which we are therefore free to
leave unspecified. Second, among the $U(1)$ extensions listed above it has
the fewest free Lagrangian parameters (just the gauge coupling $g$, the
quartic coupling $\lambda$, and the VEV $v$), which keeps the sampling
priors, and hence the prior-robustness study that is the focus of this
paper, most transparent. We stress, however, that the pipeline itself is
generic: it applies unchanged to any of the more elaborate, SM-charged
$U(1)$ extensions above, such as $B-L$, or to
kinetically-mixed dark photon scenarios. The only difference in those
cases is the additional model-building input required to specify how the
new sector couples to the SM (charge assignments or portal/kinetic-mixing
couplings) and, correspondingly, the additional experimental constraints
(collider, beam-dump, precision-electroweak, or fifth-force bounds) that
then apply alongside the GW limits derived here.

Here, we consider a generic Abelian Higgs model in a dark sector with a
complex scalar $\Phi$ charged under a gauge group $U(1)_X$ and the
corresponding gauge boson $X_\mu$. The Lagrangian is
\begin{equation}
\mathcal{L}
=
(D_\mu\Phi)^\dagger(D^\mu\Phi )
-\frac{1}{4}X_{\mu\nu}X^{\mu\nu}
-V(\Phi ),
\label{eq:U1X_Lagrangian}
\end{equation}
where
\begin{align}
D_\mu\Phi  &= (\partial_\mu+igX_\mu)\Phi ,
\label{eq:U1X_covderiv}\\
X_{\mu\nu} &= \partial_\mu X_\nu-\partial_\nu X_\mu,
\label{eq:U1X_fieldstrength}\\
V(\Phi ) &= -\mu^2\Phi ^\dagger\Phi +\lambda(\Phi ^\dagger\Phi )^2
=
\lambda\left(\Phi ^\dagger\Phi -\frac{v^2}{2}\right)^2
+\mathrm{const.}
\label{eq:U1X_potential}
\end{align}
For convenience we have chosen the $U(1)_X$ charge equal to one for the
scalar. We assume $\mu^2>0$, such that the $U(1)_X$ symmetry is broken
spontaneously by a vacuum expectation value $v=\mu/\sqrt{\lambda}$, which
yields masses $m_h=\sqrt{2\lambda}\,v$ for the dark Higgs and $m_X=g\,v$
for the gauge boson at tree level.

We assume a coupling between the extra sector and the SM particles
(visible sector) that is sufficiently strong to keep both sectors in
thermal equilibrium with a common temperature but sufficiently weak to
justify neglecting the impact of SM particles on the extra Higgs effective
potential and the phase transition. This assumption is needed here
precisely because the $U(1)_X$ we sample is secluded: it is automatically
satisfied, without any additional input, whenever the new gauge boson
couples directly to SM fermions, as in the gauged $B-L$ or $L_\mu-L_\tau$
realizations discussed above, since in those cases the new sector shares a
common thermal bath with the SM by construction.

To determine whether spontaneous symmetry breaking is
associated with a FOPT and what its thermodynamical parameters are, we
compute the finite-temperature effective potential of the dark Higgs
using dimensional reduction with next-to-leading order (NLO) matching.
For each sampled point, we determine the phase structure from the
effective potential and calculate the thermal bounce action $S_3(T)$.
The transition temperature is obtained from the percolation condition.
We calculate $\alpha$ from the effective potential difference and its
temperature derivative (see \Secref{sec:ThermoPars}) and $\beta/H_*$
from the temperature derivative of $S_3(T)/T$ at the transition
temperature.  The wall velocity is subsequently evaluated in the local
thermal equilibrium (LTE) approximation (which we validate in
\Secref{sec:LTEcheck}) using \texttt{WallGo}~\cite{Ekstedt:2024etx}. 
Finally, we obtain the GW spectrum by passing the computed thermodynamical
parameters to the DBPL template for sound waves and MHD turbulence,
fixing $\varepsilon=0.05$ for the latter.  Further details are presented
in \Appref{app-abelian-higgs-procedure}.

As we see from Eqs.~(\ref{eq:U1X_Lagrangian}--\ref{eq:U1X_potential}),
the Abelian Higgs model is fully described by the parameters
$g$, $\lambda$, and $v$.  We sample these together with the
astrophysical foreground amplitude $\Omega_\text{ref}$,
comparing uniform and log-uniform priors with identical ranges for the
model parameters, which are shown in \Tabref{tab:U1priors}.
In both cases, $\Omega_\text{ref}$ has the same log-uniform prior used
previously, $\logU(10^{-10},10^{-7})$ for O3 and
$\logU(10^{-13},10^{-6})$ for O4a.

\begin{table}
\centering
\begin{tabular}{ccc}
\hline
\textbf{Parameter} & \textbf{Uniform} & \textbf{Log-Uniform} \\
\hline
$g$ &
U$(0.3,1.9)$ &
LogU$(0.3,1.9)$ \\
\hline
$\lambda$ &
U$(10^{-4},10^{-2})$ &
LogU$(10^{-4},10^{-2})$ \\
\hline
$v/\text{GeV}$ &
U$(10^5,10^8)$ &
LogU$(10^5,10^8)$ \\
\hline
\end{tabular}
\caption{Prior specification for the data analysis in terms of
Abelian Higgs model parameters.}
\label{tab:U1priors}
\end{table}

\section{Results}
\label{sec:results}
\subsection{Results based on different priors for the geometrical parameters}
\label{sec:geometrical_results}
In this section, we present the results in terms of the geometrical
parameters
$\Omega_{\rm ref}$, $\Omega_p$, $f_p/\text{Hz}$, and $\Delta_w$,
cf.~\Secref{sec:pestimation_geometrical}.
These parameters characterise the overall amplitude and location of the
phase-transition signal in frequency space, and therefore provide the
most direct way of assessing how much information is supplied by the
data themselves.

\subsubsection{Sound waves modelled as a double broken power law}
\label{sec:dbpl_sw_results}
\begin{figure}
    \centering
    \begin{minipage}{0.49\textwidth}
        \centering
        \includegraphics[height=6.9cm]{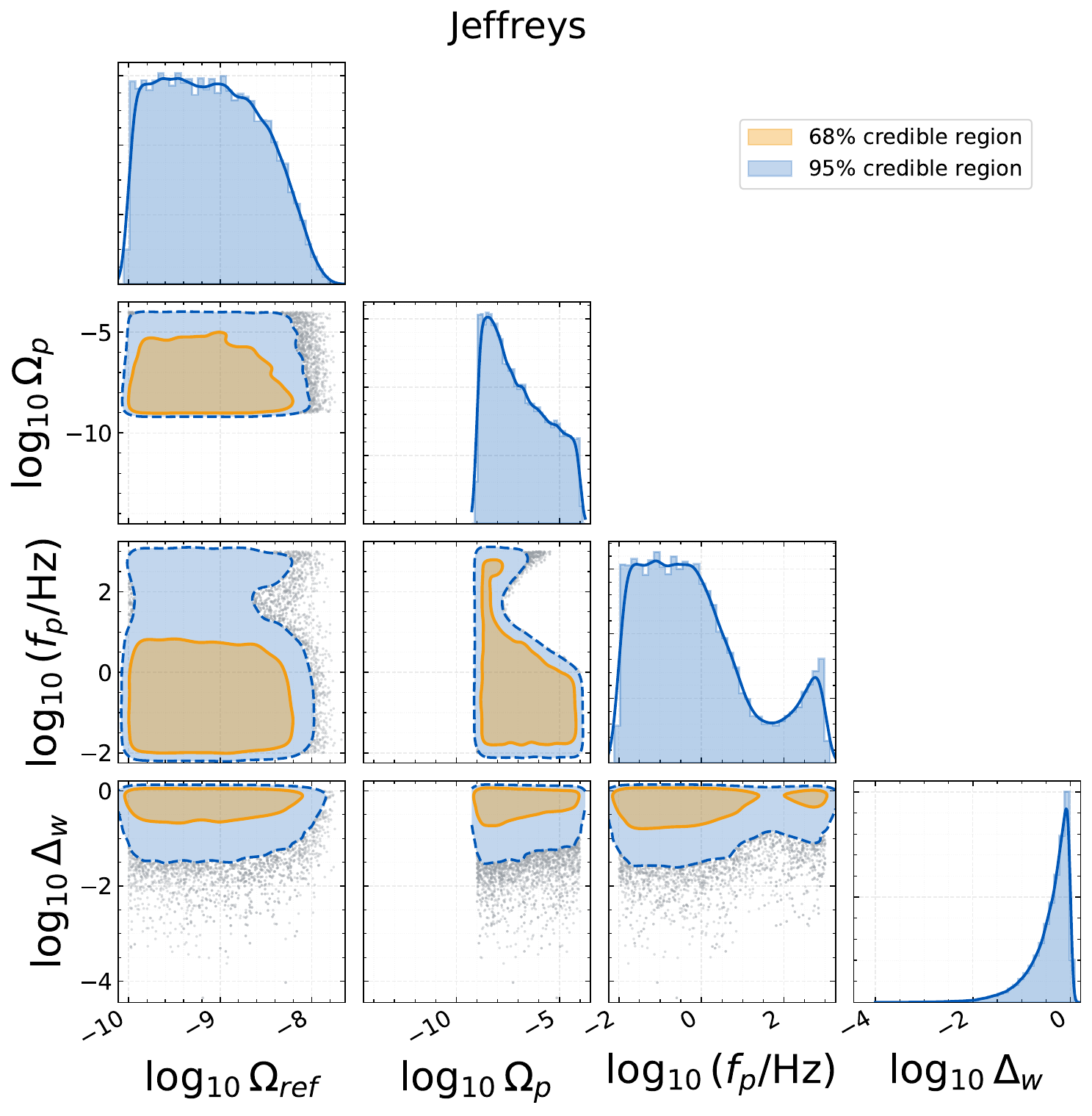}
    \end{minipage}\hfill
    \begin{minipage}{0.49\textwidth}
        \centering
        \includegraphics[height=6.9cm]{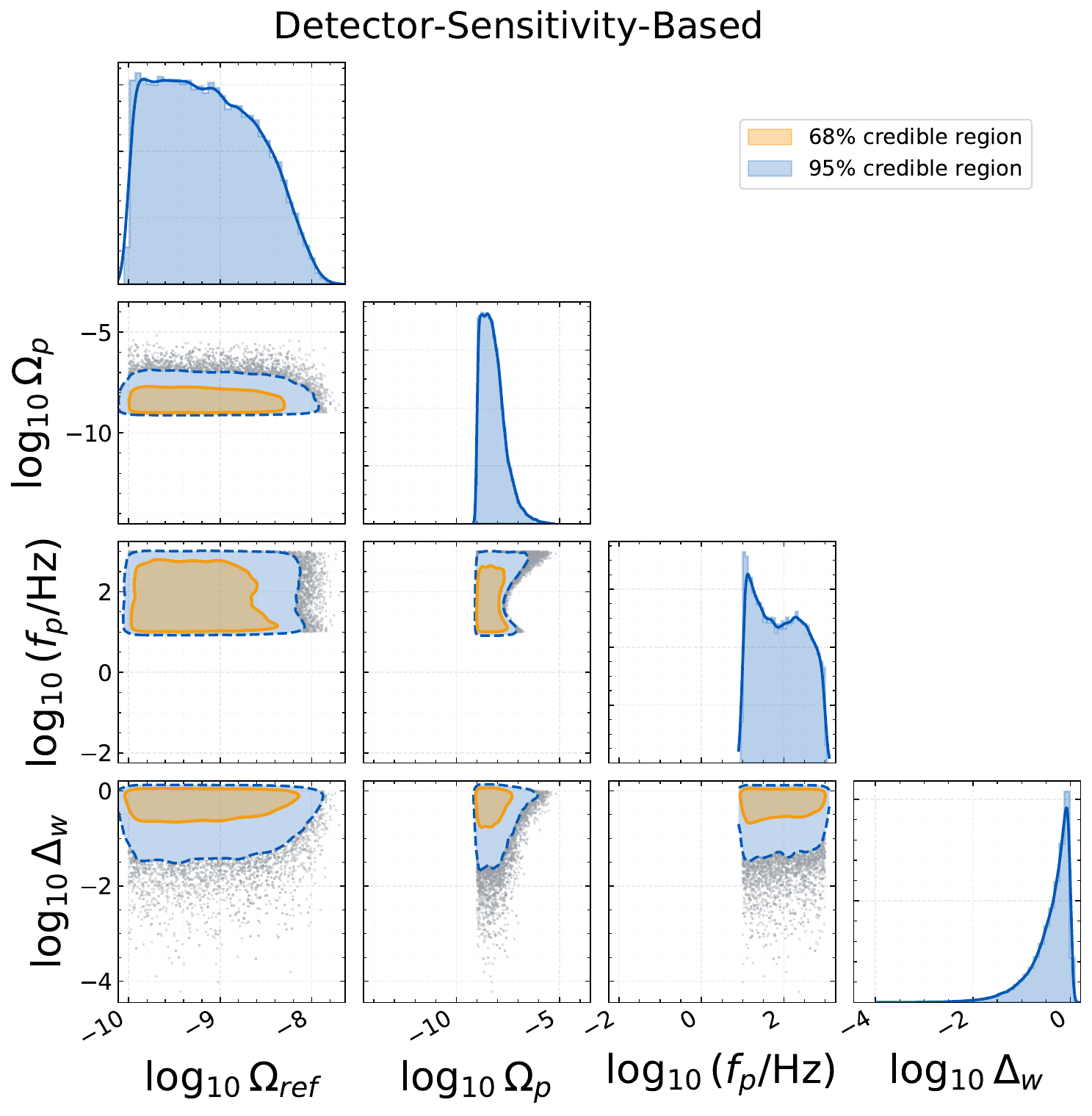}
    \end{minipage}
    \vspace{0.5cm}
    \begin{minipage}{0.49\textwidth}
        \centering
        \includegraphics[height=6.9cm]{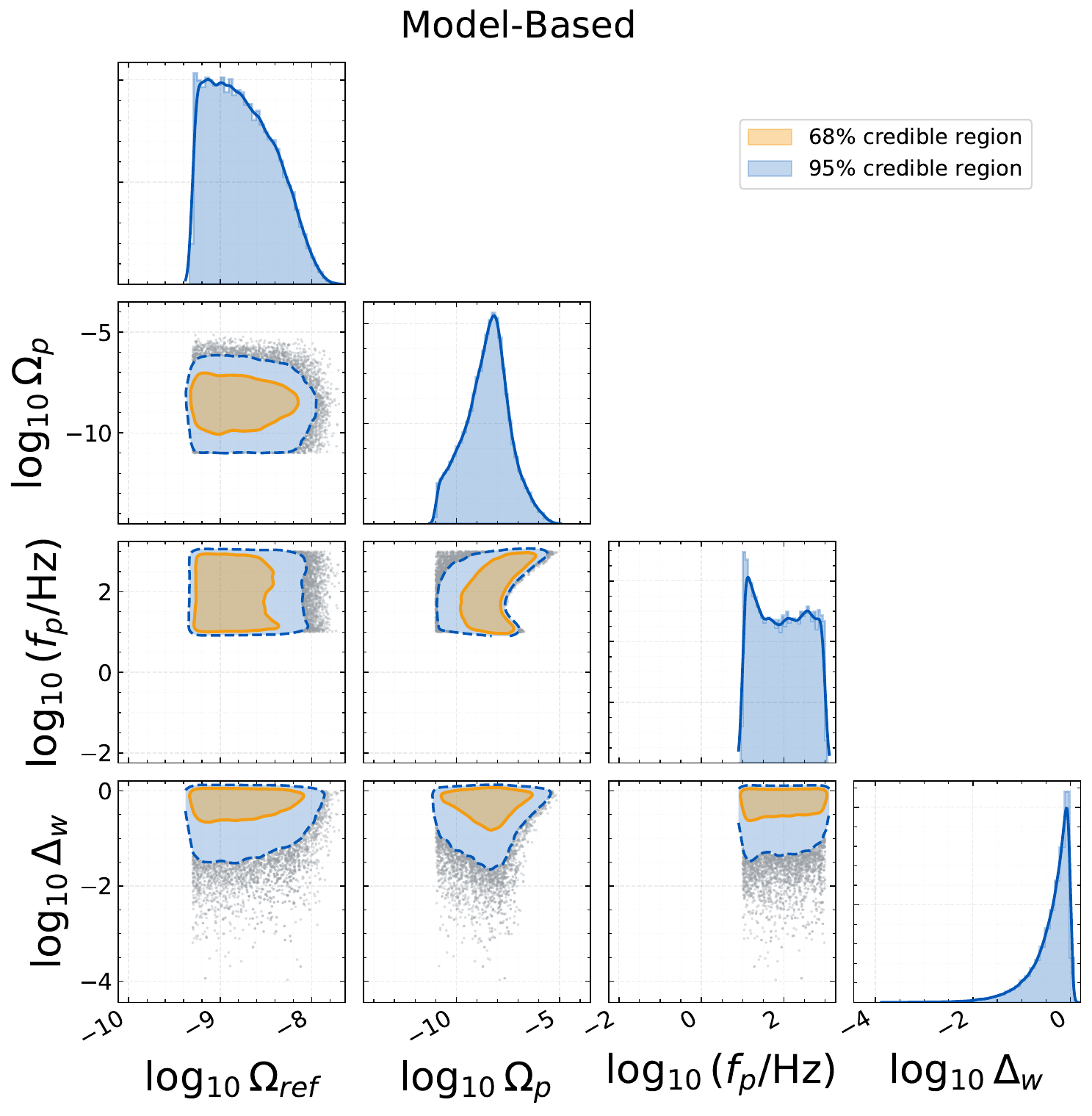}
    \end{minipage}\hfill
    \begin{minipage}{0.49\textwidth}
       \centering
        \includegraphics[height=6.9cm]{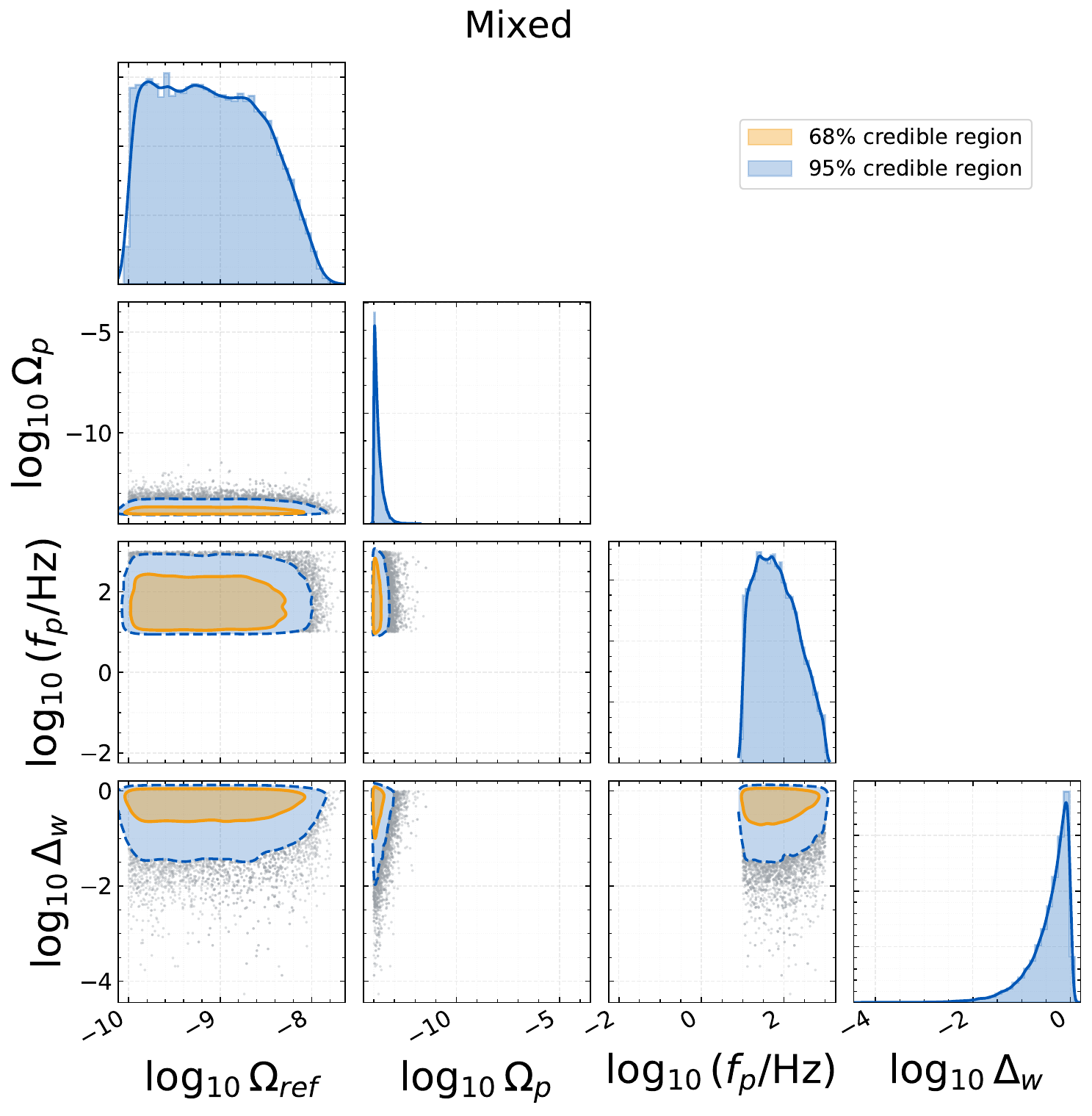}
    \end{minipage}
\caption{Posterior distributions obtained from the public O3 stochastic data for the double broken power law modelling the contribution from sound waves, using the Jeffreys, Detector-Sensitivity-Based, Model-Based, and Mixed prior, respectively. The prior definitions are given in \Tabref{tab:PriorsGeom}.}\label{fig:sw_dpbl_posterior_corner_O3}\end{figure}

\begin{figure}
    \centering
    \begin{minipage}{0.5\textwidth}
        \centering
        \includegraphics[height=6cm]{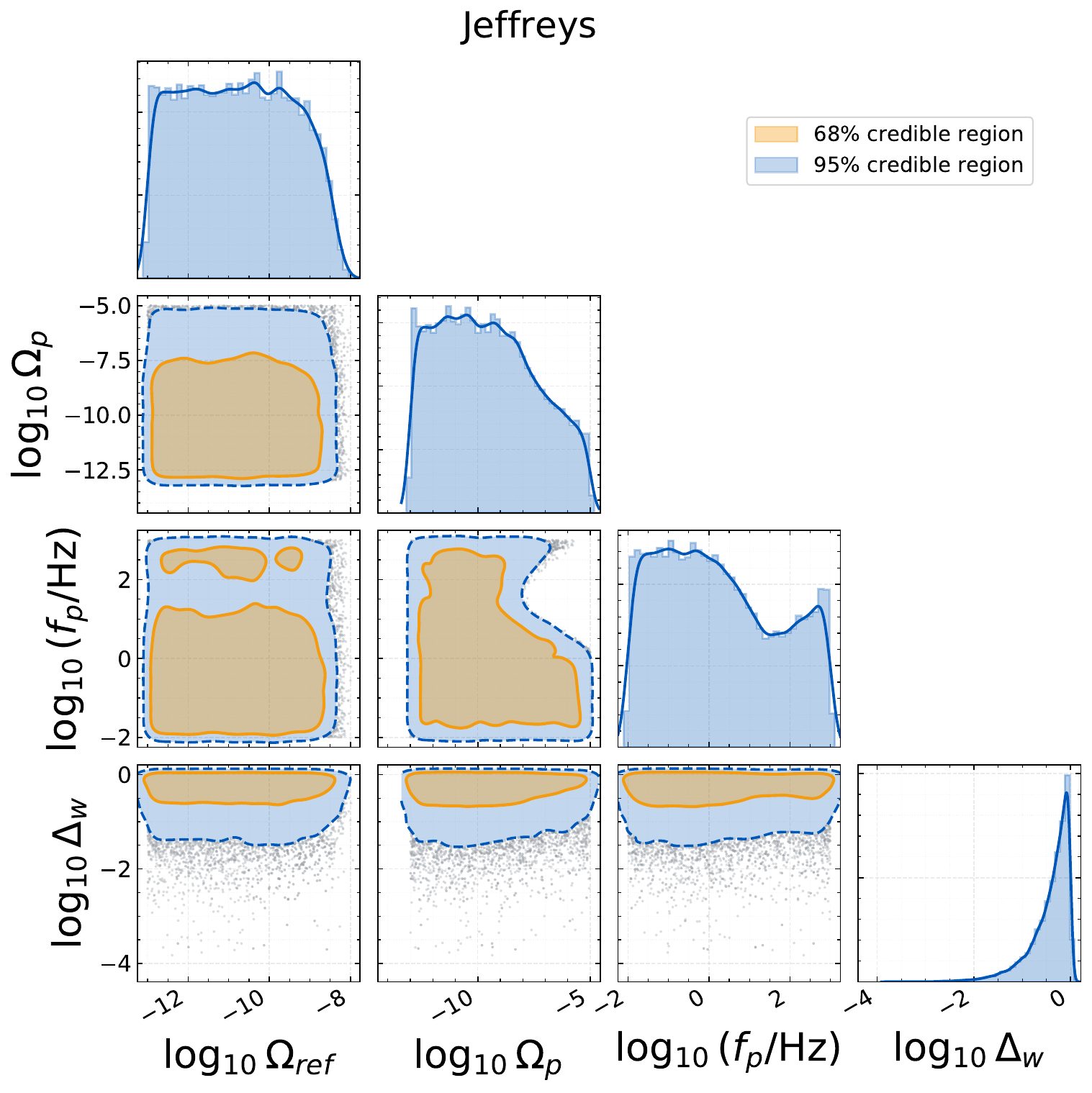}
    \end{minipage}\hfill
    \begin{minipage}{0.5\textwidth}
        \centering
        \includegraphics[height=6cm]{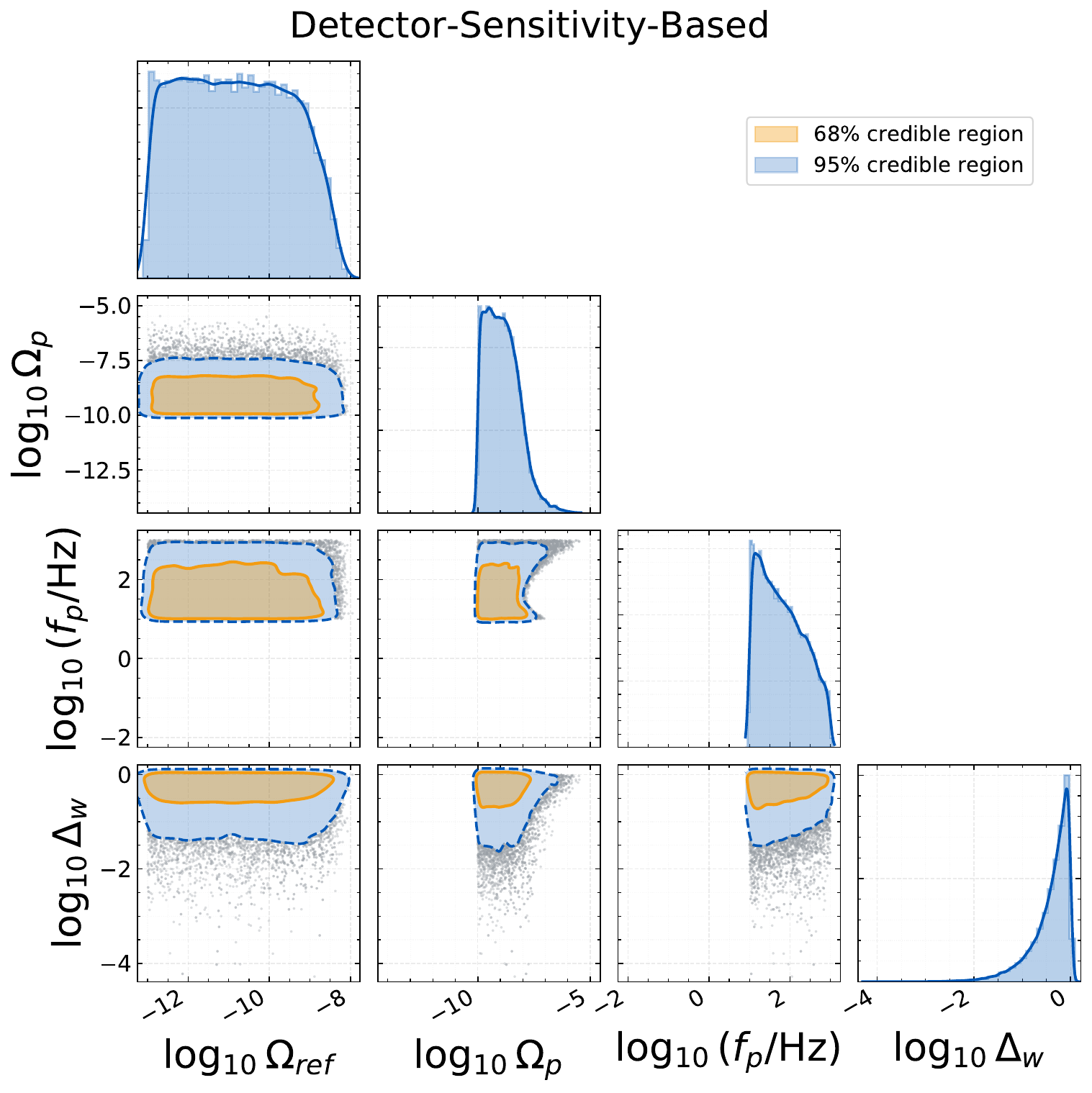}
    \end{minipage}
    \vspace{0.5cm}
    \begin{minipage}{0.5\textwidth}
        \centering
        \includegraphics[height=6cm]{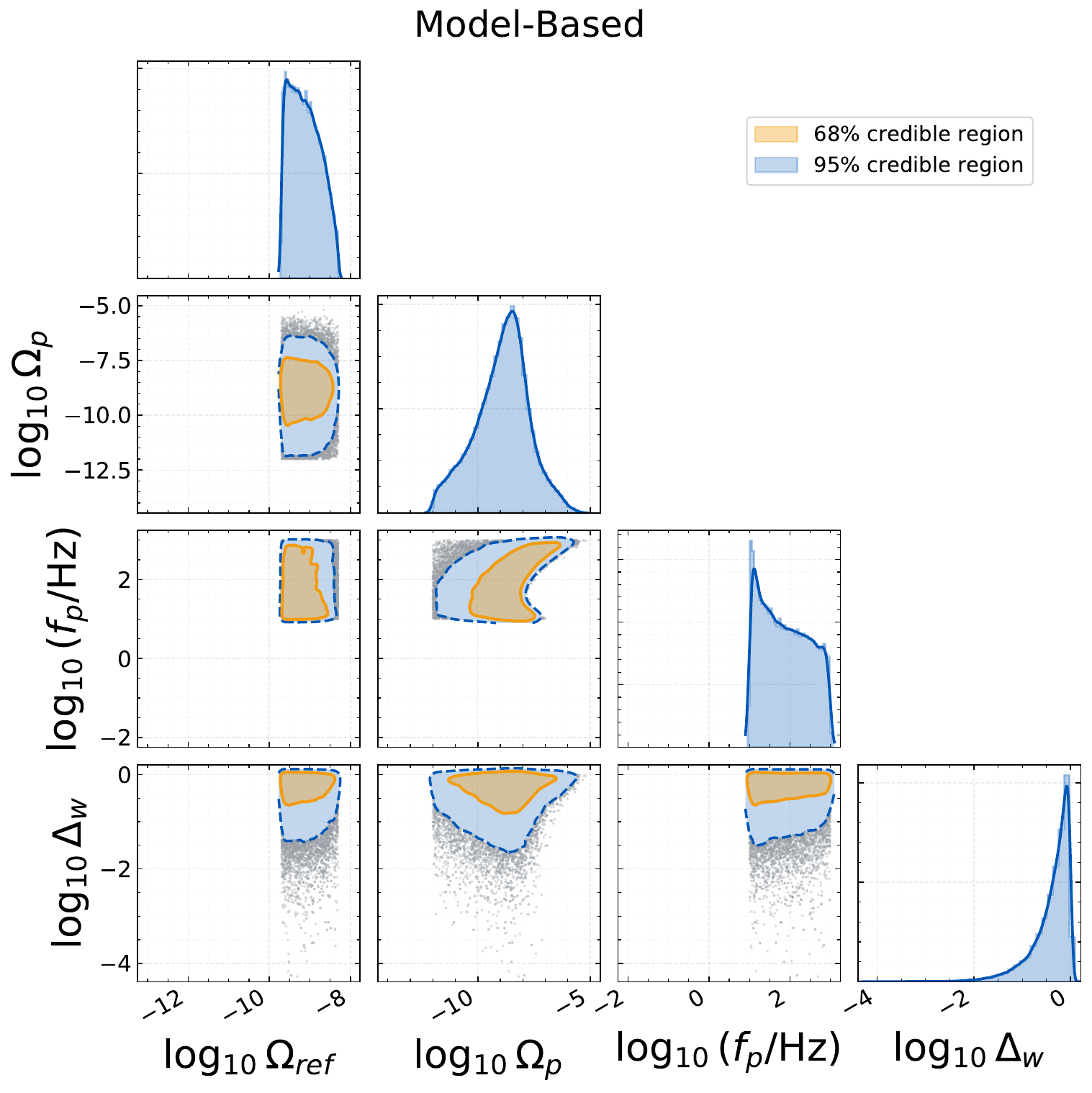}
    \end{minipage}\hfill
    \begin{minipage}{0.5\textwidth}
       \centering
        \includegraphics[height=6cm]{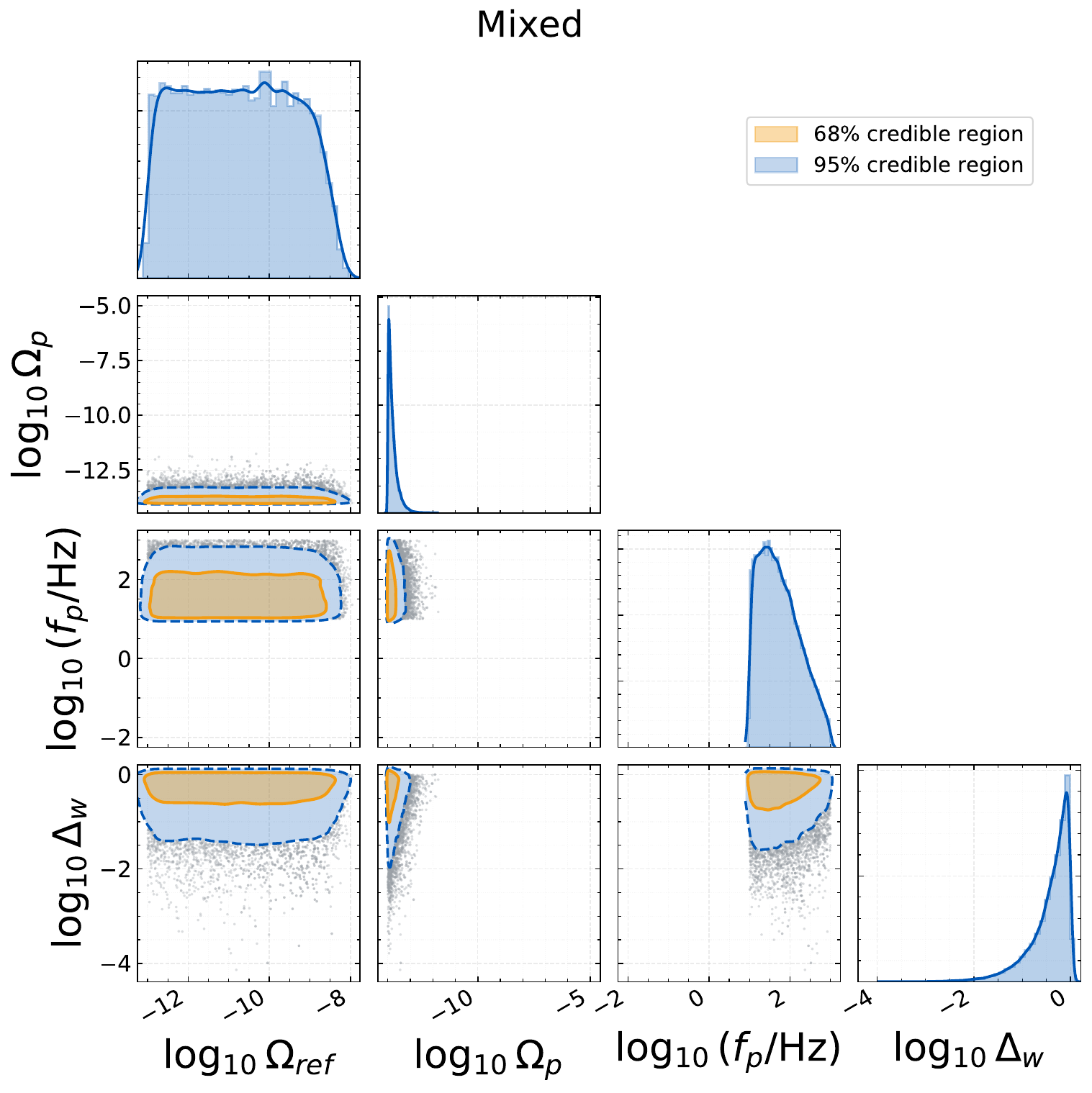}
    \end{minipage}
    \caption{Posterior distributions obtained from the O4a stochastic data for the double broken power law modelling the contribution from sound waves, using the Jeffreys, Detector-Sensitivity-Based, Model-Based, and Mixed prior, respectively.
    The prior definitions are given in \Tabref{tab:PriorGeom_O4a}.
    }
    \label{fig:sw_dpbl_posterior_corner_O4a}
\end{figure}

For the sound-wave DBPL model we retain the full O3 and O4a analyses
side by side, since both runs are sufficiently informative to be
compared quantitatively.
We are particularly interested in the dependence of the inferred SGWB on
the choice of prior, considering the four different prior prescriptions
Jeffreys, Detector-Sensitivity-Based, Model-Based, and Mixed, as defined
in Tabs.~\ref{tab:PriorsGeom} and~\ref{tab:PriorGeom_O4a}.
The same Bayesian inference procedure is applied independently to the O3
and O4a stochastic data.  The resulting posterior distributions are
shown in Figs.~\ref{fig:sw_dpbl_posterior_corner_O3} and
\ref{fig:sw_dpbl_posterior_corner_O4a} for the O3 and O4a datasets,
respectively.

We concentrate the first part of the discussion on the peak amplitude
and frequency, since we varied both prior shapes and ranges for these
parameters.
For the remaining parameters $\Omega_\text{ref}$ and $\Delta_w$, we used
the same shape for all priors and only varied the prior range in the
case of $\Omega_\text{ref}$.  Hence, the posteriors for these parameters
are not expected to vary significantly, except that the extension of the
$\Omega_\text{ref}$ posterior to small values is given by the (varying)
lower prior edge in the absence of a detection.  This expectation is
vindicated by the corresponding parts of
Figs.~\ref{fig:sw_dpbl_posterior_corner_O3} and
\ref{fig:sw_dpbl_posterior_corner_O4a}.  However, the deduced upper
limit on $\Omega_\text{ref}$ does depend on the prior range
\cite{Abbott:2021xyo}; we will return to this issue in
\Secref{sec:Omegaref}.

In order to quantify the prior-dependence of the results for $\Omega_p$
and $f_p$, we consider the robustness measures Maximin Robustness
Ratio~$\rho_{\max}$, posterior mean ratio~$\rho_\text{mean}$,
credible-region overlap~$\rho_\text{overlap}$, and Kullback-Leibler
divergence~$D_\text{KL}$.  They are defined in \Appref{app:robustness}\@,
and the values we obtain for the present analysis are given in
Tabs.~\ref{tab:dbpl_sw_rho_max} and
\ref{tab:dbpl_sw_mean_overlap_kl_robustness}\@.
Based on these measures, we classify parameters as ROBUST, MODERATE, or
NOT ROBUST under prior change, as specified in \Tabref{tab:robcriteria}
and \eq{eq:robustness_classification}.
We also show the $68\%$ credible posterior intervals in
\Tabref{tab:dbpl_sw_credible_intervals_68}.

\paragraph{O3 stochastic data}
Figure \ref{fig:sw_dpbl_posterior_corner_O3} shows the posterior
distributions obtained with the O3 stochastic data.  The corner plots
display the one-dimensional marginalised posteriors (diagonal panels)
together with the two-dimensional correlations among the geometrical
parameters.  The latter show the $68\%$ and $95\%$ credible regions.
The marginalized distributions of both $\Omega_p$ and $f_p$ change
noticeably when moving between the different prior prescriptions, while
their two-dimensional distributions with the other parameters also show
significant changes in shape and correlation structure.  This behaviour
is expected when the data do not contain sufficient information to
localise a peaked cosmological contribution independently of the prior
volume.  In such a situation, the posterior should not be interpreted as
providing a prior-independent measurement of the peak amplitude or peak frequency.

The prior-dependence of the inferred peak frequency is particularly evident in \Figref{fig:sw_dpbl_posterior_corner_O3}.
According to
\Tabref{tab:dbpl_sw_credible_intervals_68}, the 68\% credible intervals
extend from $[3.3\times10^{-2},\,18.2]\,$Hz
for the Jeffreys prior, the narrowest of the four --~the reverse of what
one might naively expect from a scale-invariant prior~--,
to $[17.9,\,459.1]\,$Hz
for the model-based prior, the widest, giving a maximum-to-minimum width
ratio $\rho_{\max}=24.2$ in \Tabref{tab:dbpl_sw_rho_max}.
The distributions also show a noticeable displacement, leading to
a vanishing credible-region overlap and $D_{\rm KL}=0.9$.
Nevertheless, the different posterior regions become substantially more
compatible when the $95\%$ credible regions are considered.  This
distinction is important: overlap at a high credibility level does not
imply that the corresponding parameter estimate is insensitive to the
prior.  Rather, it indicates that the available data still permit a
broad range of peak frequencies.

The posterior intervals for $\Omega_p$ span substantially
different regions of parameter space.  The narrowest interval is
obtained for the mixed prior,
$[1.1\times10^{-14},\,2.5\times10^{-14}]$, and the widest for the
Jeffreys prior, $[3.3\times10^{-9},\,6.0\times10^{-6}]$, corresponding
to $\rho_{\max}=4.2\times10^8$ and zero overlap.
The posterior mean under the mixed prior
lies more than eight orders of magnitude below that of the Jeffreys prior
($\rho_\text{mean}=2.8\times10^8$).
Also the Kullback-Leibler divergence reaches a very large value of
$D_\text{KL}=26.8$.
Thus, the NOT ROBUST label in
\Tabref{tab:dbpl_sw_mean_overlap_kl_robustness} reflects a genuine
disagreement in both location and width, not merely a difference in
precision.

All in all, \Tabref{tab:dbpl_sw_credible_intervals_68} provides a useful
quantitative summary of the main message of
\Figref{fig:sw_dpbl_posterior_corner_O3}.
In particular, the absence of a statistically significant
detection means that the posterior information on $\Omega_p$
and $f_p$ remains strongly influenced by the volume of
parameter space assigned by the prior.

\paragraph{O4a stochastic data}
Figure \ref{fig:sw_dpbl_posterior_corner_O4a} presents the corresponding
analysis using the O4a stochastic data.
The improved sensitivity of the
dataset generally produces more localized posterior distributions.
However, a narrower posterior does not automatically imply a more
prior-independent inference.  In other words, it is important to
distinguish between statistical precision and statistical robustness.

The cosmological peak amplitude and peak frequency
are still not well-determined by the data, and their posterior
distributions remain sensitive to the adopted prior, as demonstrated by
the robustness measures listed in the O4a blocks of
Tabs.~\ref{tab:dbpl_sw_rho_max} and
\ref{tab:dbpl_sw_mean_overlap_kl_robustness}.
Most of these measures show an improvement compared to their O3
counterparts.  For example, $\rho_{\max}$ for $f_p$ falls significantly
from $24.2$ to $3.1$.  The relative $\rho_\text{max}$ change for
$\Omega_p$ is even larger, yet at huge absolute values.  We do not find
any case where the status of a parameter changes, i.e., it remains NOT
ROBUST for both parameters according to all measures.  The
Kullback-Leibler divergence for $\Omega_p$ even increases very slightly.

According to \Tabref{tab:dbpl_sw_credible_intervals_68}, it is the same
priors as before that yield the broadest and narrowest credible
intervals.  The Jeffreys and the model-based priors yield the broadest
intervals $\Omega_p \in [1.2\times10^{-12},\,8.1\times10^{-8}]$ and $f_p
\in [16.4,\,390.0]\,$Hz, respectively.  The narrowest interval for
$\Omega_p$, $[1.1\times10^{-14},\,2.5\times10^{-14}]$, again results
from the mixed prior, while the Jeffreys prior gives the narrowest
interval for $f_p$, $[4.7\times10^{-2},\,118.9]\,$Hz; the latter is
considerably broadened compared to O3, though, resulting in a non-zero
$\rho_\text{overlap}$.

Taken together, Figs.~\ref{fig:sw_dpbl_posterior_corner_O3} and
\ref{fig:sw_dpbl_posterior_corner_O4a}, and
Tabs.~\ref{tab:dbpl_sw_rho_max}--\ref{tab:dbpl_sw_credible_intervals_68}
demonstrate that increased detector sensitivity and prior robustness
are distinct concepts. The O4a dataset generally reduces the volume of
the allowed parameter space.
The peak amplitude and peak frequency
remain sensitive to the prior even when their
posterior distributions become narrower. Therefore, a narrower
credible interval should not by itself be interpreted as evidence for
a more robust constraint.

\begin{table}
\centering
\begin{tabular}{lcccc}
\hline\hline
Parameter & $\rho_{\max}$ & $\max_{\pi}(\mathrm{Width}(\pi))$ & $\min_{\pi}(\mathrm{Width}(\pi))$ & Status \\
\hline
\multicolumn{5}{c}{O3 data} \\
\hline
$\Omega_{p}$ & $4.2 \times 10^{8}$ & $6.0 \times 10^{-6}$ & $1.4 \times 10^{-14}$ & NOT ROBUST \\
$f_{p}/\mathrm{Hz}$ & $24.2$ & $441.2$ & $18.2$ & NOT ROBUST \\
\hline
\multicolumn{5}{c}{O4a data} \\
\hline
$\Omega_{p}$ & $5.9 \times 10^{6}$ & $8.1 \times 10^{-8}$ & $1.4 \times 10^{-14}$ & NOT ROBUST \\
$f_{p}/\mathrm{Hz}$ & $3.1$ & $373.6$ & $118.8$ & NOT ROBUST \\
\hline\hline
\end{tabular}
\caption{Maximin Robustness Ratio $\rho_{\max}$
(\eq{eq:rho_max_general}) for the geometrical parameters of the
sound-wave DBPL model, O3 and O4a datasets.}
\label{tab:dbpl_sw_rho_max}
\end{table}

\begin{table}
\centering
\begin{tabular}{lcccc}
\hline\hline
Parameter & $\rho_{\mathrm{mean}}$ & $\rho_{\mathrm{overlap}}$ &
$D_{\mathrm{KL}}$ & Status \\
\hline
\multicolumn{5}{c}{O3 data} \\
\hline
$\Omega_{p}$ & $2.8 \times 10^{8}$ & $0.0$ & $26.8$ & NOT ROBUST \\
$f_{p}/\mathrm{Hz}$ & $3.7$ & $0.0$ & $0.9$ & NOT ROBUST \\
\hline
\multicolumn{5}{c}{O4a data} \\
\hline
$\Omega_{p}$ & $1.5 \times 10^{7}$ & $0.0$ & $26.9$ & NOT ROBUST \\
$f_{p}/\mathrm{Hz}$ & $2.3$ & $0.3$ & $0.5$ & NOT ROBUST \\
\hline\hline
\end{tabular}
\caption{Posterior mean ratio
$\rho_{\mathrm{mean}}=\max_{\pi}\langle\theta\rangle_{\pi}/\min_{\pi}\langle\theta\rangle_{\pi}$,
the ratio of the largest to the smallest posterior mean over the 4
priors (independent of their order), credible-region overlap and
Kullback-Leibler divergence (worst case over the pairs of priors) for
the geometrical parameters of the sound-wave DBPL model, O3 and O4a
datasets. The robustness status is obtained from $\rho_{\max}$, $\rho_{\mathrm{overlap}}$ and $D_{\mathrm{KL}}$: ROBUST if at least two of them are ROBUST, MODERATE if one is, NOT ROBUST if none is.}
\label{tab:dbpl_sw_mean_overlap_kl_robustness}
\end{table}

\begin{table}[t!]
\centering
\begin{tabular}{lcc}
\hline\hline
\multicolumn{3}{c}{O3 data} \\
\hline
Parameter & Detector-Sensitivity-Based & Jeffreys \\
\hline
$\Omega_{p}$ & $[1.7 \times 10^{-9}, 2.0 \times 10^{-8}]$ & $[3.3 \times 10^{-9}, 6.0 \times 10^{-6}]$ \\
$f_{p}/\mathrm{Hz}$ & $[17.3, 390.8]$ & $[3.3 \times 10^{-2}, 18.2]$ \\
\hline
Parameter & Model-Based & Mixed \\
\hline
$\Omega_{p}$ & $[2.1 \times 10^{-10}, 3.1 \times 10^{-8}]$ & $[1.1 \times 10^{-14}, 2.5 \times 10^{-14}]$ \\
$f_{p}/\mathrm{Hz}$ & $[17.9, 459.1]$ & $[19.0, 250.0]$ \\
\hline
\multicolumn{3}{c}{O4a data} \\
\hline
Parameter & Detector-Sensitivity-Based & Jeffreys \\
\hline
$\Omega_{p}$ & $[2.1 \times 10^{-10}, 6.8 \times 10^{-9}]$ & $[1.2 \times 10^{-12}, 8.1 \times 10^{-8}]$ \\
$f_{p}/\mathrm{Hz}$ & $[16.4, 280.7]$ & $[4.7 \times 10^{-2}, 118.9]$ \\
\hline
Parameter & Model-Based & Mixed \\
\hline
$\Omega_{p}$ & $[6.9 \times 10^{-11}, 1.4 \times 10^{-8}]$ & $[1.1 \times 10^{-14}, 2.5 \times 10^{-14}]$ \\
$f_{p}/\mathrm{Hz}$ & $[16.4, 390.0]$ & $[16.9, 200.9]$ \\
\hline\hline
\end{tabular}
\caption{$68\%$ credible intervals for the geometrical parameters of the
sound-wave DBPL model under the four priors of
Tabs.~\ref{tab:PriorsGeom} and \ref{tab:PriorGeom_O4a}.}
\label{tab:dbpl_sw_credible_intervals_68}
\end{table}

\subsubsection{Bubble collisions modelled as a broken power law}
\label{sec:bpl_bc_results}
\begin{figure}
    \centering
    \begin{minipage}{0.48\textwidth}
        \centering
        \includegraphics[height=6.9cm]{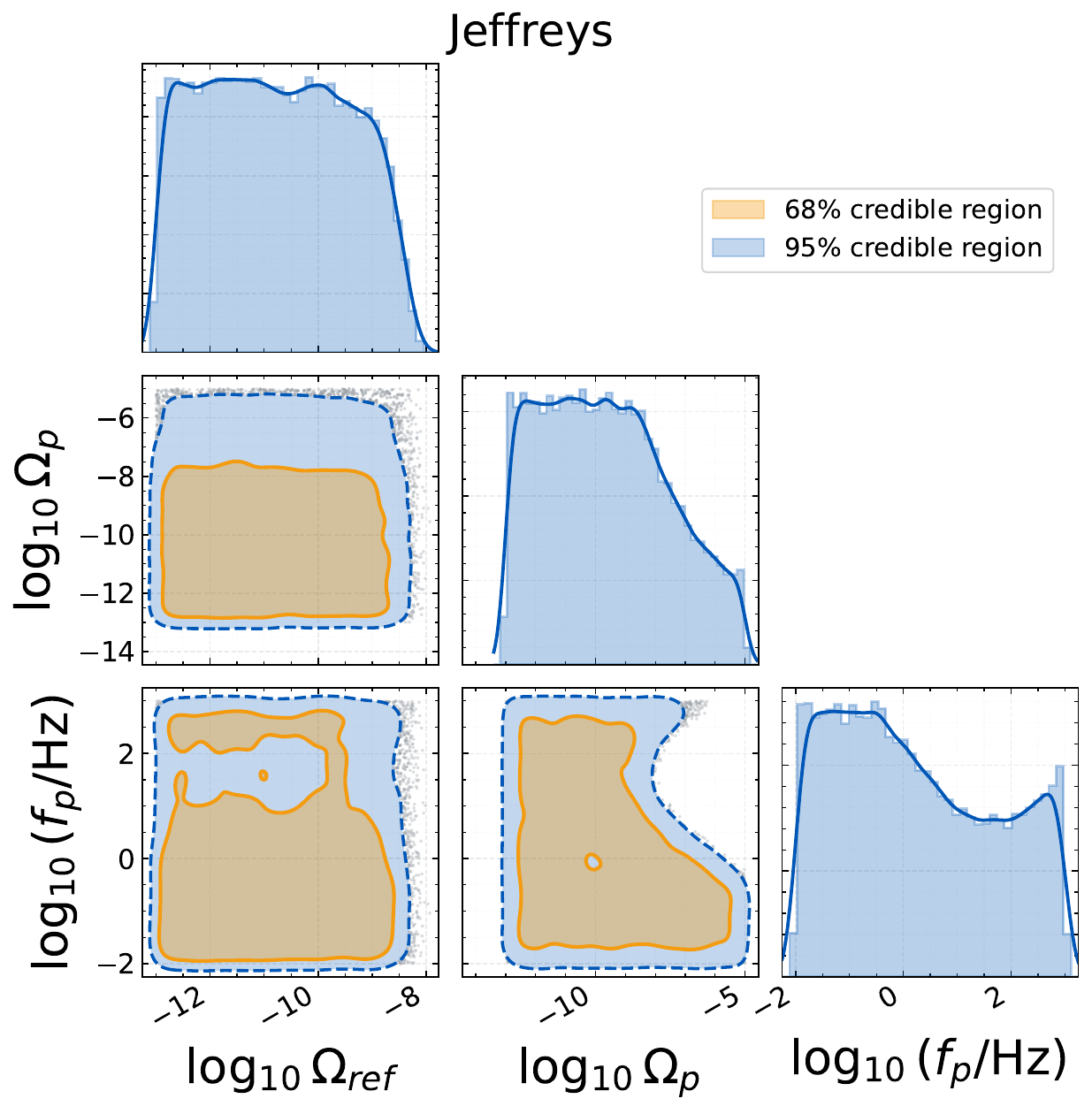}
    \end{minipage}\hfill
    \begin{minipage}{0.48\textwidth}
        \centering
        \includegraphics[height=6.9cm]{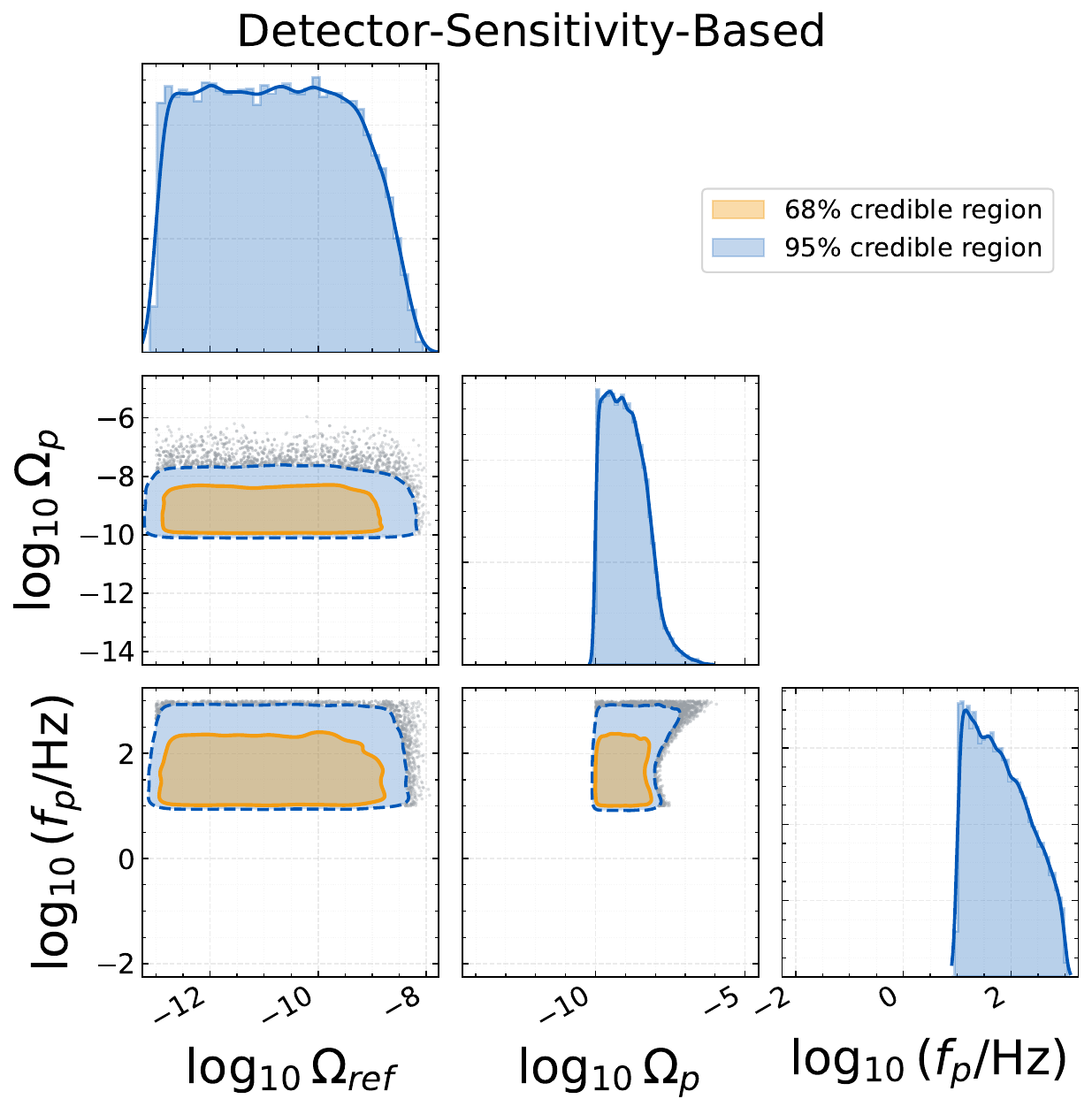}
    \end{minipage}
    \vspace{0.5cm}
    \begin{minipage}{0.48\textwidth}
        \centering
        \includegraphics[height=6.9cm]{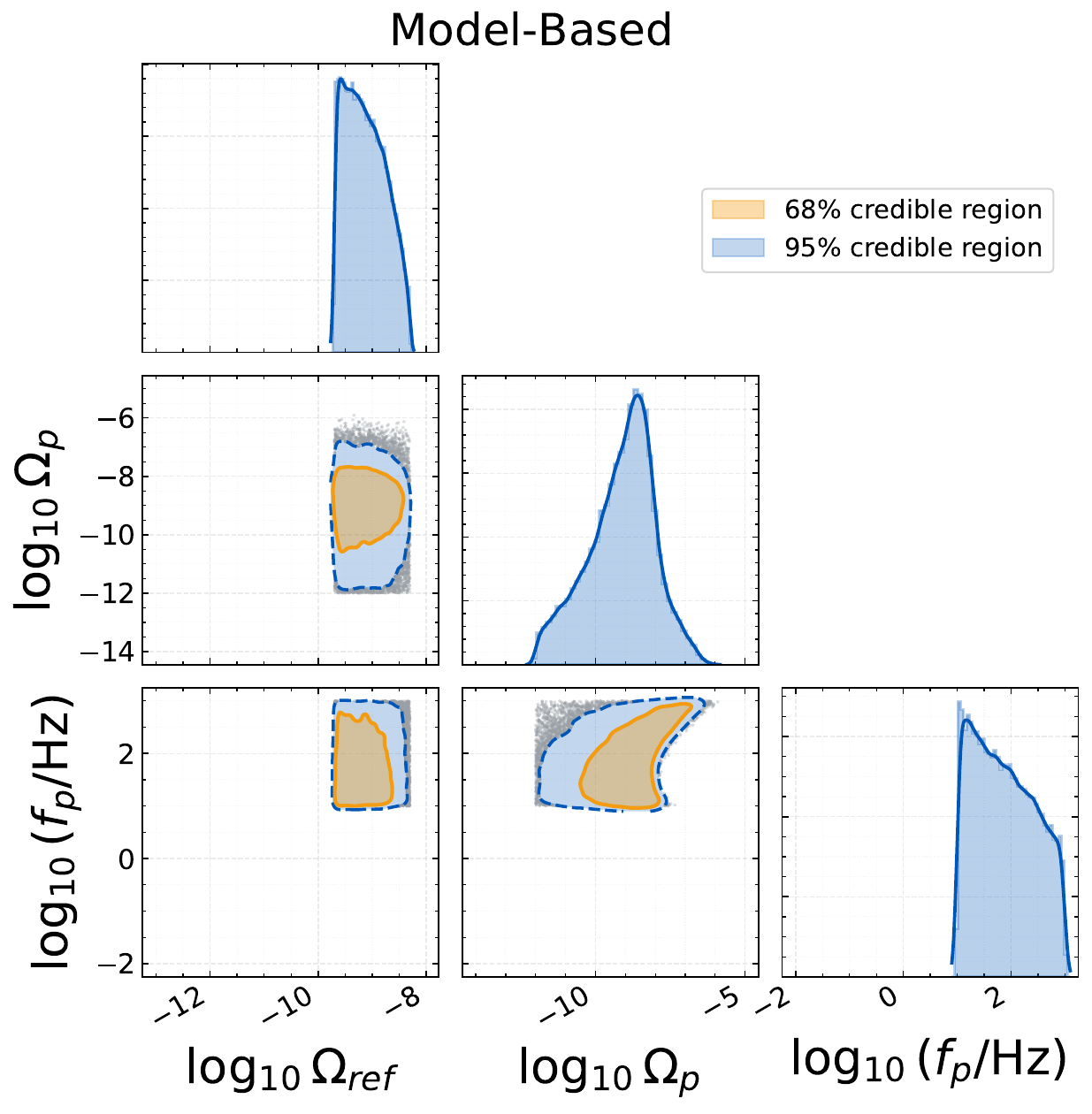}
    \end{minipage}\hfill
    \begin{minipage}{0.48\textwidth}
        \centering
        \includegraphics[height=6.9cm]{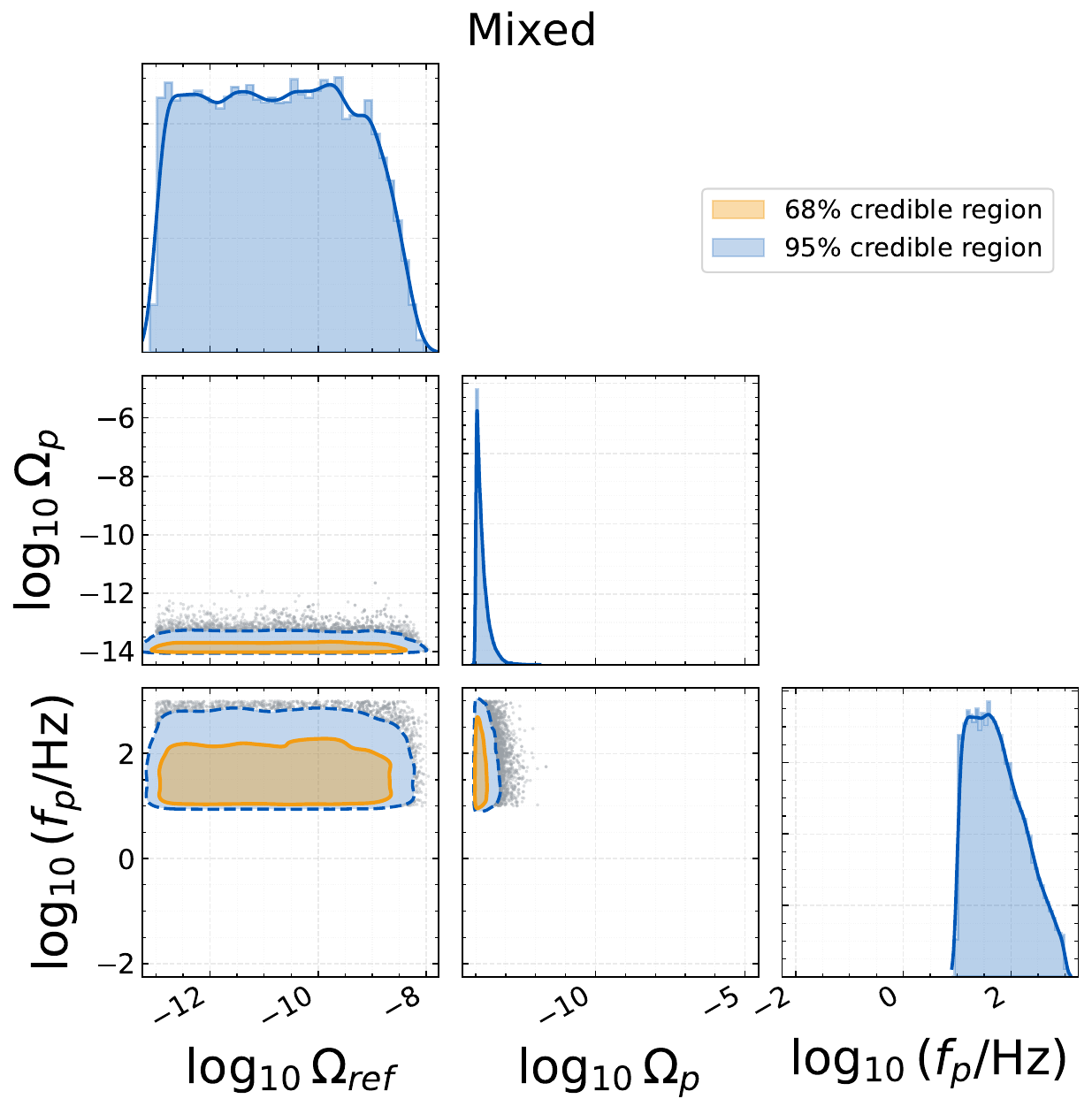}
    \end{minipage}
    \caption{Posterior distributions obtained from the O4a stochastic
    data for the simple broken power law modelling the contribution from
    bubble collisions, using the Jeffreys, Detector-Sensitivity-Based,
	Model-Based, and Mixed prior, respectively.  The different priors
    are defined in \Tabref{tab:PriorGeom_O4a}.
    }
    \label{fig:bc_bl_posterior_corner_O4}
\end{figure}

We now turn to the BPL model describing GW from bubble collisions.
In this case we present the O4a results in full, including the
robustness diagnostics of \Appref{app:robustness} for $\Omega_p$ and
$f_p$, and only briefly comment on the O3 results, as the two datasets lead to the same
conclusions and the O4a data are more constraining.
Figure~\ref{fig:bc_bl_posterior_corner_O4} shows the posterior
distributions for the geometrical parameters governing the bubble
collision contribution.  The four prior prescriptions are the same as
for the sound-wave contribution (except for $\Delta_w$, which is not
needed here), providing an important cross-check because the BPL
parameterization differs from the DBPL description of the sound-wave
contribution.

Table \ref{tab:bpl_bc_mean_overlap_kl_robustness_O4a} summarises the
robustness diagnostics for the O4a dataset, and
\Tabref{tab:bpl_bc_credible_intervals_68_O4a} gives the corresponding
$68\%$ credible intervals.  As for the DBPL analysis, the improved data
sensitivity generally leads to more localized posterior regions compared
to O3. The peak amplitude is strongly prior-dependent
with $\rho_{\max}=3.1\times10^{6}$ and $\rho_{\rm overlap}=0$ (NOT
ROBUST).  The Maximin Robustness Ratio for the peak frequency improves
markedly in absolute terms relative to O3, falling from
$\rho_{\max}=19.8$ to $2.9$.  Also the credible-region overlap
improves from $\rho_\text{overlap}=3.3\times10^{-3}$ to $0.3$.  However,
neither improvement is sufficient to change the classification and it
remains NOT ROBUST\@.
The one-dimensional posterior for the peak frequency and its
correlations with the amplitude parameters vary visibly between the four
choices, and the differences are especially apparent when comparing the
broad Jeffreys posterior with the more concentrated Model-Based and
Mixed posteriors.

The comparison between the BPL results here and the DBPL results in
\Secref{sec:dbpl_sw_results} is useful for separating two sources
of uncertainty.  Some differences between the posterior distributions arise
from the parameterisation used to describe the GW
spectrum, while others arise directly from the adopted prior.  The
persistence of qualitatively similar prior sensitivity across the two
phenomenological descriptions -- and across both the O3 and O4a datasets --
indicates that this effect is not restricted to a particular choice of
BPL template.

\begin{table}
\centering
\begin{tabular}{lcccc}
\hline\hline
Parameter & $\rho_{\max}$ & $\max_{\pi}(\mathrm{Width}(\pi))$ & $\min_{\pi}(\mathrm{Width}(\pi))$ & Status \\
\hline
$\Omega_{p}$ & $3.1 \times 10^{6}$ & $4.4 \times 10^{-8}$ & $1.4 \times 10^{-14}$ & NOT ROBUST \\
$f_{p}/\mathrm{Hz}$ & $2.9$ & $337.2$ & $118.2$ & NOT ROBUST \\
\hline
\end{tabular}
\\
\medskip
\begin{tabular}{lcccc}
\hline
Parameter & $\rho_{\mathrm{mean}}$ & $\rho_{\mathrm{overlap}}$ &
$D_{\mathrm{KL}}$ & Status \\
\hline
$\Omega_{p}$ & $1.2 \times 10^{7}$ & $0.0$ & $26.8$ & NOT ROBUST \\
$f_{p}/\mathrm{Hz}$ & $2.2$ & $0.3$ & $0.5$ & NOT ROBUST \\
\hline\hline
\end{tabular}
\caption{Maximin Robustness Ratio, posterior mean ratio, credible-region overlap, and
Kullback-Leibler divergence for the geometrical parameters of the bubble-collision BPL model, O4a
dataset.}
\label{tab:bpl_bc_mean_overlap_kl_robustness_O4a}
\end{table}

\begin{table}[t!]
\centering
\begin{tabular}{lcc}
\hline\hline
Parameter & Detector-Sensitivity-Based & Jeffreys \\
\hline
$\Omega_{p}$ & $[2.0 \times 10^{-10}, 4.9 \times 10^{-9}]$ & $[1.0 \times 10^{-12}, 4.4 \times 10^{-8}]$ \\
$f_{p}/\mathrm{Hz}$ & $[16.8, 273.7]$ & $[4.3 \times 10^{-2}, 118.2]$ \\
\hline
Parameter & Model-Based & Mixed \\
\hline
$\Omega_{p}$ & $[4.9 \times 10^{-11}, 7.6 \times 10^{-9}]$ & $[1.1 \times 10^{-14}, 2.5 \times 10^{-14}]$ \\
$f_{p}/\mathrm{Hz}$ & $[17.5, 354.7]$ & $[17.0, 200.1]$ \\
\hline\hline
\end{tabular}
\caption{$68\%$ credible intervals for the geometrical parameters of the
bubble-collision BPL model under the four priors of \Tabref{tab:PriorGeom_O4a}.}
\label{tab:bpl_bc_credible_intervals_68_O4a}
\end{table}

\subsubsection{Implications for the interpretation of Bayesian limits}
\label{sec:implications}

The results presented in this section have important consequences for the
interpretation of Bayesian upper limits and credible regions.  When the
likelihood is sufficiently informative, reasonable prior choices lead to
similar posterior distributions and the resulting limits can be regarded
as approximately data dominated.  Conversely, when the likelihood provides
only weak information about a parameter, the posterior can retain a
significant memory of the prior.

This distinction is clearly visible in the present analysis.  The peak
parameters $\Omega_p$ and, most notably, $f_p$ show large variations.
Therefore, they cannot be assigned a
single apparently precise constraint without also specifying the prior
assumptions that produced it.

We consequently regard the comparison between priors as an integral part
of the inference rather than as an auxiliary consistency check.  A
constraint is considered more reliable when its posterior support remains
substantially unchanged under reasonable variations of the prior.  When
different priors produce substantially different credible regions, the
appropriate conclusion is not that one of the priors is
incorrect, but rather that the available data are insufficient to determine
that parameter independently of prior assumptions.

This point is particularly important for searches for cosmological
SGWB in the absence of a significant
detection.  In such a regime, quoting a single Bayesian limit without
testing its stability against reasonable prior choices can give an
\emph{incomplete picture of the information supplied by the data}. Fitting the geometrical parameters directly has the advantage that it is not necessary to specify how the underlying phase-transition dynamics determine the peak amplitude and frequency of the GW signal. However,
this approach still relies on an assumed spectral template and on priors for its parameters. It therefore does not, by itself, guarantee prior-independent limits on a cosmological SGWB.
 The results of this section therefore motivate the more quantitative robustness measures
defined in \Appref{app:robustness} and applied throughout, where the overlap and relative-width
variations of the posterior distributions are used to distinguish robust
constraints from parameters that remain prior dominated.

\subsubsection{Upper limit on the astrophysical foreground}
\label{sec:Omegaref}
As mentioned earlier, the posterior distributions of the astrophysical
foreground amplitude $\Omega_\text{ref}$ do not vary significantly under
the considered priors for this parameter, which all have the same shape
but different ranges.  In this section, we discuss the effect of the
prior range on the 95\% CL upper limit on $\Omega_\text{ref}$.  We
restrict ourselves to the Jeffreys prior, for which we have already used
different ranges in the O3 and O4a analyses, see Tabs.~\ref{tab:PriorsGeom}
and~\ref{tab:PriorGeom_O4a}\@.  In addition, we consider different
combinations of datasets, prior ranges, and GW spectra as well as fit
parameters for the cosmological signal.  The results are shown in
\Tabref{tab:94_CL_Omegaref}.

\begin{table}
  \centering
  \small
  \setlength{\tabcolsep}{5pt}
  \renewcommand{\arraystretch}{1.15}
  \begin{tabular}{llllc}
    \hline
    \hline
    \textbf{Dataset} & \textbf{Fit parameters} & \textbf{GW spectrum} &
    $\bm{\Omega_\text{ref}}$ \textbf{prior} & \textbf{95\% CL limit} \\
    \hline
    \multirow{5}{*}{O3}
    & Geometrical      & Bubble collisions (BPL)  & $\logU(10^{-10},10^{-7})$ & $6.05\times10^{-9}$ \\
    & Geometrical      & Bubble collisions (BPL)  & $\logU(10^{-13},10^{-6})$ & $3.5\times10^{-9}$ \\
    & Geometrical      & Sound waves (DBPL)       & $\logU(10^{-10},10^{-7})$ & $5.99\times10^{-9}$ \\
    & Thermodynamical  & Bubble collisions (BPL)  & $\logU(10^{-10},10^{-7})$ & $6.21\times10^{-9}$ \\
    & Thermodynamical  & Sound waves + MHD (DBPL) & $\logU(10^{-10},10^{-7})$ & $6.17\times10^{-9}$ \\
    \hline
    \multirow{6}{*}{O4a}
    & Geometrical      & Bubble collisions (BPL)  & $\logU(10^{-13},10^{-6})$ & $1.95\times10^{-9}$ \\
    & Geometrical      & Bubble collisions (BPL)  & $\logU(10^{-10},10^{-7})$ & $2.00\times10^{-9}$ \\
    & Geometrical      & Sound waves (DBPL)       & $\logU(10^{-13},10^{-6})$ & $1.97\times10^{-9}$ \\
    & Geometrical      & Sound waves (DBPL)       & $\logU(10^{-10},10^{-7})$ & $2.0\times10^{-9}$ \\
    & Thermodynamical  & Bubble collisions (BPL)  & $\logU(10^{-13},10^{-6})$ & $2.01\times10^{-9}$ \\
    & Thermodynamical  & Sound waves + MHD (DBPL) & $\logU(10^{-13},10^{-6})$ & $1.99\times10^{-9}$ \\
    \hline
    \hline
  \end{tabular}
  \caption{95\% CL upper limit on the amplitude $\Omega_\text{ref}$ of the
  astrophysical foreground for the Jeffreys prior.  The fit parameters are
  either the geometrical ones of \Secref{sec:pestimation_geometrical} or the
  thermodynamical ones of \Secref{sec:pestimation_thermo}.  In both cases
  the GW spectrum of the phase transition is a BPL for bubble collisions and
  a DBPL for sound waves, where the thermodynamical analysis also includes
  MHD turbulence.  The fourth column gives the prior range for
  $\Omega_\text{ref}$.  The default ranges are those of
  \Tabref{tab:PriorsGeom} for O3 and \Tabref{tab:PriorGeom_O4a} for O4a;
  for comparison, we also give the limits for O3 data with the O4a range
  and for O4a data with the O3 range.}
  \label{tab:94_CL_Omegaref}
\end{table}

We observe that our results are well compatible with those of previous
studies in the cases where the same dataset \emph{and prior range} is
used.  Specifically, for the O3 data, the limits using the range
$(10^{-13},10^{-7})$ agree with the results of Ref.~\cite{Badger:2022jzg},
which used the same range.  However, they differ from the limit 
$\Omega_\text{ref} < 3.4\times10^{-9}$ reported by the LVK collaboration
\cite{Abbott:2021xyo} by almost a factor of $2$.  That LVK limit is
based on a prior range with a much smaller lower edge of $10^{-13}$.%
\footnote{We were unable to find a specification of the upper edge in
Ref.~\cite{Abbott:2021xyo}.}
If we perform our analysis with the same lower prior edge, the derived
upper limit agrees well with the LVK result.

A dependence of the upper limit on the prior range can be attributed to
the fact that the likelihood becomes nearly flat below the detector
sensitivity.  Thus, decreasing the lower prior bound increases the
posterior probability assigned to this region.  Consequently, the $95\%$
upper credible limit can be pulled towards smaller values.

Interestingly, we do not observe such a dependence using the O4a dataset.
All our analyses yield a limit that is virtually identical to the value
of $2.0\times10^{-9}$ reported by LVK~\cite{LVK:2025O4a}, independently
of whether we use the same lower edge of $10^{-13}$ as LVK or the larger
value of $10^{-10}$ from the O3 prior set.

Thus, we find that the dependence of a limit on the prior range can be
very different for different datasets.  Presumably, it can only be
determined by repeating the analysis with different prior ranges.

\subsection{Results based on different priors for the thermodynamical parameters}
\label{sec:thermodynamical_results}

The analysis in the previous section demonstrated that
the inferred geometrical parameters of the stochastic GW
spectrum exhibit a significant dependence on the prior.
However, the geometrical parameters are not the fundamental
quantities describing the underlying phase transition.  In order to assess
how the prior dependence propagates to the physical properties of the
transition, we next perform the inference directly in terms of the
thermodynamical phase-transition parameters.
As in \Secref{sec:geometrical_results}, we present the O4a analysis in full for both templates considered below.

In contrast to the geometrical analysis, sampling in the thermodynamical
parameter space involves a larger number of parameters and therefore
exposes additional degeneracies.  Several combinations of thermodynamical
parameters can produce very similar GW spectra.  
Consequently, a constraint on one thermodynamic quantity should not be
interpreted independently of the allowed ranges for the other parameters.
This makes the dependence on the adopted prior especially relevant when
interpreting posterior constraints on the underlying phase transition.

\subsubsection{Bubble-collision template}
\label{sec:template_bc_results}
\begin{figure}
    \centering
    \begin{minipage}{0.48\textwidth}
        \centering
        \includegraphics[height=6.9cm]{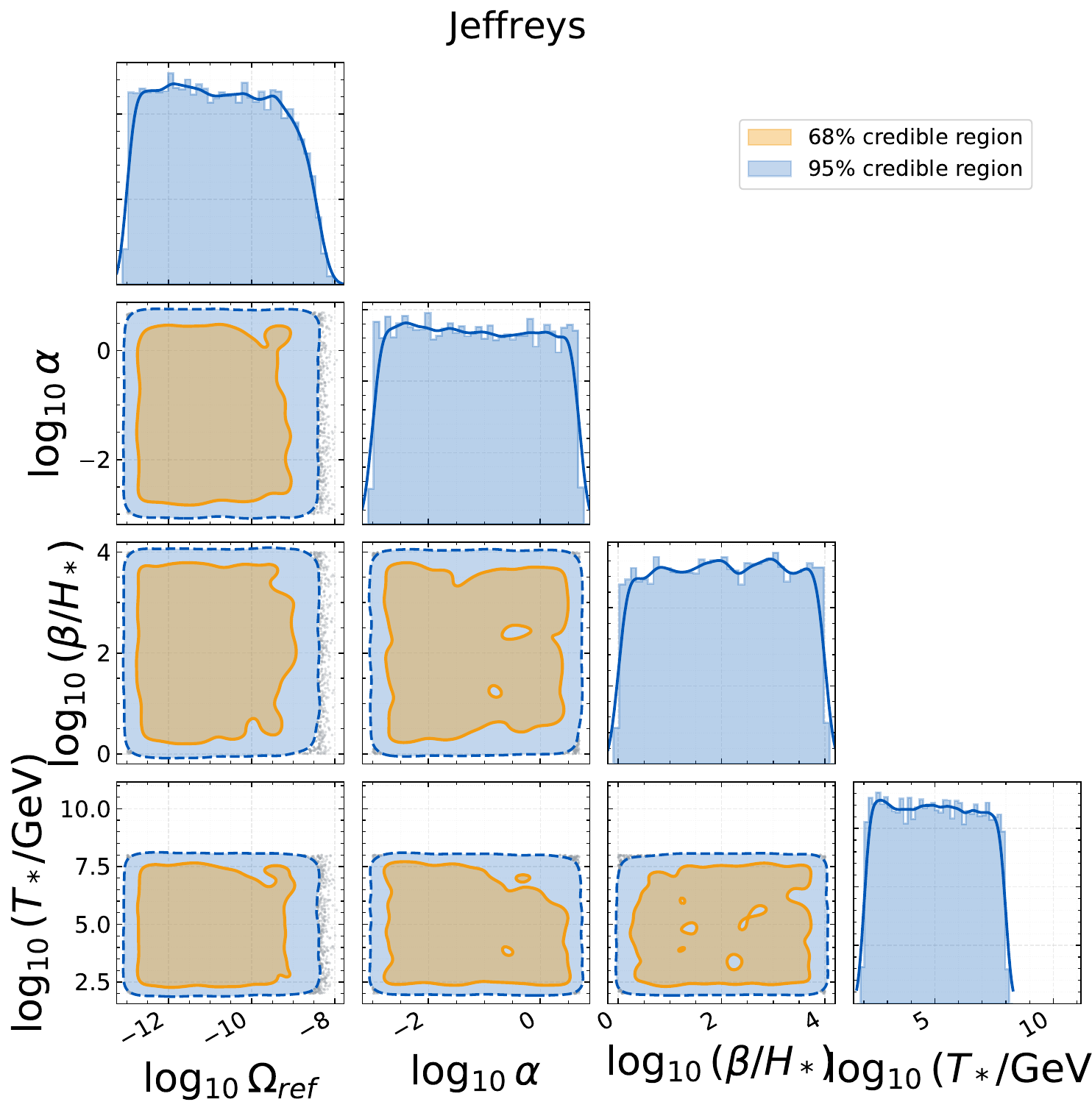}      
    \end{minipage}\hfill
    \begin{minipage}{0.48\textwidth}
        \centering
        \includegraphics[height=6.9cm]{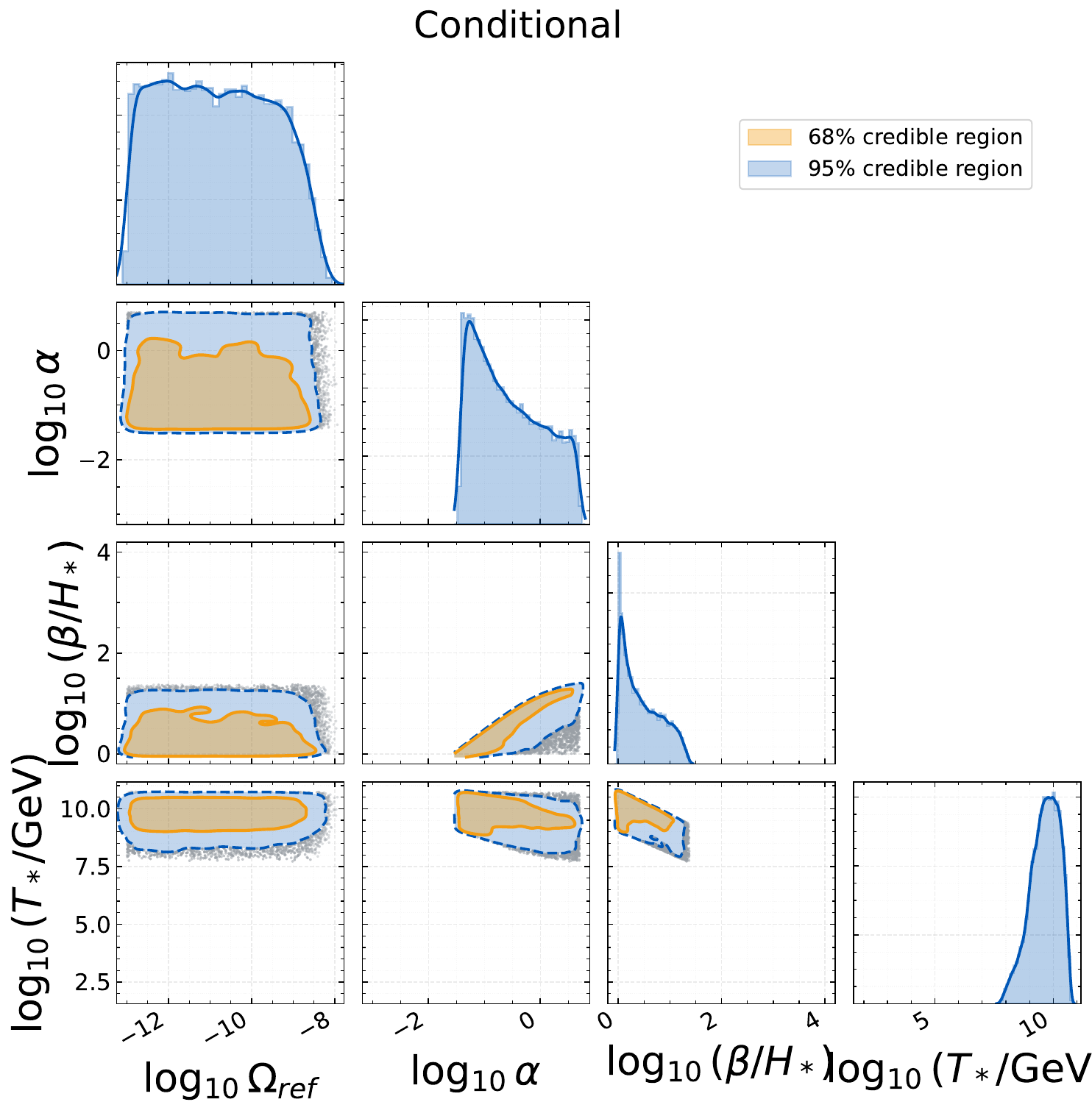}
    \end{minipage}
    \caption{Posterior distributions for the thermodynamical parameters
    determining the GW spectrum from bubble collisions, using the O4a
    stochastic data and two different priors, Jeffreys and Conditional.}
    \label{fig:template_bc_posterior_corner_O4a}
\end{figure}

For the bubble-collision contribution, we consider the parameters
$\left\{ \alpha,\, \beta/H_\ast,\, T_\ast \right\}$, see \Secref{sec:ThermoPars}.
The priors are specified in \Tabref{tab:analysis_nucl_O4a}\@.
Figure~\ref{fig:template_bc_posterior_corner_O4a} shows the posterior
distributions obtained from the O4a stochastic data.  The two corner
plots correspond to the Jeffreys and the Conditional prior, respectively.
The corresponding robustness diagnostics are collected in
Tabs.~\ref{tab:template_bc_mean_overlap_kl_robustness_O4a} and~\ref{tab:template_bc_credible_intervals_68_O4a}.

The comparison between the Jeffreys and Conditional priors provides a direct
test of the robustness of the thermodynamical inference.  The
transition strength $\alpha$ is ROBUST with $\rho_{\max}=1.2$,
$\rho_{\rm overlap}=0.8$, and $D_{\rm KL}=9.1\times10^{-2}$.
This classification is driven by the upper edge of its $68\%$ credible
interval in \Tabref{tab:template_bc_credible_intervals_68_O4a}, which is similar for the two priors ($1.3$ for Jeffreys and $1.6$ for
Conditional), whereas the lower edge still
moves with the prior, from $3.8\times10^{-3}$ (Jeffreys) to $5.9\times10^{-2}$
(Conditional).  In
contrast, $\beta/H_\ast$ and $T_\ast$ are both NOT ROBUST with respect
to all measures.  For these two
parameters the posterior distributions obtained with the two priors overlap
only marginally or not at all, and both the extent and the location of the
allowed regions change between the two prior choices.
This is particularly important for parameters for which the likelihood does
not strongly localize the posterior.  In such cases, the posterior
distribution retains a significant contribution from the prior volume.

The temperature $T_\ast$ is also strongly correlated with the other
transition parameters and the least robust quantity due to
$\rho_{\max}=1.8\times10^{3}$ and $\rho_{\rm overlap}=0$, although
$D_{\rm KL}=2.4$ is a bit smaller than the value of $6.5$ obtained for
$\beta/H_*$.  Its $68\%$ credible intervals do not overlap at all: the
one for the Jeffreys prior, $[858.9,\,1.0\times10^{7}]$\,GeV,
lies entirely below the one for the Conditional prior,
$[1.0\times10^{9},\,1.9\times10^{10}]$\,GeV\@.  This behaviour is expected
from the relation \eqref{eq:BPL_frequency}
between the characteristic frequency of the GW
signal, the redshifted Hubble scale at the transition (which is
proportional to $T_*$), and the inverse duration $\beta/H_*$.  A change in
$T_\ast$ can be compensated, over part of the parameter space, by
a corresponding change in $\beta/H_\ast$, for which
$\rho_{\max}=320.2$ and $\rho_{\rm overlap}=1.5\times10^{-3}$.
Going from the Jeffreys to the Conditional prior moves $T_\ast$ to higher
values and $\beta/H_\ast$ to lower ones, from $[4.8,\,2.3\times10^{3}]$ to
$[1.2,\,8.3]$.  The resulting
degeneracy is visible in the two-dimensional posterior distributions.

Similarly, the transition strength $\alpha$ and the inverse duration
$\beta/H_\ast$ are not independently determined by the stochastic data.
The amplitude of the bubble-collision contribution depends on the amount of
energy released by the transition and on the characteristic duration of
the transition.  As a consequence, different combinations of $\alpha$ and
$\beta/H_\ast$ can lead to comparable GW amplitudes.

Figure \ref{fig:template_bc_posterior_corner_O4a} illustrates an important difference between constraining
a phenomenological spectrum and constraining the underlying phase
transition.  Even when the geometrical parameters of the spectrum appear
reasonably localised, the inverse mapping to thermodynamical parameters can
introduce substantial degeneracies.  The resulting posterior sensitivity
to the prior must consequently be taken into account before interpreting
the inferred values of $\alpha$, $\beta/H_\ast$, or $T_\ast$ as quantitative
measurements of the phase transition.

\begin{table}
\centering
\begin{tabular}{lcccc}
\hline
\textbf{Parameter} & $\rho_{\max}$ & $\max_\pi(\mathrm{Width}(\pi))$ & $\min_\pi(\mathrm{Width}(\pi))$ & \textbf{Status} \\
\hline
$\alpha$ & $1.2$ & $1.5$ & $1.3$ & ROBUST \\
$\beta/H_\ast$ & $320.2$ & $2.3\times10^{3}$ & $7.1$ & NOT ROBUST \\
$T_\ast/\mathrm{GeV}$ & $1.8\times10^{3}$ & $1.8\times10^{10}$ & $1.0\times10^{7}$ & NOT ROBUST \\
\hline
\end{tabular}
\\
\medskip
\begin{tabular}{lcccc}
\hline
\textbf{Parameter} & $\rho_{\rm mean}$ & $\rho_{\rm overlap}$ & $D_{\rm KL}$ & \textbf{Status} \\
\hline
$\alpha$ & $1.3$ & $0.8$ & $9.1\times10^{-2}$ & ROBUST \\
$\beta/H_\ast$ & $251.7$ & $1.5\times10^{-3}$ & $6.5$ & NOT ROBUST \\
$T_\ast/\mathrm{GeV}$ & $1.3\times10^{3}$ & $0.0$ & $2.4$ & NOT ROBUST \\
\hline
\end{tabular}
\caption{Maximin Robustness Ratio, posterior mean ratio, credible-region
overlap, and Kullback-Leibler divergence for the thermodynamical
parameters of the bubble-collision template, O4a dataset.}
\label{tab:template_bc_mean_overlap_kl_robustness_O4a}
\end{table}

\begin{table}[t!]
\centering
\begin{tabular}{lcc}
\hline
\textbf{Parameter} & Conditional & Jeffreys \\
\hline
$\alpha$ & $[5.9\times10^{-2},\,1.6]$ & $[3.8\times10^{-3},\,1.3]$ \\
$\beta/H_\ast$ & $[1.2,\,8.3]$ & $[4.8,\,2.3\times10^{3}]$ \\
$T_\ast/\mathrm{GeV}$ & $[1.0\times10^{9},\,1.9\times10^{10}]$ & $[858.9,\,1.0\times10^{7}]$ \\
\hline
\end{tabular}
\caption{$68\%$ credible intervals for the thermodynamical parameters of
the bubble-collision template, O4a dataset.}
\label{tab:template_bc_credible_intervals_68_O4a}
\end{table}

\subsubsection{Sound waves and MHD turbulence}
\label{sec:template_swmhd_results}
\begin{figure}
    \centering
    \begin{minipage}{0.48\textwidth}
        \centering
        \includegraphics[height=6.9cm]{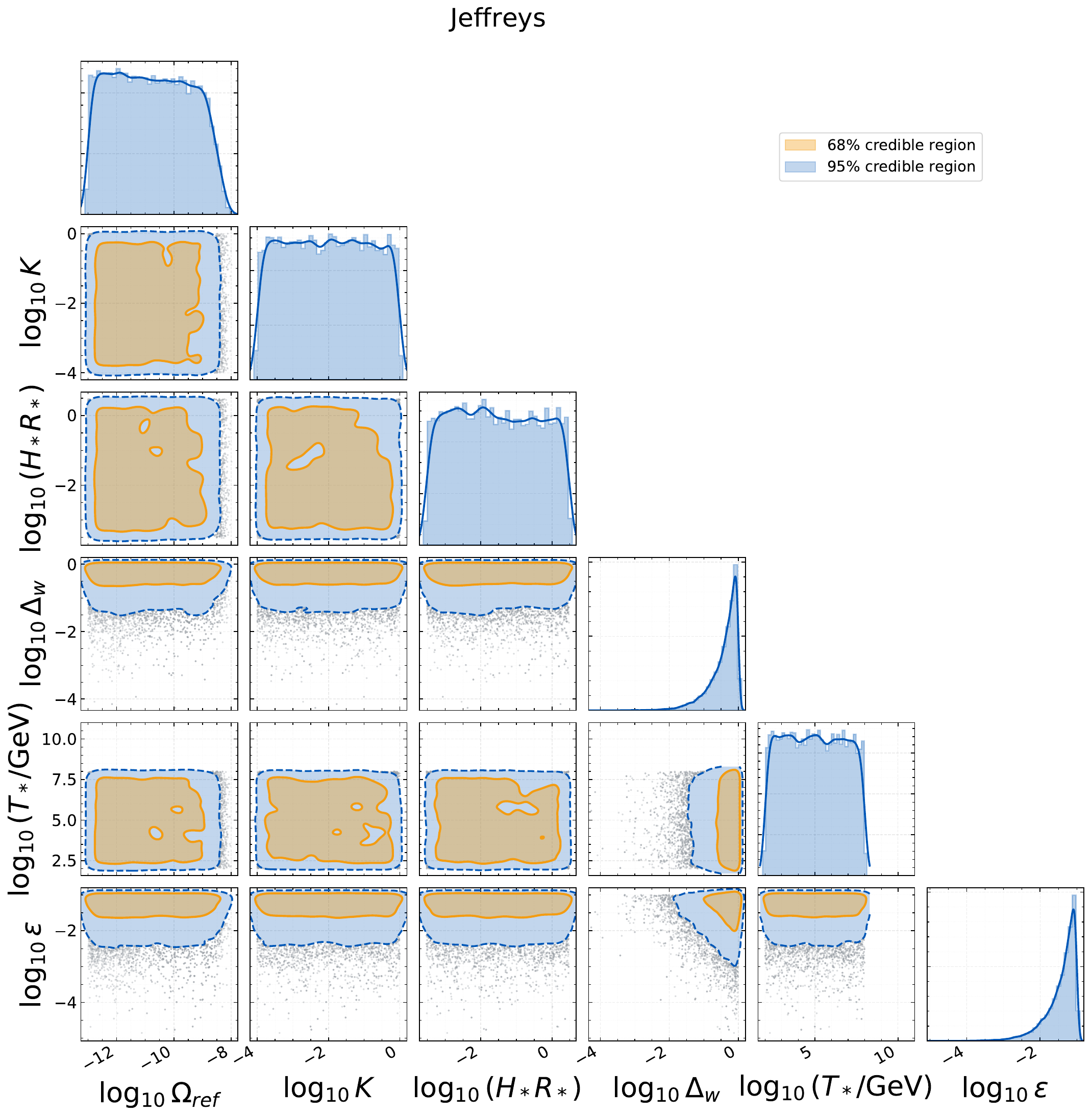}      
    \end{minipage}\hfill
    \begin{minipage}{0.48\textwidth}
        \centering
        \includegraphics[height=6.9cm]{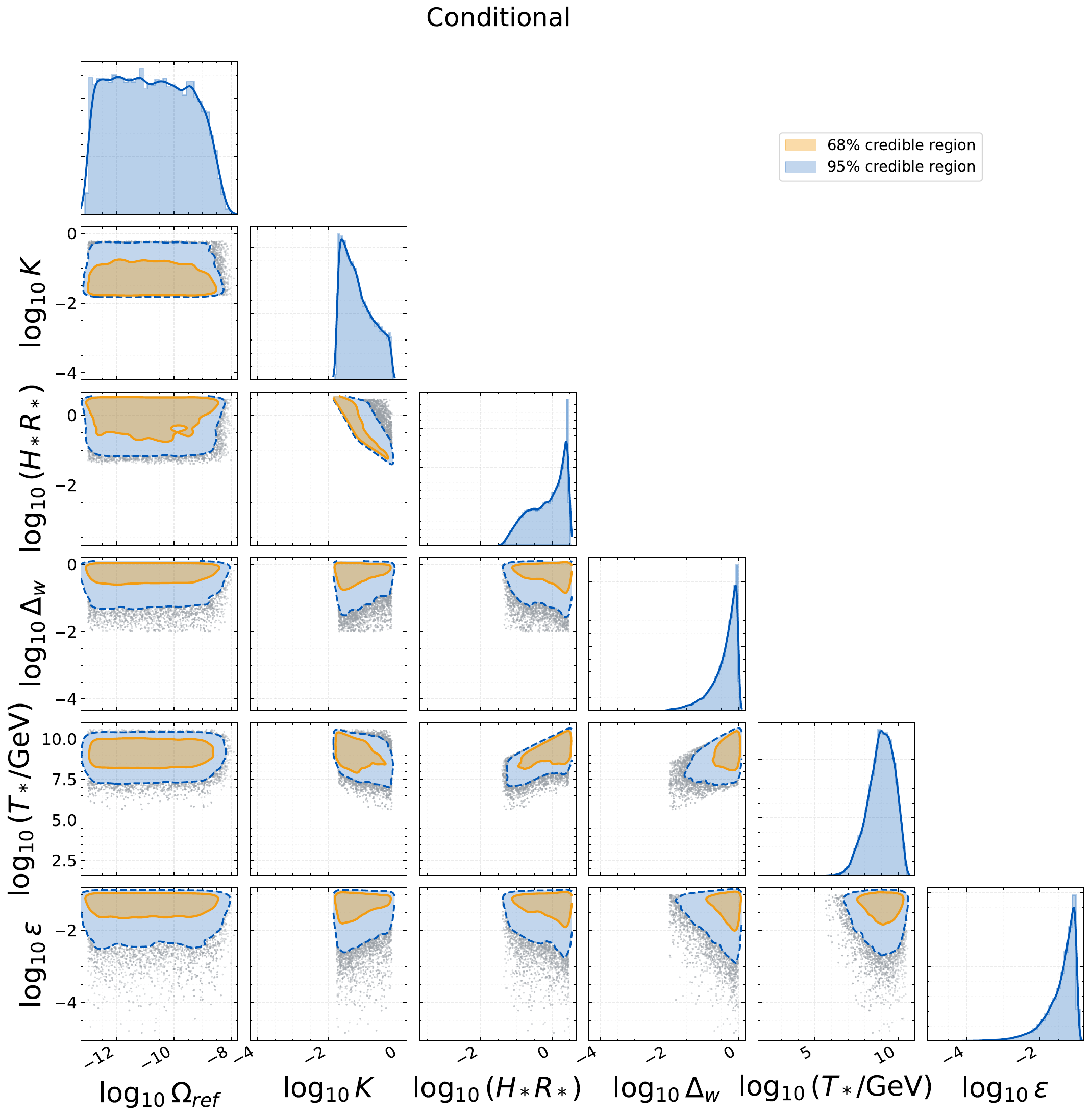}
    \end{minipage}
    \caption{Posterior distributions for the thermodynamical parameters
    determining the GW spectrum from a combination of sound waves and
    MHD turbulence, modelled by a double broken power law.  The results
	stem from the O4a stochastic data and two different priors, Jeffreys
    and Conditional, defined in \Tabref{tab:O4_KR}.}
    \label{fig:sw_mhd_mpbl_posterior_corner_O4a}
\end{figure}

For the combined sound-wave and MHD contribution, it is convenient to
use the variables
$\left\{ K,\, H_\ast R_\ast,\, T_\ast,\, \Delta_w,\, \varepsilon \right\}$,
as discussed in \Secref{sec:pestimation_thermo}.
The priors for these parameters are specified in \Tabref{tab:O4_KR}.
Figure \ref{fig:sw_mhd_mpbl_posterior_corner_O4a} shows the results of
the corresponding analysis based on the O4a dataset.
The increased dimensionality of this parameter space leads to a richer
correlation structure than in the bubble-collision analysis. In particular,
the posterior distributions reflect that the same amplitude of the
GW signal can be obtained through different combinations of
the bulk kinetic-energy fraction $K$, the characteristic bubble separation
$H_\ast R_\ast$, the sound-shell parameter $\Delta_w$, and the turbulence fraction $\varepsilon$.

The quantitative robustness
diagnostics under a change from Jeffreys to Conditional prior are
collected in
Tabs.~\ref{tab:template_sw_mhd_mean_overlap_kl_robustness_O4a} and~\ref{tab:template_sw_mhd_credible_intervals_68_O4a}.
As in \Secref{sec:dbpl_sw_results}, we do not present the robustness
measures for $\Omega_\text{ref}$ and $\Delta_w$, whose priors we did not
vary.  For the same reason, the measures for the turbulent energy
fraction~$\epsilon$ are omitted.

The bulk kinetic-energy fraction $K$ has a fairly large Kullback-Leibler
divergence $D_{\rm KL}=0.9$ but is nevertheless classified as ROBUST due
to the remaining diagnostics $\rho_{\max}=1.2$, $\rho_{\rm
overlap}=0.9$, and $\rho_\text{mean}=1.0$.  The lower edge of its $68\%$
credible interval moves from $4.3\times10^{-4}$ (Jeffreys) to
$2.4\times10^{-2}$ (Conditional), while the upper edge, $0.2$, is the
same for both priors.  The mean bubble separation
$H_\ast R_\ast$ is NOT ROBUST ($\rho_{\max}=3.3$, $\rho_{\rm overlap}=0.2$,
$D_{\rm KL}=0.9$), with $68\%$ credible intervals $[1.3\times10^{-3},\,0.7]$
(Jeffreys) and $[0.2,\,2.4]$ (Conditional).  The
corresponding degeneracies are visible in the two-dimensional posterior
distributions of \Figref{fig:sw_mhd_mpbl_posterior_corner_O4a}.

The transition temperature $T_\ast$ again appears in correlations with the
parameters controlling the characteristic frequency and amplitude of the
signal, and it is by far the least robust parameter of this analysis
with $\rho_{\max}=567.1$, $\rho_{\rm overlap}=0$ and $D_{\rm KL}=2.4$.
Its $68\%$ credible intervals, $[886.5,\,1.0\times10^{7}]$\,GeV for the
Jeffreys prior and $[1.6\times10^{8},\,5.9\times10^{9}]$\,GeV for the
Conditional prior, do not overlap.
This reflects the fact that the redshifted Hubble scale at the transition
enters directly into the mapping between the phase-transition parameters
and the GW spectrum.

\begin{table}
\centering
\begin{tabular}{lcccc}
\hline
\textbf{Parameter} & $\rho_{\max}$ & $\max_\pi(\mathrm{Width}(\pi))$ & $\min_\pi(\mathrm{Width}(\pi))$ & \textbf{Status} \\
\hline
$K$ & $1.2$ & $0.2$ & $0.2$ & ROBUST \\
$H_\ast R_\ast$ & $3.3$ & $2.2$ & $0.7$ & NOT ROBUST \\
$T_\ast/\mathrm{GeV}$ & $567.1$ & $5.8\times10^{9}$ & $1.0\times10^{7}$ & NOT ROBUST \\
\hline
\end{tabular}
\\
\medskip
\begin{tabular}{lcccc}
\hline
\textbf{Parameter} & $\rho_{\rm mean}$ & $\rho_{\rm overlap}$ & $D_{\rm
KL}$ & \textbf{Status} \\
\hline
$K$ & $1.0$ & $0.9$ & $0.9$ & ROBUST \\
$H_\ast R_\ast$ & $3.7$ & $0.2$ & $0.9$ & NOT ROBUST \\
$T_\ast/\mathrm{GeV}$ & $438.6$ & $0.0$ & $2.4$ & NOT ROBUST \\
\hline
\end{tabular}
\caption{Robustness measures for the thermodynamical parameters of the
sound-wave $+$ MHD template, O4a dataset.}
\label{tab:template_sw_mhd_mean_overlap_kl_robustness_O4a}
\end{table}

\begin{table}[t!]
\centering
\begin{tabular}{lcc}
\hline
\textbf{Parameter} & Conditional & Jeffreys \\
\hline
$K$ & $[2.4\times10^{-2},\,0.2]$ & $[4.3\times10^{-4},\,0.2]$ \\
$H_\ast R_\ast$ & $[0.2,\,2.4]$ & $[1.3\times10^{-3},\,0.7]$ \\
$T_\ast/\mathrm{GeV}$ & $[1.6\times10^{8},\,5.9\times10^{9}]$ & $[886.5,\,1.0\times10^{7}]$ \\
\hline
\end{tabular}
\caption{$68\%$ credible intervals for the thermodynamical parameters of
the sound-wave $+$ MHD template, O4a dataset.}
\label{tab:template_sw_mhd_credible_intervals_68_O4a}
\end{table}

\subsubsection{Comparison of the O3 and O4a thermodynamical analyses}
\label{sec:thermo_comparison}
Figures \ref{fig:template_bc_posterior_corner_O4a} and \ref{fig:sw_mhd_mpbl_posterior_corner_O4a} provide two complementary descriptions of the O4a
constraints.
The former 
considers the bubble-collision contribution, while
the latter 
includes the sound-wave and MHD contributions.  Despite the different
spectral models and parameterizations, both analyses illustrate the same
general feature: the absence of a statistically significant cosmological
stochastic background implies that the likelihood does not completely
localize the underlying phase-transition parameters, for either run.

This is particularly relevant for the parameters $\alpha$, $\beta/H_\ast$,
$T_\ast$, $K$, and $H_\ast R_\ast$.  These quantities enter the
GW spectrum through combinations rather than independently,
and therefore broad degeneracy directions can remain even when the
GW amplitude itself is constrained.  A narrow posterior in
one direction of parameter space should therefore not be interpreted as a
model-independent determination of the corresponding physical quantity.

The comparison of the Jeffreys and Conditional posteriors provides a
conservative diagnostic of this effect.  Parameters for which the two
posterior regions overlap substantially can be regarded as comparatively
stable with respect to the choice of prior, whereas parameters for which the
posterior regions are displaced or have substantially different widths
should be regarded as prior sensitive.

As emphasised in \Secref{sec:template_bc_results} and \Secref{sec:template_swmhd_results},
moving from O3 to O4a does not uniformly improve the robustness
classification of a parameter, even though it typically narrows the
individual posteriors: $\beta/H_\ast$ and $H_\ast R_\ast$ remain firmly NOT
ROBUST in both runs despite a $\sim\!25$--$50\%$ reduction in $\rho_{\max}$,
while $T_\ast$ is, if anything, marginally less robust in O4a than in O3 in
both templates.  
Increased statistical sensitivity is therefore not itself a
test of prior independence: a reduction in $\rho_{\max}$ or in credible-interval
width reflects the likelihood becoming more informative, while whether the
posterior actually becomes prior independent is a separate question that
must be answered with the overlap- and divergence-based diagnostics of
\Appref{app:robustness}, applied to each run individually.

This distinction is important because the transformation from geometrical
to thermodynamical parameters is not innocuous from a statistical
perspective.  A prior that is relatively uninformative in one set of
variables need not remain uninformative after a nonlinear transformation.
Thus, prior robustness should ideally be tested in the physical parameter
space in which the final constraints are quoted, rather than inferred only
from the apparent stability of the geometrical posterior, or assumed to
improve automatically as more sensitive data become available.

Overall, the thermodynamic analysis reinforces the conclusion obtained
from the geometrical parameter study: Bayesian limits on an SGWB
should be interpreted together with an
explicit, run-by-run assessment of prior dependence.  This is particularly
important when translating the absence of a detected signal into constraints on fundamental properties of a FOPT.

\subsection{Results for the Abelian Higgs model}
\label{sec:U1_results}
The phenomenological analysis presented in the previous sections
constrains the parameters describing the GW spectrum and, subsequently,
the thermodynamical properties of the phase transition.  We now apply
the same inference framework directly to a microscopic realisation of a
FOPT, the minimal Abelian Higgs model with gauge group $U(1)_X$
described in \Secref{sec:U1_direct_sampling}.  This setup provides a
simple microscopic framework that determines the phase-transition
properties and the resulting GW spectrum by the underlying model
parameters.  Thus, the constraints from SGWB searches can be propagated
to fundamental particle-physics parameters, allowing us to investigate
the dependence of the inferred particle-physics parameters on the choice
of prior.

\begin{figure}
    \centering
    \includegraphics[width=0.45\linewidth]{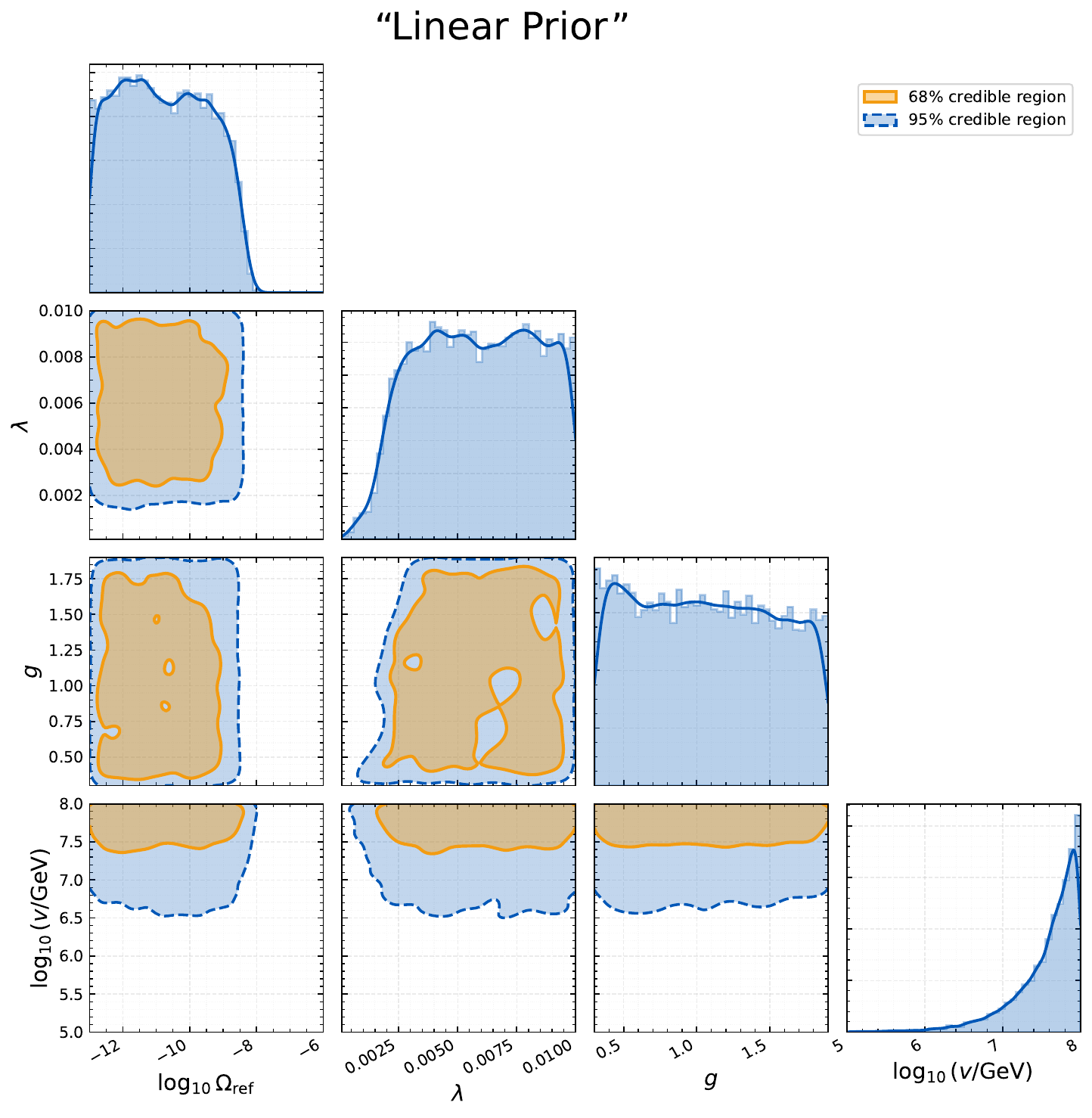}
    \includegraphics[width=0.45\linewidth]{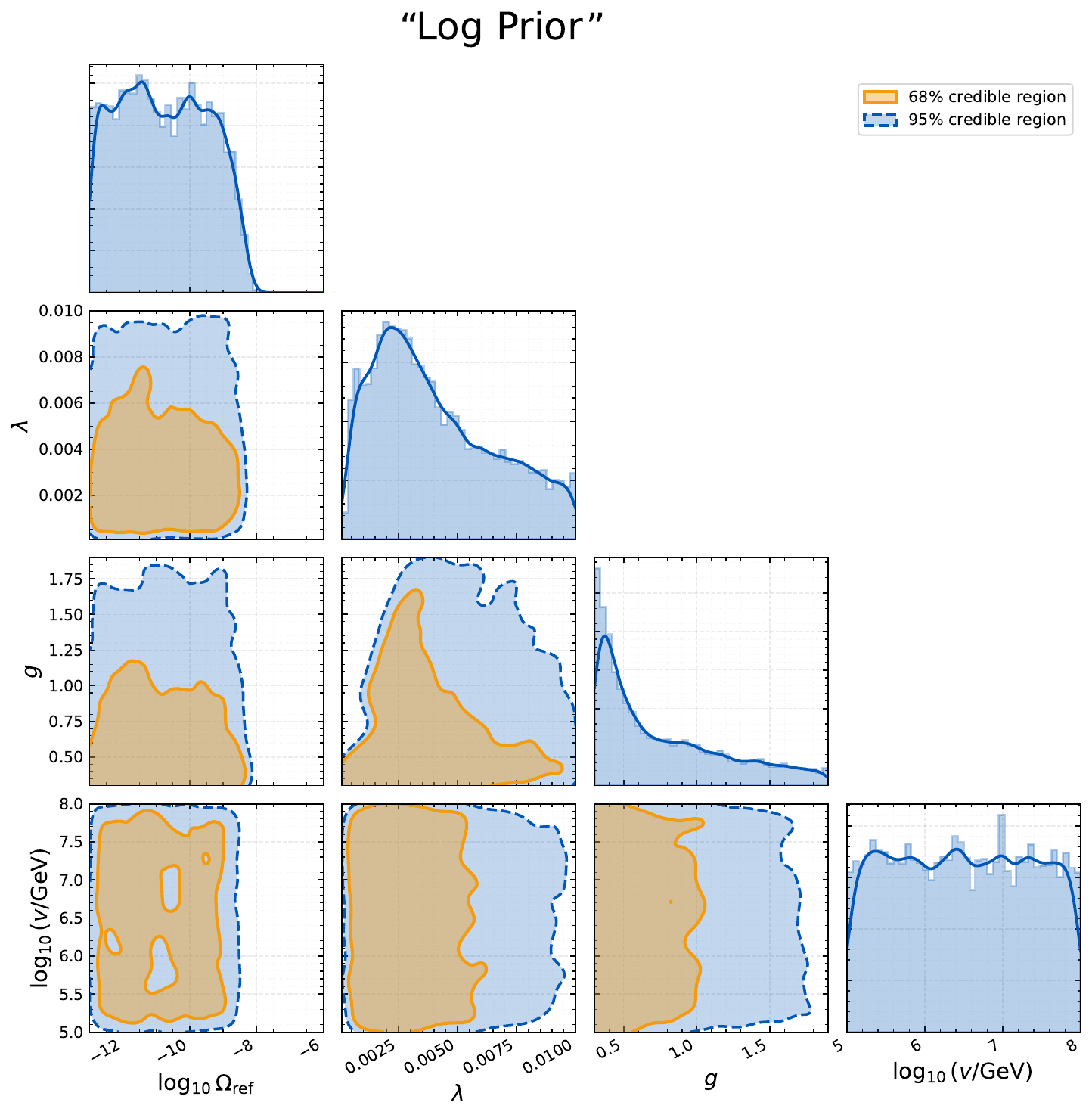}
    \caption{Results for the direct sampling of fundamental parameters of the $U(1)$ model for the O4a dataset. The plot on the left shows the results for a linear prior and the one on the right those for a logarithmic prior. We have used the LTE approximation in \texttt{WallGo} for this case.}
    \label{fig:abelian_res_O4a}
\end{figure}

We analyse the O4a stochastic data by sampling $\theta=\{\Omega_{\rm
ref},g,\lambda,v\}$.  We compare uniform and log-uniform priors for $g$,
$\lambda$, and $v$, which are given in \Tabref{tab:U1priors} and
referred to as \emph{linear} and \emph{log} prior, respectively, in the
following.  In \Figref{fig:abelian_res_O4a}, we present the resulting
marginalised posterior distributions and the correlations between the
sampled parameters for the two priors.  Also in this analysis, the
distributions of the model parameters change with the prior choice, with
the log-uniform prior shifting the posterior support towards smaller
values, particularly for $\lambda$ and $v$. In contrast, the posterior
for $\Omega_{\rm ref}$ remains similar between the two analyses, for
which its prior is unchanged.

Since these results rely on the LTE approximation
for the wall velocity, we first compare this
prescription with full collision-based \texttt{WallGo}~\cite{Ekstedt:2024etx}
calculations at selected parameter points.
We then examine the $68\%$ credible intervals
under the two priors and quantify their differences using the robustness diagnostics
introduced previously.

\subsubsection{Representative full wall-dynamics cross-checks}
\label{sec:LTEcheck}
The model-level posterior figures in \Figref{fig:abelian_res_O4a} are based on the LTE
wall-velocity prescription. To assess its behaviour in selected
regions of the NLO dimensionally reduced parameter space, we
performed independent, off-equilibrium \texttt{WallGo} calculations
with collision integrals for nine representative points drawn from
successful LTE scan evaluations. The calculations were carried out
at the archived transition temperature $T_\ast$ for each point,
without adjusting the temperature to seek a particular wall
solution. The native solver classifications and velocities are
collected in \Tabref{tab:u1_fullwallgo_validation}.

\begin{table}[t]
\centering
\setlength{\tabcolsep}{5pt}
\begin{tabular}{@{}lcccl@{}}
\toprule
Point & $v_w^{\rm LTE}$ & $v_w^{\rm full}$ & $N$ & Native status \\
\midrule
A & $0.580640$ & $0.580179$ & $5$ & Deflagration \\
B & $0.583337$ & $0.582916$ & $5$ & Deflagration \\
C & $1$ & --- & $5,7$ & \texttt{RUNAWAY} \\
D & $0.592595$ & $0.591981$ & $5$ & Deflagration \\
E & $1$ & $0.588903$ & $5,7$ & Near-Jouguet candidate \\
F & $1$ & --- & $5$ & \texttt{RUNAWAY} \\
G & $0.577516$ & $0.577120$ & $5$ & Deflagration \\
H & $1$ & --- & $5$ & \texttt{RUNAWAY} \\
I & $1$ & --- & $5$ & \texttt{RUNAWAY} \\
\bottomrule
\end{tabular}
\caption{Representative comparisons between the LTE prediction
and full collision-based \texttt{WallGo} calculations. $N$ denotes
the momentum grid size. A dash in $v_w^{\rm full}$ indicates that
no stationary velocity was returned; RUNAWAY is the native
solver classification, not a measured terminal speed. Point E is a
near-Jouguet candidate: its reported velocity equals the
Jouguet boundary at both resolutions and must not be treated as an
independently converged root. Other single-grid stationary solutions
are provisional with respect to grid convergence.}
\label{tab:u1_fullwallgo_validation}
\end{table}

For the four representative subluminal LTE solutions A, B, D
and G, the relative differences between the full and LTE velocities
are approximately $0.079\%$, $0.072\%$, $0.104\%$ and $0.069\%$,
respectively. An independent stationary reference point, checked
at $N=5$, $7$ and $9$, likewise shows a stable full-wall velocity
near $0.5883$. These comparisons indicate small off-equilibrium
corrections to the wall speed in the tested stationary regime, but
are not a systematic error estimate for the entire posterior.

Point E is qualitatively different: its LTE prescription returns
$v_w^{\rm LTE}=1$, whereas the full solver returns the
DEFLAGRATION classification with
$v_w^{\rm full}=0.588903$ at both $N=5$ and $N=7$. This reported
value coincides exactly with the Jouguet velocity $v_J$.
At $N=7$ the wall pressure changes sign close to $v_J$. However,
the identical returned values should not be interpreted as
proof of velocity convergence: the solution lies at a hydrodynamic
branch boundary and may reflect the numerical stopping criterion.
We therefore regard E as a near-Jouguet stationary candidate, not
as a precision determination of a terminal wall velocity.

By contrast, C, F, H and I are classified RUNAWAY in both
the deflagration/hybrid and detonation searches. For C this
classification persists from $N=5$ to $N=7$, with negative wall
pressures across the sampled detonation velocities. For F, H and I
only the $N=5$ result has been checked. These classifications are
conditional on the included particle content, collision treatment,
and approximations of the thermal effective description. In
particular, the collision integrals here use leading-order thermal
mass inputs, while the equilibrium potential is obtained by NLO
dimensional reduction. A native RUNAWAY classification
should not be identified with a stationary solution at $v_w=1$ or
a complete prediction for the division of released energy between
plasma motion and scalar-field gradients.

The validation therefore supports the LTE velocity approximation
for the tested subluminal stationary examples, but also shows that
$v_w^{\rm LTE}=1$ alone does not uniquely identify the outcome of
the off-equilibrium wall calculation. We consequently retain the
LTE-based posterior analysis as the principal result and present
these tests as limited wall-dynamics benchmarks; a complete
collision-based reanalysis of the GW spectra is beyond the scope
of the present study.

\subsubsection{Robustness results \label{subsec:abelian_robustness}}

The most direct way to see the effect of the prior is to put the $68\%$ credible intervals side by side (\Tabref{tab:swmhd_ci68_O4a}), because the different
credible regions show up immediately.
For all parameters except $\Omega_\text{ref}$, whose interval is the same under both priors, the log prior moves the whole interval towards smaller values, as expected from a prior that assigns comparatively
more weight to the small-value region than a flat one. The shift, however, is far from a rigid translation. For $g$ the lower edge falls by a factor of about $1.5$ and the upper edge by a factor of about $1.3$; for $\lambda$ the lower edge falls by about a
factor of $2$ ($3.1\times10^{-3}\to1.5\times10^{-3}$) while the upper
edge drops by only a factor of $1.2$
($8.7\times10^{-3}\to7.2\times10^{-3}$); for $v/\text{GeV}$ the lower edge falls by a factor of about $50$ ($1.6\times10^{7}\to3.0\times10^{5}$) and the upper edge by a factor of $2.6$ ($8.4\times10^{7}\to3.2\times10^{7}$). The \emph{lower tails} are therefore more sensitive to the prior than the upper ones, strikingly so for~$v$, which suggests that the likelihood bounds these parameters mainly from above, whereas the small-value side is largely shaped by the prior.

\begin{table}[tbp]
\centering
\begin{tabular}{lll}
\toprule
Parameter & Linear & Log \\
\midrule
$g$ & $[0.5,\,1.6]$ & $[0.4,\,1.3]$ \\
$\lambda$ & $[3.1 \times 10^{-3},\,8.7 \times 10^{-3}]$ & $[1.5 \times 10^{-3},\,7.2 \times 10^{-3}]$ \\
$v/\mathrm{GeV}$ & $[1.6 \times 10^{7},\,8.4 \times 10^{7}]$ & $[3.0 \times 10^{5},\,3.2 \times 10^{7}]$ \\
\bottomrule
\end{tabular}
\caption{Central $68\%$ credible intervals of the Abelian Higgs model parameters under the linear and the log prior.}
\label{tab:swmhd_ci68_O4a}
\end{table}

\begin{table}[tbp]
\centering
\small
\begin{tabular}{lllll}
\toprule
Parameter & $\rho_{\max}$ & $\max_{\pi}(\mathrm{Width}(\pi))$ & $\min_{\pi}(\mathrm{Width}(\pi))$ & Status \\
\midrule
$g$ & $1.2$ & $1.1$ & $0.9$ & ROBUST \\
$\lambda$ & $1.0$ & $5.6 \times 10^{-3}$ & $5.6 \times 10^{-3}$ & ROBUST \\
$v/\mathrm{GeV}$ & $2.1$ & $6.8 \times 10^{7}$ & $3.2 \times 10^{7}$ & NOT ROBUST \\
\bottomrule
\end{tabular}
\caption{Maximin Robustness Ratio $\rho_{\max}$ of the $68\%$ credible intervals of
the Abelian model, together with the largest and smallest interval
width over the two priors. Only $v/\mathrm{GeV}$ exceeds 2, the reference used in
\Tabref{tab:rhomax_overlap_criteria}; see \Appref{app:robustness} for the general
definition of $\rho_{\max}$ (\eq{eq:rho_max_general}), of which this table
is the two-prior, direct-sampling analogue.}
\label{tab:swmhd_rho_max_O4a}
\end{table}

Judging by the width in \Tabref{tab:swmhd_rho_max_O4a} alone, we can see that no interval changes in size by more than a factor $2.1$. The
couplings $g$ and $\lambda$ are essentially unaffected
($\rho_{\max}=1.2$ and $1.0$, respectively), and only $v$
exceeds the reference value of $2$ for a \textsc{robust} parameter ($\rho_{\max}=2.1$), its interval being about twice as wide under the linear prior
as under the log prior.

This diagnostic has a blind spot, though: it measures the \emph{size} of
an interval and is completely insensitive to its \emph{position}. The
self-coupling $\lambda$ is
the clearest example, since its two intervals have practically identical
widths ($5.6\times10^{-3}$ under both priors) and yet sit at visibly
different places on the axis, see \Tabref{tab:swmhd_ci68_O4a}. A small $\rho_{\max}$ is thus
necessary for robustness, but by no means sufficient, which is why the position-sensitive diagnostics below are indispensable.

The posterior means in \Tabref{tab:swmhd_overlap_kl_O4a} tell the same story as the interval edges, but with a sharper contrast.
For $g$ and $\lambda$ the mean is lowered by roughly $30\%$ when going
from the linear to the log prior ($\rho_{\rm mean}=1.4$, $\Delta_{\rm
rel}=0.3$ and $0.4$, respectively), which puts both in the
\textsc{moderate} band between \textsc{robust} and \textsc{not robust}. For $v/\mathrm{GeV}$, by contrast, the mean is reduced by about $70\%$ ($\rho_{\rm mean}=3.5$, $\Delta_{\rm rel}=1.1$), more than twice the level $\Delta_{\rm rel}=0.5$ beyond which we declare a parameter \textsc{not robust}.

Table \ref{tab:swmhd_overlap_kl_O4a} also adds two diagnostics that look at the posteriors beyond their mean: the interval overlap and the Kullback-Leibler divergence, the latter being sensitive to the full shape of the posteriors.
The robustness status deduced from these measures coincides with that
from $\rho_\text{mean}$.

\begin{table}[tbp]
\centering
\begin{tabular}{lllll}
\toprule
Parameter & $\rho_{\mathrm{mean}}$ & $\rho_{\mathrm{overlap}}$ &
$D_{\mathrm{KL}}$ & Status \\
\midrule
$g$ & $1.4$ & $0.6$ & $0.2$ & MODERATE \\
$\lambda$ & $1.4$ & $0.6$ & $0.3$ & MODERATE \\
$v/\mathrm{GeV}$ & $3.5$ & $0.2$ & $1.0$ & NOT ROBUST \\
\bottomrule
\end{tabular}
\caption{Posterior-mean ratio, credible-interval overlap and symmetrised
Kullback-Leibler divergence between the linear and the log prior for the
Abelian model. The last column is the final status obtained by
combining all three diagnostics.}
\label{tab:swmhd_overlap_kl_O4a}
\end{table}

\begin{table}[tbp]
\centering
\begin{tabular}{llllllll}
\toprule
Parameter & $\rho_{\max}$ & $\max_{\pi}(\mathrm{Width}(\pi))$ & $\min_{\pi}(\mathrm{Width}(\pi))$ & $\rho_{\mathrm{overlap}}$ & $\Delta_{\mathrm{rel}}$ & $D_{\mathrm{KL}}$ & Final Status \\
\midrule
$g$ & $1.2$ & $1.1$ & $0.9$ & $0.6$ & $0.3$ & $0.2$ & MODERATE \\
$\lambda$ & $1.0$ & $5.6 \times 10^{-3}$ & $5.6 \times 10^{-3}$ & $0.6$ & $0.4$ & $0.3$ & MODERATE \\
$v/\mathrm{GeV}$ & $2.1$ & $6.8 \times 10^{7}$ & $3.2 \times 10^{7}$ & $0.2$ & $1.1$ & $1.0$ & NOT ROBUST \\
\bottomrule
\end{tabular}
\caption{Summary of all robustness diagnostics for the Abelian Higgs model
(O4a data). The final status counts how many of the four diagnostics
$\rho_{\max}$, $\rho_{\mathrm{overlap}}$,
$\Delta_{\mathrm{rel}}=2\,|\langle\theta\rangle_{\rm linear}-\langle\theta\rangle_{\rm log}|/(\langle\theta\rangle_{\rm linear}+\langle\theta\rangle_{\rm log})$,
and
$D_{\mathrm{KL}}$ satisfy the robustness criteria discussed in \Appref{app:robustness}.}
\label{tab:swmhd_summary_O4a}
\end{table}

Table \ref{tab:swmhd_summary_O4a} collects everything in one place. The
couplings $g$ and $\lambda$ each pass only one test, the Maximin
Robustness Ratio,
while $v/\mathrm{GeV}$ passes none. For $g$ the Kullback-Leibler divergence, $D_{\rm
KL}=0.24$ (rounded to $0.2$ in the table), lies just above the value of $0.2$ below which it counts as passed. The parameters
therefore fall into a hierarchy of prior dependence, in which $g$ and $\lambda$ are comparably affected by the prior, with $g$ marginally less so in all three position-sensitive diagnostics (overlap, mean shift and divergence), while $v/\mathrm{GeV}$ is the parameter that is largely determined by the prior.

The label \textsc{moderate} for $g$ and $\lambda$ rests on a single
passed test, the Maximin Robustness Ratio, which, as discussed above,
cannot see shifts. The three position-sensitive diagnostics place both
parameters in the intermediate band: none of them is \textsc{robust},
but none is in the \textsc{not robust} regime either, which is precisely
the situation that the label \textsc{moderate} is meant to describe. For
$v/\mathrm{GeV}$, instead, all four diagnostics flag a strong dependence
on the prior. However, as we can see from the analysis of Secs.~\ref{sec:geometrical_results} and~\ref{sec:thermodynamical_results}, the robustness tests are much more favourable in this case.

The O4a data thus constrain the Abelian model to a certain extent.
The couplings $g$ and $\lambda$ are \emph{moderately prior-dependent}:
their posterior means shift by about $30\%$ between the two priors,
which is visible but below the level at which conclusions would change. The parameter $v/\mathrm{GeV}$ is \emph{prior-dependent}, with the lower tail of its posterior driven largely by the prior (the lower edge of the interval moves by a factor of about $50$) while the upper edge is comparatively stable (a factor of $2.6$); the same pattern, much weaker, is seen for $\lambda$.

In practice we therefore recommend to state the prior explicitly whenever results on $g$, $\lambda$ and, in particular, $v/\mathrm{GeV}$ are quoted, and to report
them together with the corresponding result under the alternative prior rather than as a single interval; claims that rest on the upper edges of the
intervals are the least exposed to this ambiguity. Since all diagnostics are marginal, a prior dependence of the joint posterior that is not visible in any
single parameter cannot be excluded by the present analysis, and the quantitative classification above is meant as a guide to the relative sensitivity of the parameters rather than as a sharp statistical test.

\section{Conclusions and future directions} \label{sec:Conclusions}

We have re-examined the Bayesian inference of cosmological FOPT
parameters from the LVK stochastic searches in the observing runs O1--O3
(``O3'') and O1--O4a (``O4a''), with the dependence of the results on the
prior as our central concern.  The inference was set up at three levels:
in terms of the geometrical parameters of the spectral templates
($\Omega_{\rm ref}$, $\Omega_p$, $f_p$ and, for sound waves, $\Delta_w$),
in terms of the thermodynamical parameters of the transition
($\alpha$, $\beta/H_*$, $T_*$ for bubble collisions; $K$, $H_*R_*$,
$\Delta_w$, $T_*$, $\varepsilon$ for sound waves plus MHD turbulence), and
in terms of the Lagrangian parameters of a dark Abelian Higgs model.  At
each level we compared at least two priors, including the Jeffreys prior,
priors informed by the detector sensitivity, and priors motivated by the
expected distribution of FOPT signals, which we specified in
Tabs.~\ref{tab:PriorsGeom}--\ref{tab:PriorGeom_O4a} for the geometrical
parameters and in Tabs.~\ref{tab:analysis_nucl_O4a}--\ref{tab:O4_KR} for the
thermodynamical parameters.  The agreement between the resulting posteriors
was quantified by the Maximin Robustness Ratio $\rho_{\max}$, the
posterior-mean ratio $\rho_\text{mean}$, the credible-region
overlap $\rho_{\rm overlap}$ and the Kullback-Leibler divergence
$D_\text{KL}$, which are defined in \Appref{app:robustness}.

On the technical side, we derived a refined approximation for the peak
frequency of the double broken power law, \eq{eq:fpapprox}, and the
corresponding peak amplitude, \eq{eq:OmegaP2}, which is needed to relate
these parameters to the break frequencies and amplitudes of the
template.  The new approximation reproduces the exact peak position to
about $2\%$ or better for the sound-wave template, compared to
deviations of up to $10\%$ for the zeroth-order result.

\emph{Geometrical parameters.}
No cosmological SGWB is detected.  The 95\% upper limits on the
amplitude of the astrophysical foreground for the Jeffreys prior are
$\Omega_{\rm ref}<6\times10^{-9}$ for O3 and
$\Omega_\text{ref}<2\times10^{-9}$ for O4a, as reported in
\Tabref{tab:94_CL_Omegaref}\@. They are essentially independent of the
template used for the FOPT component.  However, we emphasise that, for a
log-uniform prior, such limits also depend on the prior range assumed
for $\Omega_{\rm ref}$.  For the O3 dataset, this causes a variation of
the limit by a factor of almost $2$, whereas we find a negligible
variation for the O4a data.  

The parameters of the cosmological signal are not
robust with respect to change of prior.  For the peak amplitude we find $\rho_{\max}>10^6$.
The posteriors do not overlap and their means differ by orders
of magnitude between the priors.  The peak frequency improves when going
from O3 to O4a data but not sufficiently to reach prior-independence.
For example, $\rho_{\max}$ drops from $24.2$ to $3.1$ for sound waves.
The credible-region overlap increases only to $\rho_\text{overlap}=0.3$
for both sound waves and bubble collisions.  The Jeffreys posterior,
although the narrowest, extends to frequencies of order
$(0.01$--$0.1)$\,Hz where the detectors are insensitive, i.e., it is set
by the prior to a large extent.  This is expected in the absence of a
detection: the likelihood does not localise a peak independently of the
prior volume.

\emph{Thermodynamical parameters.}
Mapping the data onto the characteristic parameters of the phase
transition allows for an interpretation in terms of parameters related
to the physical GW source but suffers from degeneracies, since different
combinations of thermodynamical parameters can produce very similar GW
spectra.  For GW produced by bubble-collisions and the O4a dataset, the
phase transition strength $\alpha$ turns out to be robustly constrained
with, for example, a small Kullback-Leibler divergence $D_\text{KL}
\simeq 0.1$.  The inverse duration $\beta/H_*$ and the transition
temperature $T_*$ are strongly prior-dependent and affected by the
degeneracy between $T_*$ and $\beta/H_*$ in the peak frequency and
between $\alpha$ and $\beta/H_*$ in the peak amplitude.  Their posterior
distributions obtained with the two priors considered overlap only
marginally or not at all, and both the extent and the location of the
allowed regions change drastically.

For sound waves plus MHD turbulence, we use different thermodynamical
parameters that simplify the analysis and avoid case distinctions.  Also
in this case, the parameter determining the phase transition strength,
$K$, is robustly constrained by the O4a data, while no firm conclusions
are possible for the mean bubble separation $H_*R_*$ and the temperature
$T_*$ due to a strong prior-dependence of their posteriors.

Generally, parameters that directly determine the dominant amplitude of
the spectrum are constrained most efficiently, whereas parameters tied
to the peak frequency, subdominant contributions, or degenerate
combinations remain sensitive to the prior, independently of the LVK
run.

\emph{Abelian Higgs model.}
Specifying a benchmark particle physics model in which we computed
the phase transition parameters allowed us to map the SGWB search data
onto the microphysical parameters (gauge and quartic scalar coupling,
vacuum expectation value) appearing in the Lagrangian.  We found that
the data allow to draw limited conclusions about the couplings, which
are moderately prior-dependent, while the posterior for the vacuum
expectation value, which sets the energy scale of the phase transition and
thus the peak frequency of the GW spectrum, is not robust under a prior
change.

\emph{Sensitivity versus prior-independence.}
One important conclusion is that smaller statistical uncertainty does
not equate to less prior-dependence.  The increased sensitivity obtained
from adding the data from observing run O4a narrows most posteriors, but
the robustness measures for most fit parameters remain in the
\textsc{not robust} regime.  Thus, prior-independence is a separate
property that has to be assessed run by run, for example, using the
overlap- and divergence-based diagnostics we discussed here.  We
therefore advocate treating the comparison of priors as an integral part
of the inference, in particular for searches without a significant
detection, where a single Bayesian limit or credible region can give an
incomplete picture of the information supplied by the data.

Moreover, even for parameters such as the astrophysical foreground
amplitude, which are robustly constrained by the data, an upper
limit can depend on the prior range used.
As a consequence, limits from Bayesian analyses should always be
presented together with the employed prior, specifying both its shape
and its range.

\emph{Direct sampling versus two-step reconstruction.}
Another, complementary approach is to reconstruct thermodynamical
parameters from the geometrical parameters, a technique considered by the
LISA Cosmology Working Group~\cite{Caprini:2015zlo}.
Direct sampling from the thermodynamical parameter space represents the
conceptually straightforward approach, which we have used.  However, this
method suffers from significant computational overhead in resolving
parameter degeneracies and from the mismatch between parameter space
dimensions, since there are 3 (bubble collisions) to 5 (sound waves and
turbulence) thermodynamical parameters vs.\ 2 to 3 geometrical parameters.

The other method corresponds to a \emph{two-step} reconstruction:
(i) constraining the geometrical parameters peak amplitude, peak
frequency, and (for DBPL templates) knee parameter, and
(ii) converting the reconstructed geometrical parameters to
thermodynamical parameters via transformation equations, applying the
transformation to each point in the produced geometrical ensemble and
taking into account Jacobian factors.
This method is preferred for practical implementation due to the
\emph{degeneracy-free sampling} (geometrical parameters exhibit no
significant degeneracies, unlike their thermodynamical counterparts),
\emph{computational efficiency} (running geometrical parameter inference
first substantially reduces computational cost) and finally
\emph{dimension matching} (the parameter space dimension mismatch
inherent in direct thermodynamical sampling is avoided).
The two-step approach technically employs different prior assumptions
compared to direct sampling due to the variable transformation.  However,
the practical impact is expected to be negligible because the employed
priors are flat, the parameter relationships in logarithmic space are
predominantly linear, and any prior transformation effects are minimal.
As the main scope of this paper is to confront the validity of sampling
with only one prior, we leave an implementation of this approach for
future work.

\emph{Outlook.}
LVK stochastic searches already constrain portions of the parameter space
for extremely strong supercooled FOPT, though theoretical uncertainties
remain significant: the conversion from microphysics to $\Omega_{\rm GW}(f)$
involves assumptions on the bubble wall velocity, the partition of energy
between scalar gradients and bulk motion, the lifetime of the sound-wave
source, and reheating and dilution effects, and foreground subtraction and
non-Gaussian detector noise can bias the inference if not carefully
controlled.  Improved templates, joint astrophysical/cosmological
inference, and upcoming detectors (LISA for lower frequencies, ET/CE for
higher sensitivity) will dramatically improve the reach.  Careful
modelling -- including sound-wave lifetime, bubble dynamics, and reheating
-- is necessary to reliably map GW constraints to particle physics, and
the prior-sensitivity analysis advocated here should accompany any such
mapping.

\addtocontents{toc}{\protect\setcounter{tocdepth}{-1}}
\section*{Acknowledgements}

We thank Jorinde van der Vis, Philipp Schicho, Kylar Greene, and Thomas Konstandin for helpful discussions. 
The work of L.V.S.\ is supported by the National Research Foundation of Korea (NRF), grant no.\ RS-2023-00273508.
J.K.\ was supported by the NRF's \emph{Brain Pool} program under grant no.\ RS-2023-00283129 and the KIAS visitor program.
AI tools (Claude, Anthropic) were used for editing tables and formatting in the text; the authors take full responsibility for the content of this paper.

\appendix
\renewcommand{\thesubsection}{\thesection.\arabic{subsection}}
\renewcommand{\thesubsubsection}{\thesubsection.\arabic{subsubsection}}
\addtocontents{toc}{\protect\newpage}
\addtocontents{toc}{\protect\setcounter{tocdepth}{3}}

\section{Calculation of the 95\% upper limit on \texorpdfstring{$\Omega_{\rm ref}$}{Omega_ref}}
\label{app:CalcLimit}
The 95\% credible upper limits on $\Omega_{\rm ref}$ quoted in this paper
(e.g., in \Tabref{tab:94_CL_Omegaref}) are computed from the output of the
nested sampling runs with \texttt{dynesty}.  A nested sampling run does
not return samples drawn from the posterior but a list of $N$ points
$\{x_i\}_{i=1}^N$ (here the values of $\Omega_{\rm ref}$), each with an
importance weight
\begin{equation} \label{eq:Weights}
w_i=\frac{\mathcal{L}_i\,\Delta X_i}{\mathcal{Z}}\,,
\qquad
\sum_{i=1}^N w_i=1\,,
\end{equation}
where $\mathcal{L}_i$ is the likelihood at the point, $\Delta X_i$ the
prior-volume shell associated with it, and $\mathcal{Z}$ the evidence; in
\texttt{dynesty}, $w_i=\exp(\texttt{logwt}_i-\texttt{logz}_N)$.  Posterior
statistics must therefore be computed with these weights.

\subsection{Weighted percentile}
Let $x_{(1)}\leq\dots\leq x_{(N)}$ denote the sorted points, $w_{(i)}$ the
correspondingly ordered weights, and
\begin{equation}
\tilde{C}_k=\frac{\sum_{i\leq k}w_{(i)}}{\sum_{i=1}^{N}w_{(i)}}
\end{equation}
the normalised cumulative weight.  The 95\% credible upper limit
$x_{95}$ is the smallest value for which the posterior probability of
$\Omega_{\rm ref}\leq x_{95}$ reaches $0.95$,
\begin{equation}
\sum_{\{i\,|\,x_i\leq x_{95}\}} w_i\ \geq\ 0.95\,.
\end{equation}
This is evaluated as follows:
\begin{enumerate}
    \item Remove non-finite values by applying the same mask to the
    points and the weights.
    \item Sort the points in ascending order and apply the same ordering
    to the weights.
    \item Compute the cumulative sum of the sorted weights and normalise
    it to $[0,1]$.
    \item Find the smallest index $k^*$ such that $\tilde{C}_{k^*}\geq0.95$.
    \item The upper limit is $x_{95}=x_{(k^*)}$.
\end{enumerate}
  The result is deterministic: for a given nested sampling run, it
  is reproduced exactly, in contrast to the resampling approach. In this sense, the limit is taken at a sample point rather than interpolated between points.  For the large numbers of live points used, the spacing of neighbouring points is small and the difference from an interpolated
  quantile is negligible. The remaining uncertainty is the statistical error of the nested
  sampling run itself, which depends on the number of live points.  It can
  be assessed by repeating a run with different random seeds; it is not
  included in the values quoted in the tables. The upper limit is a quantile of the \emph{posterior}.  It
  therefore depends on the prior and for the log-uniform prior used for $\Omega_{\rm ref}$ also on the range of the prior (in particular the lower edge, since the likelihood is flat for amplitudes far below the
  sensitivity).  Limits must thus always be quoted together with the  prior specified in Tabs.~\ref{tab:PriorsGeom}--\ref{tab:PriorGeom_O4a} and \Tabref{tab:O4_KR}.

\section{Measures of robustness}
\label{app:robustness}
\addtocontents{toc}{\protect\setcounter{tocdepth}{1}}

Let $\boldsymbol{\theta}$ denote the model parameters and let
$\pi(\boldsymbol{\theta})$ be a prior distribution. Given data $D$, the
posterior is, \cite{bayes1763,gelman2013},
\begin{equation}
p(\boldsymbol{\theta}\mid D,\pi)
\propto
\mathcal{L}(D\mid\boldsymbol{\theta})\,\pi(\boldsymbol{\theta}) .
\label{eq:posterior_def}
\end{equation}

For two priors, $\pi_1$ and $\pi_2$, we obtain
\begin{equation}
p_1(\boldsymbol{\theta}\mid D)
=
p(\boldsymbol{\theta}\mid D,\pi_1),
\qquad
p_2(\boldsymbol{\theta}\mid D)
=
p(\boldsymbol{\theta}\mid D,\pi_2).
\end{equation}

The purpose of the robustness analysis is not to decide which prior is
``correct'', but to determine whether the inferred constraints are stable
when the prior is changed \cite{berger1985,berger1990,berger1994}. Posterior samples are stored as
$
\{\boldsymbol{\theta}^{(k)}\}_{k=1}^{N}
$,
where each row is a posterior sample and each column corresponds to one
parameter. Below we present the most well-known statistical measures \cite{gelman2013,berger1990,berger1994}
that can be implemented. In \Secref{sec:geometrical_results} and
\Secref{sec:thermodynamical_results} we applied these measures to the
geometrical and thermodynamic parameters, respectively, while in
\Secref{sec:U1_results} the measures were extended to the fundamental parameters of the Abelian $U(1)$ model.
All robustness quantities that we present here are statements about the
posterior obtained with each prior and computed with the importance
weights from \eq{eq:Weights} taken into account.  They are built from
the credible intervals, the posterior means, and the posterior densities.

Note that the corner plots in \Secref{sec:results} show the posterior
distributions on logarithmic scales, whereas the credible intervals,
posterior means, and resulting robustness measures are given in terms of
the fit parameters themselves (i.e., on a linear scale).  An alternative
approach would be to consider the credible intervals etc.\ for the
logarithms of the parameters, which would naturally result in different
values for the robustness measures.

\subsection*{$68\%$ credible intervals}

For a parameter $\theta_i$, the analysis uses the central equal-tail credible
interval \cite{edwards1963,gelman2013}. For credibility level $\alpha$,
\begin{equation}
\theta_{i,\mathrm{low}}
=Q_{(1-\alpha)/2}(\theta_i),
\qquad
\theta_{i,\mathrm{high}}
=Q_{(1+\alpha)/2}(\theta_i).
\end{equation}

At $68\%$ credibility \cite{hyndman1996},
\begin{equation}
\mathcal{C}_{68}(\theta_i)
=
[Q_{0.16}(\theta_i),Q_{0.84}(\theta_i)] .
\label{eq:C68_definition}
\end{equation}

For every parameter and every prior, we record
\begin{align}
W_i &= \theta_{i,\mathrm{high}}-\theta_{i,\mathrm{low}},
\label{eq:width_definition}\\
\bar{\theta}_i &= \frac{1}{N}\sum_{k=1}^{N}\theta_i^{(k)},\\
\theta_{i,\mathrm{med}} &= Q_{0.5}(\theta_i),
\end{align}
as well as the posterior standard deviation.

\subsection*{Credible-region overlap ratio}

For two credible regions, define \cite{jaccard1901,jaccard1912}
\begin{equation}
\rho_{\mathrm{overlap}}
=
\frac{
|\mathcal{C}_{\alpha}(\pi_1)\cap
\mathcal{C}_{\alpha}(\pi_2)|
}{
|\mathcal{C}_{\alpha}(\pi_1)\cup
\mathcal{C}_{\alpha}(\pi_2)|
} .
\label{eq:rho_overlap_general}
\end{equation}

For the parameter-by-parameter implementation, the regions are intervals,
$\mathcal{C}_1=[L_1,U_1]$, $\mathcal{C}_2=[L_2,U_2]$. Their intersection has length
\begin{equation}
I=
\max\left[0,\min(U_1,U_2)-\max(L_1,L_2)\right],
\end{equation}
and their union has length
\begin{equation}
U=
\max(U_1,U_2)-\min(L_1,L_2).
\end{equation}

Hence
\begin{equation}
\rho_{\mathrm{overlap}}=\frac{I}{U}.
\end{equation}

The ratio satisfies
$0\leq\rho_{\mathrm{overlap}}\leq1.$

The limiting cases are
\begin{align}
\rho_{\mathrm{overlap}}=1
&\quad\Longrightarrow\quad
\text{identical credible intervals},\\
\rho_{\mathrm{overlap}}=0
&\quad\Longrightarrow\quad
\text{non-overlapping credible intervals}.
\end{align}

\subsection*{Comparison of several priors}
For priors
$\pi_1,\pi_2,\ldots,\pi_M$, 
the overlap ratio is evaluated for every unique pair $(\pi_a,\pi_b)$, $a<b$. For each parameter $\theta_i$ this gives the pairwise overlap
\begin{equation}
\rho_{\mathrm{overlap}}^{ab}(\theta_i)
\equiv
\rho_{\mathrm{overlap}}(\theta_i;\pi_a,\pi_b)
=
\frac{
\left|\mathcal{C}_{68}(\pi_a)\cap \mathcal{C}_{68}(\pi_b)\right|
}{
\left|\mathcal{C}_{68}(\pi_a)\cup \mathcal{C}_{68}(\pi_b)\right|
},
\label{eq:rho_overlap_pairwise}
\end{equation}
which is simply \eq{eq:rho_overlap_general} evaluated at the $68\%$
credibility level for the specific pair $(\pi_a,\pi_b)$. Values close to
unity indicate that the inferred parameter region is stable under the
change of prior, whereas small values indicate that the posterior support
changes substantially. Equivalently, for more than two priors one may
construct a symmetric overlap matrix
\begin{equation}
R^{(i)}_{ab}
=
\rho_{\mathrm{overlap}}(\theta_i;\pi_a,\pi_b),
\end{equation}
with
\begin{equation}
R^{(i)}_{aa}=1,
\qquad
R^{(i)}_{ab}=R^{(i)}_{ba}.
\end{equation}

This pairwise approach avoids privileging one prior as a reference. In
\Secref{sec:thermodynamical_results} and \Secref{sec:U1_results}
only two priors are compared at a time (Jeffreys versus Conditional \cite{jeffreys1946,jeffreys1961,kass1996}, or
Jeffreys versus the model's Conditional prior), so
\eq{eq:rho_overlap_pairwise} reduces to a single number per parameter; in
\Secref{sec:geometrical_results}, where four priors (Jeffreys, Detector-Based,
Model-Based, Mixed) are compared, the tabulated $\rho_{\rm overlap}$ is the
minimum of $R^{(i)}_{ab}$ over all six pairs, so that a parameter is only
reported as robust if \emph{every} pair of priors agrees.

\subsection*{Maximin Robustness Ratio}

The overlap ratio probes agreement between the locations and extents of the
credible intervals. A complementary diagnostic,
\begin{equation}
\rho_{\max}
=
\frac{\max_{\pi}W_i(\pi)}
{\min_{\pi}W_i(\pi)},
\label{eq:rho_max_general}
\end{equation}
measures changes in their
widths (\eq{eq:width_definition}) \cite{berger1990}.
Ratios $> 2.0$ indicate substantial prior-dependent uncertainty. 

\begin{table}
\centering
\addtolength{\tabcolsep}{0.2ex}
\begin{tabular}{llcc}
\toprule
Quantity & Measures & ROBUST & NOT ROBUST \\
\midrule
$\rho_{\max}$ & Change in interval width & $\simeq 1$ & $>2$\\
$\rho_{\mathrm{overlap}}$ & Agreement of credible regions & $\simeq 1$ & $\leq0.4$\\
\bottomrule
\end{tabular}
\caption{Criteria for robustness for $\rho_{\max}$ and $\rho_{\mathrm{overlap}}$. \label{tab:robcriteria} }
\label{tab:rhomax_overlap_criteria}
\end{table}

These diagnostics are complementary. Two priors can produce similarly
located intervals but substantially different widths, or intervals of similar
width that are shifted relative to one another. This is why, throughout
\Secref{sec:results}, an improvement in $\rho_{\max}$ between the O3 and O4a
runs (i.e., a narrower posterior under the more sensitive dataset) is never
by itself taken as evidence of improved prior independence: only a joint
improvement in $\rho_{\max}$ \emph{and} $\rho_{\rm overlap}$ (and, where
available, in the Kullback-Leibler divergence \cite{kullback1951} introduced below) supports
that conclusion.

For the adopted diagnostic classification,
\begin{equation}
\begin{array}{ccl}
\rho_{\mathrm{overlap}}>0.7
&\Rightarrow& \mathrm{ROBUST},\\[0.2em]
0.4<\rho_{\mathrm{overlap}}\leq0.7
&\Rightarrow& \mathrm{MODERATE},\\[0.2em]
\rho_{\mathrm{overlap}}\leq0.4
&\Rightarrow& \mathrm{NOT\ ROBUST}.
\end{array}
\label{eq:robustness_classification}
\end{equation}

These thresholds are diagnostic criteria for the analysis rather than
universal statistical thresholds. The implemented overlap ratio is a \emph{one-dimensional,
parameter-by-parameter} measure. It compares the marginal $68\%$ credible
intervals of an individual parameter under two priors. It should therefore not be interpreted as the volume overlap of the full
multidimensional $68\%$ posterior regions. Such a quantity would require
constructing the corresponding joint credible sets and evaluating their
multidimensional volumes.

The one-dimensional measure is nevertheless useful for reporting the
robustness of individual parameter constraints: it is symmetric, transparent,
easy to reproduce, and directly connected to the posterior intervals that
are reported in the analysis.

The prior-robustness analysis combines two complementary pieces of
information. First, the $68\%$ credible interval describes the constraint
obtained under each prior, \eq{eq:C68_definition}. Second, the overlap
ratio \eq{eq:rho_overlap_pairwise} quantifies the common part of two such
constraints. Together with the interval-width ratio $\rho_{\max}$ of
\eq{eq:rho_max_general}, these quantities provide
a compact assessment of whether the posterior constraints are stable against
reasonable changes of prior. Robust inference is indicated by substantial
overlap between credible intervals and by limited changes in their widths.

\subsection*{Posterior mean discrepancy and Kullback-Leibler divergence}

In order to complement the width-based quantities $\rho_{\rm overlap}$
and $\rho_{\max}$, we consider several further diagnostics.
The relative discrepancy between posterior means under two priors
$\pi_1$ and $\pi_2$ is
\begin{equation}
\Delta_{\rm rel}(\theta_i)
=
\left|
1-\frac{\mathbb{E}[\theta_i\mid D,\pi_1]}{\mathbb{E}[\theta_i\mid D,\pi_2]}
\right|.
\label{eq:delta_rel_definition}
\end{equation}
It probes the stability of the posterior \emph{location} independently
of its width, and can therefore disagree with $\rho_{\max}$ for a given
parameter (see \Secref{subsec:abelian_robustness}).

For an arbitrary number of priors, $\rho_{\rm mean}$ is the ratio of the largest to the smallest posterior mean
\begin{equation}
\rho_{\rm mean}=\max_{\pi}E[\theta_i|D,\pi]\,/\min_{\pi}E[\theta_i|D,\pi]\ge1.
\end{equation}
It does not depend on the order of the priors.

Finally, the symmetrised
Kullback-Leibler divergence between two posteriors \cite{kullback1951,jeffreys1946,kullback1959,cover2006},
\begin{equation}
D_{\rm KL}^{\rm sym}(\pi_1,\pi_2)
=
\frac{1}{2}
\left[
\mathrm{KL}(p_1\Vert p_2)+\mathrm{KL}(p_2\Vert p_1)
\right],
\label{eq:kl_sym_definition}
\end{equation}
is a full-posterior-shape diagnostic that is sensitive to differences in
location, width, and higher moments simultaneously (see
\eq{eq:kl_sym_definition}). As with $\rho_{\rm overlap}$ and $\rho_{\max}$,
neither $\Delta_{\rm rel}$ nor $D_{\rm KL}^{\rm sym}$ should be used in
isolation: a combined classification, as in \Tabref{tab:swmhd_summary_O4a},
requires all four diagnostics together.

\addtocontents{toc}{\protect\setcounter{tocdepth}{3}}
\section{General redshifting relations \label{sec:redshift}}

The present-day frequency and energy density of GW produced in a FOPT
are related to their values at production by
\cite{Caprini:2024hue}
\begin{eqnarray}
f &=& \frac{a_*}{a_0} f_* = H_{*,0}\,\frac{f_*}{H_*},
\label{eq:redshift_frequency}
\\
h^2\Omega_{\rm GW}(f) &=& h^2 F_{\rm GW,0}\,\Omega_*(f_*).
\label{eq:redshift_energy}
\end{eqnarray}
The redshifted Hubble rate and energy-density factor are
\begin{eqnarray}
H_{*,0} &\simeq& 1.65 \times 10^{-5}\,{\rm Hz}
\left(\frac{g_*(T_*)}{100}\right)^{1/6}
\left(\frac{T_*}{100\,\text{GeV}}\right),
\label{eq:Hstar0}
\\
h^2 F_\text{GW,0} &=&
h^2 \left(\frac{a_*}{a_0}\right)^4 \left(\frac{H_*}{H_0}\right)^2 \simeq
1.64 \times 10^{-5}
\left(\frac{100}{g_*(T_*)}\right)^{1/3}.
\label{eq:FGW0}
\end{eqnarray}

\section{Details of spectrum templates} \label{app:templates}
Here we collect all details regarding the GW templates from
\cite{Caprini:2024hue} that have not been given in the main text.

\subsection{Template I: Broken power law} \label{sec:templateI}
The BPL template is given by \eq{eq:BPL}.
The peak frequency and peak amplitude are
\begin{eqnarray}
f_p &=& f_b\left(-\frac{n_1}{n_2}\right)^{1/a_1} ,
\label{eq:BPL_fp}
\\
\Omega_p &=&
\Omega_b
\left[
\frac{1}{2}
\left(-\frac{n_2}{n_1}\right)^{\frac{n_1}{n_1-n_2}}
+
\frac{1}{2}
\left(-\frac{n_1}{n_2}\right)^{\frac{-n_2}{n_1-n_2}}
\right]^{-\frac{n_1-n_2}{a_1}} .
\label{eq:BPL_Omegap}
\end{eqnarray}
For strong transitions, modelled after bubble collisions, the relevant energy fraction is
\begin{equation}
\tilde K \equiv \frac{\alpha}{1+\alpha}.
\label{eq:Ktilde}
\end{equation}
The peak amplitude and frequency today read
\begin{eqnarray}
h^2 \Omega_p &=&
h^2 F_{\rm GW,0}\,
A_{\rm str}\,
\tilde K^2
\left(\frac{H_*}{\beta}\right)^2,
\label{eq:BPL_amplitude}
\\
f_p &\simeq& 0.11\,H_{*,0}\,\frac{\beta}{H_*},
\label{eq:BPL_frequency}
\end{eqnarray}
where
\begin{equation}
A_{\rm str} \simeq 0.05,
\label{eq:Astrong}
\end{equation}
and the rest of the parameters, corresponding to the redshifting
parameters, are defined in App.~\ref{sec:redshift}.

\subsection{Template II: Double broken power law}
\label{sec:templateII}

\subsubsection{General form}
\label{subsec:DBPL_general}

The double broken power law (DBPL) spectrum is given by
Eqs.~(\ref{eq:DBPL},\,\ref{eq:Sf}).
Depending on the production mechanism, the relevant parameters and
relations between them vary.
In all cases, the mean bubble separation, $R_*$ at the end of the PT is
relevant and estimated through \cite{Caprini:2019egz}
\begin{equation} 
\label{eq:H_star_R_star}
R_* H_*= (8\pi)^{1/3}\,\max(v_w,c_s)\,\frac{H_*}{\beta},
\end{equation}
where $c_s=1/\sqrt{3}$ is the sound speed. Note that we write $R_* H_*$ as this combination appears explicitly in various fits.

\subsubsection{Sound-wave contribution}
\label{sec:soundwaves}

The frequency breaks are
\begin{eqnarray}
f_1 &\simeq& 0.2\,H_{*,0}\,(H_*R_*)^{-1},
\label{eq:SW_f1}
\\
f_2 &\simeq& 0.5\,H_{*,0}\,\Delta_w^{-1}(H_*R_*)^{-1},
\label{eq:SW_f2}
\end{eqnarray}
where the bubble wall velocity $v_w$ enters via the parameter $\Delta_w$
defined in \eq{eq:DefDeltaw}.
We use
\begin{equation}
h^2\Omega_{\rm int}
=
h^2 F_\text{GW,0}\,
A_{\rm sw}\,
K^2\,
(H_*\tau_{\rm sw})\,
(H_*R_*)
\label{eq:SW_Omegint}
\end{equation}
for the integrated GW energy density
$\Omega_\text{int} = \int \Omega_\text{GW}^\text{DBPL}(f) \, d\log f$,
where \cite{Caprini:2019egz,Jinno:2022mie}
\begin{align}
A_\text{sw} &\simeq 0.11,
\\
K &\simeq 0.6 \, \kappa(\alpha,v_w) \, \frac{\alpha}{1+\alpha} ,
\label{eq:eff-factor_K}\\
H_* \tau_\text{sw} &\simeq \min\big( 1 , 2 H_*R_*/\sqrt{3K} \big).
\label{eq:tausw}
\end{align}
The integrated energy density is related to the parameter $\Omega_2$ in
\eq{eq:DBPL} by \cite{Caprini:2024hue} 
\begin{equation} \label{eq:AuxOmegas}
	\Omega_2 =
	\frac{1}{\pi} \left( \sqrt{2} + \frac{2 \, f_2/f_1}{1 + f_2^2/f_1^2} \right)
	\Omega_\text{int} .
\end{equation}

The stochastic GW signal produced by sound waves depends on the fraction
of the released vacuum energy that is converted into bulk kinetic energy
of the plasma. This fraction is parameterized by the hydrodynamic
efficiency factor $\kappa(\alpha,v_w)$ in \eq{eq:eff-factor_K}.
The factor $0.6$ reflects the correction due to the quantity $K/0.6$
representing the contribution from a single bubble. Unlike the early
Jouguet approximation, which assumes that the wall propagates at the
Jouguet velocity, $v_w=v_J(\alpha)$, modern treatments regard the wall
velocity as an independent hydrodynamic variable determined by
microscopic friction. Consequently, $\kappa$ must be evaluated
throughout the entire hydrodynamic parameter space.  This treatment is
suitable for parameter inference and Bayesian analyses in which the wall
velocity is sampled independently.  The Jouguet velocity is given by
\begin{align}
v_J(\alpha)
=
\frac{
\sqrt{\alpha\left(\frac23+\alpha\right)}
+\frac1{\sqrt3}
}
{1+\alpha}.
\end{align}
Following the fitting formulae of Ref.~\cite{Espinosa:2010hh}, calibrated to numerical hydrodynamic solutions, we define the auxiliary quantities
\begin{align}
\kappa_A
&=
\frac{6.9\,\alpha}
{1.36-0.037\sqrt{\alpha}+\alpha},
\\
\kappa_B
&=
\frac{\alpha^{2/5}}
{0.017+\left(0.997+\alpha\right)^{2/5}},
\\
\kappa_C
&=
\frac{\sqrt{\alpha}}
{0.135+\sqrt{0.98+\alpha}},
\\
\kappa_D
&=
\frac{\alpha}
{0.73+0.083\sqrt{\alpha}+\alpha},
\end{align}
together with
\begin{align}
\delta\kappa
=
-0.9
\ln
\left(
\frac{\sqrt{\alpha}}
{1+\sqrt{\alpha}}
\right).
\end{align}
The efficiency factor is then evaluated continuously over the three hydrodynamic regimes.

For subsonic deflagrations, we have
\begin{align}
v_w<c_s,
\end{align}
and the efficiency factor is
\begin{align}
\kappa
=
\frac{
c_s^{11/5}
\kappa_A
\kappa_B
}
{
\left(c_s^{11/5}-v_w^{11/5}\right)\kappa_B
+
v_w c_s^{6/5}\kappa_A
}.
\end{align}

For supersonic deflagrations, 
\begin{align}
c_s\le v_w < v_J,
\end{align}
we consider the hybrid solutions, for which we introduce
\begin{align}
x
=
\frac{v_w-c_s}
{v_J-c_s},
\end{align}
and interpolate according to
\begin{align}
\kappa
&=
\kappa_B
+
(v_w-c_s)\delta\kappa
+
x^3
\left[
\kappa_C
-
\kappa_B
-
(v_J-c_s)\delta\kappa
\right].
\end{align}
This interpolation provides a smooth transition between the deflagration and detonation branches and avoids introducing artificial discontinuities near the Jouguet point.

For detonations,
\begin{align}
v_w\ge v_J,
\end{align}
and the efficiency factor becomes
\begin{align}
\kappa
=
\frac{
(v_J-1)^3
v_J^{5/2}
v_w^{-5/2}
\kappa_C
\kappa_D
}
{
\left[
(v_J-1)^3
-
(v_w-1)^3
\right]
v_J^{5/2}
\kappa_C
+
(v_w-1)^3
\kappa_D
}.
\end{align}

\subsubsection{Magnetohydrodynamic turbulence}
\label{sec:MHD}
MHD turbulence is expected to be subdominant but not necessarily negligible.
In this case, the spectrum parameters listed in
\Tabref{tab:spectral_indices} were taken from Ref.~\cite{Caprini:2024hue}, where
they were determined based on the simulations in Ref.~\cite{RoperPol:2022iel}.
The turbulent energy fraction is
\begin{equation}
\Omega_s = \varepsilon K \,,
\label{eq:Omegas}
\end{equation}
where $\varepsilon$ represents an unknown fraction of overall kinetic energy in bulk motion that is converted to MHD turbulence.
The break frequencies are then given by
\begin{align}
f_1 &=
\frac{\sqrt{3\Omega_s}}{2{\mathcal{N}}} \, H_{*,0}\,(H_*R_*)^{-1}
\,,
\label{eq:MHD_f1}
\\
f_2 &\simeq 2.2\,H_{*,0}\,(H_*R_*)^{-1} \,,
\label{eq:MHD_f2}
\end{align}
where $\mathcal{N} \simeq 2$ \cite{RoperPol:2023bqa}.

The amplitude at the second break is
\begin{equation}
h^2\Omega_2 =
h^2F_{\rm GW,0}\,
A_{\rm MHD}\,
\Omega_s^2\,
(H_*R_*)^2,
\label{eq:MHD_Omega2}
\end{equation}
where
\begin{equation}
A_{\rm MHD} \simeq 4.37 \times 10^{-3}.
\label{eq:AMHD}
\end{equation}

\begin{table}
\centering
\begin{tabular}{lccccc}
\toprule
Source & $n_1$ & $n_2$ & $n_3$ & $a_1$ & $a_2$ \\
\midrule
Bubble collisions \cite{Lewicki:2022pdb} & 2.4 & $-2.4$ & -- & 1.2 & -- \\
Sound waves \cite{Jinno:2022mie} & 3 & 1 & $-3$ & 2 & 4 \\
MHD turbulence \cite{Caprini:2024hue} & 3 & 1 & $-8/3$ & 4 & 2.15 \\
\bottomrule
\end{tabular}
\caption{Spectral slopes and smoothness parameters used in
Eqs.~\eqref{eq:BPL} and \eqref{eq:Sf}.
}
\label{tab:spectral_indices}
\end{table}

\section{Numerical procedure for the Abelian Higgs phase transition}
\label{app-abelian-higgs-procedure}

In this Appendix we briefly summarize the numerical procedure used to compute the
thermodynamic and dynamical quantities entering the first-order phase transition
analysis. The calculation is based on the dimensionally reduced Abelian Higgs model
and combines three ingredients: the finite-temperature effective theory, the thermal
nucleation calculation, and the wall-velocity calculation with \texttt{WallGo}~\cite{Ekstedt:2024etx}.

\subsection{Four-dimensional theory}

The four-dimensional Abelian Higgs model is defined in
Eqs.~\eqref{eq:U1X_Lagrangian}--\eqref{eq:U1X_potential} of
\Secref{sec:U1_direct_sampling}, in terms of a complex scalar $\Phi$
of unit $U(1)_X$ charge, a gauge field $X_\mu$, the gauge coupling $g$,
and the scalar-potential parameters $\mu^2$ and $\lambda$.
For the dimensional-reduction formulae in this appendix, we use the
equivalent notation
$A_\mu\equiv-X_\mu$ and $m^2\equiv\mu^2$.
Consequently, the covariant derivative is
$D_\mu=\partial_\mu-igA_\mu=\partial_\mu+igX_\mu$.
The minus sign in the $A_\mu$ convention is therefore just a field
redefinition; the value and sign of the gauge coupling $g$ are unchanged.
We use $A_\mu$ and $m^2$ consistently in the remainder of this appendix.

To calculate the effective potential, we introduce the real scalar
background $\phi$ by writing $\Phi_{\rm bg}=\phi/\sqrt{2}$; the
background phase can be removed by a $U(1)_X$ transformation.
Fluctuations about this background are written as
$\Phi=(\phi+h+iG)/\sqrt{2}$, where $h$ is the real (Higgs) mode and
$G$ is the Goldstone mode. We retain $G$ explicitly because the
calculation is performed in Landau gauge rather than in unitary gauge.
Thus, $\Phi$ is the full complex scalar field and $\phi$ its canonically
normalised real background value.

In terms of $\phi$, the potential and squared vacuum
expectation value at zero-temperature are
\begin{equation}\label{4d-potential-appendix}
V_{4D}(\phi)=-\frac{1}{2}m^2\phi^2+\frac{\lambda}{4}\phi^4,
\qquad v^2=\frac{m^2}{\lambda}=\frac{\mu^2}{\lambda}.
\end{equation}
The sign convention for the quadratic term differs from that of
Ref.~\cite{Ekstedt:2022bff}; here $m^2=\mu^2>0$ is the
symmetry-breaking mass parameter introduced in
\eq{eq:U1X_potential}. The field-dependent masses of the Higgs
mode, Goldstone mode and gauge boson in Landau gauge are
\begin{equation}
m_h^2(\phi)=-m^2+3\lambda\phi^2,\qquad
m_G^2(\phi)=-m^2+\lambda\phi^2,\qquad
m_A^2(\phi)=g^2\phi^2.
\end{equation}
At $\phi=v$, these reproduce $m_h^2=2\lambda v^2$ and
$m_A^2=g^2v^2$, as in \Secref{sec:U1_direct_sampling}.

\subsection{Dimensional reduction and two-loop thermal masses}
\label{app:DR}

At high temperature the non-zero Matsubara modes, with masses of order $\pi T$, are
integrated out~\cite{Ginsparg:1980ef,Appelquist:1981vg}. We follow the standard matching
procedure~\cite{Farakos:1994kx,Kajantie:1995dw,Ekstedt:2022bff}, assuming the power counting
$\lambda\sim g^2\ll 1$, $m^2\sim g^2T^2$. Scalar masses are matched at two loops and
couplings at one loop. The resulting three-dimensional theory is
\begin{equation}
\mathcal{L}_{\rm 3D}
= \tfrac14 F_{ij}F_{ij} + (D_i\Phi)^\dagger(D_i\Phi)
- m_3^2\,\Phi^\dagger\Phi + \lambda_3(\Phi^\dagger\Phi)^2
+ \tfrac12(\partial_i A_0)^2 + \tfrac12 m_D^2 A_0^2
+ \tfrac12 h_3\, A_0^2\,\Phi^\dagger\Phi + \dots
\label{eq:L3D}
\end{equation}
where $D_i=\partial_i-ig_3A_i$, $A_0$ is the temporal gauge component, which acquires
the Debye mass $m_D$, and the ellipsis denotes the $A_0^4$ coupling and higher-dimensional
operators, which are neglected. The sign of the mass term is chosen such that $m_3^2$ has
the same sign convention as $m^2$ in Eq.~\eqref{4d-potential-appendix}. At leading order
$h_3=2g^2T$, so that $\tfrac12 h_3 A_0^2\Phi^\dagger\Phi = g^2T\,A_0^2\Phi^\dagger\Phi$.

We define
\begin{equation}
\mathrm{L}_b \equiv 2\ln\frac{\bar\mu\,e^{\gamma_E}}{4\pi T}\,,
\label{eq:Lb}
\end{equation}
where $\bar\mu$ is the $\overline{\rm MS}$ renormalisation scale and $m^2,\lambda,g^2$ are
$\overline{\rm MS}$ parameters at $\bar\mu$. In Landau gauge we find
\begin{align}
g_3^2 &= g^2T\left[1-\frac{g^2}{48\pi^2}\,\mathrm{L}_b\right],
\label{eq:g3sq}\\[1mm]
\lambda_3 &= T\left\{\lambda+\frac{1}{16\pi^2}\Big[\big(6g^2\lambda-10\lambda^2-3g^4\big)\mathrm{L}_b+2g^4\Big]\right\},
\label{eq:lam3}\\[1mm]
m_D^2 &= \frac{g^2T^2}{3}+\frac{g^2T^2}{144\pi^2}\Big[(7-\mathrm{L}_b)\,g^2+12\lambda\Big]-\frac{g^2m^2}{4\pi^2},
\label{eq:mD}\\[1mm]
h_3 &= g^2T\left[2+\frac{\lambda}{\pi^2}-\frac{g^2(\mathrm{L}_b-4)}{24\pi^2}\right].
\label{eq:h3}
\end{align}
The scalar mass parameter is determined at two loops,
\begin{equation}
\begin{split}
m_3^2 ={}& m^2\left[1+\left(\frac{3g^2}{16\pi^2}-\frac{\lambda}{4\pi^2}\right)\mathrm{L}_b\right]
-\left(\frac{g^2}{4}+\frac{\lambda}{3}\right)T^2\\
&-\frac{T^2}{576\pi^2}\Big\{24\lambda^2\big(\mathrm{L}_b-6\gamma_E+72\ln\mathcal{A}\big)
-24g^2\lambda\big(3\mathrm{L}_b-1-6\gamma_E+72\ln\mathcal{A}\big)\\
&\hspace{2.6cm}+g^4\big(69\,\mathrm{L}_b-8-108\gamma_E+1296\ln\mathcal{A}\big)\Big\}\\
&-\frac{1}{16\pi^2}\Big[4g_3^4+\tfrac12 h_3^2-8g_3^2\lambda_3+8\lambda_3^2\Big]\ln\frac{\bar\mu_3}{\bar\mu}\,,
\end{split}
\label{eq:m3sq}
\end{equation}
where $\mathcal{A}\simeq1.2824$ is the Glaisher--Kinkelin constant ($\ln\mathcal{A}\simeq0.24875$)
and $\bar\mu_3$ is the renormalisation scale of the three-dimensional theory. The last line
of Eq.~\eqref{eq:m3sq} is the two-loop running of $m_3^2$ inside the 3D theory. It is the only
parameter that runs in 3D, with
\begin{equation}
\bar\mu_3\frac{{\rm d}m_3^2}{{\rm d}\bar\mu_3}
=-\frac{1}{16\pi^2}\Big[4g_3^4+\tfrac12 h_3^2-8g_3^2\lambda_3+8\lambda_3^2\Big]
\;\simeq\;-\frac{T^2}{16\pi^2}\big(6g^4-8g^2\lambda+8\lambda^2\big),
\label{eq:m3run}
\end{equation}
where the last step holds to the order considered. The expressions
\eqref{eq:g3sq}--\eqref{eq:m3sq} are independent of $\bar\mu$ at the order considered, as
follows from the one-loop $\beta$-functions
\begin{equation}
\mu\frac{{\rm d}g^2}{{\rm d}\mu}=\frac{g^4}{24\pi^2},\qquad
\mu\frac{{\rm d}\lambda}{{\rm d}\mu}=\frac{20\lambda^2-12g^2\lambda+6g^4}{16\pi^2},\qquad
\mu\frac{{\rm d}m^2}{{\rm d}\mu}=\frac{(8\lambda-6g^2)\,m^2}{16\pi^2}.
\label{eq:beta}
\end{equation}
Choosing $\bar\mu=4\pi e^{-\gamma_E}T$ removes $\mathrm{L}_b$. At leading order the matching reduces to
$g_3^2=g^2T$, $\lambda_3=\lambda T$, $m_D^2=g^2T^2/3$ and $m_3^2=m^2-(\lambda/3+g^2/4)T^2$.

The field $A_0$ has mass $m_D\sim gT$ and is integrated out in the next step of the hierarchy.
Expanding its one-loop contribution $-\frac{1}{12\pi}\big(m_D^2+h_3\Phi^\dagger\Phi\big)^{3/2}$ to
quadratic order in $\Phi^\dagger\Phi$ shifts the parameters of the remaining theory by
\begin{equation}
\bar m_3^2 = m_3^2+\frac{h_3\,m_D}{8\pi}\,,\qquad
\bar\lambda_3 = \lambda_3-\frac{h_3^2}{32\pi\, m_D}\,,
\label{eq:softstep}
\end{equation}
which reduce to $\bar m_3^2=m_3^2+g_3^2m_D/(4\pi)$ and $\bar\lambda_3=\lambda_3-g_3^4/(8\pi m_D)$
for $h_3=2g_3^2$. The three-dimensional fields are related to the four-dimensional ones by
$\phi_3^2=\phi^2(1+\delta Z)/T$ with $\delta Z=-\tfrac{3g^2}{16\pi^2}\mathrm{L}_b$.

\subsection{Effective potential}
\label{app:Veff}
The phase-transition dynamics is controlled by the effective potential, which we write in
four-dimensional normalisation as $V_{\rm eff}(\phi,T)=T\,V_{\rm 3D}(\phi_3)$ with
$\phi=\sqrt{T}\,\phi_3$. At leading order in the three-dimensional theory,
\begin{equation}
V_{\rm 3D}(\phi_3)
= -\tfrac12 m_3^2\,\phi_3^2+\tfrac14\lambda_3\,\phi_3^4
-\frac{\big(g_3^2\phi_3^2\big)^{3/2}}{6\pi}
-\frac{1}{12\pi}\Big(m_D^2+\tfrac12 h_3\,\phi_3^2\Big)^{3/2},
\label{eq:V3D}
\end{equation}
that is,
\begin{equation}
V_{\rm eff}(\phi,T)
= -\tfrac12 m_3^2\,\phi^2+\frac{\lambda_3}{4T}\,\phi^4
-\frac{g_3^3}{6\pi\sqrt{T}}\,\phi^3
-\frac{T}{12\pi}\Big(m_D^2+\frac{h_3\,\phi^2}{2T}\Big)^{3/2},
\label{eq:Veff}
\end{equation}
where $m_3^2,\lambda_3,g_3^2,m_D^2$, and $h_3$ are given in Eqs.~\eqref{eq:g3sq}--\eqref{eq:m3sq}.
The cubic term arises from the two transverse polarisations of the three-dimensional gauge
field, each contributing $-(g_3\phi_3)^3/(12\pi)$, and the last term from the temporal
component $A_0$, whose mass $m_D^2+\tfrac12h_3\phi_3^2$ depends on the background field.
At leading order, $V_{\rm eff}\supset-\frac{g^3T}{6\pi}\phi^3-\frac{g^3T}{12\pi}\big(\phi^2+T^2/3\big)^{3/2}$.
Scalar loops are not included, which is justified for $\lambda\ll g^2$.

For $\phi^2\ll T^2/3$ the $A_0$ term may be expanded and absorbed into shifted parameters,
Eq.~\eqref{eq:softstep}, which yields the polynomial form
$V_{\rm eff}=A\phi^2+B\phi^3+C\phi^4$ with $A=-\tfrac12\bar m_3^2$,
$B=-g_3^3/(6\pi\sqrt T)$ and $C=\bar\lambda_3/(4T)$. This expansion fails in the supersoft
regime ($\lambda\lesssim0.07\,g^3$) and for $\phi\gtrsim T/\sqrt3$, so we do not use it
for the results in this paper; the critical temperature is instead obtained by
solving $V_{\rm eff}(\phi_{\rm true},T_c)=V_{\rm eff}(0,T_c)$ numerically from Eq.~\eqref{eq:Veff}.

\subsection{Critical and spinodal temperatures}
\label{app:Tc}
We normalise the potential of Eq.~\eqref{eq:Veff} to vanish in the symmetric phase,
\begin{equation}
\tilde V(\phi,T)\equiv V_{\rm eff}(\phi,T)-V_{\rm eff}(0,T)\,,
\end{equation}
so that $\phi_{\rm false}=0$ at all temperatures. Its derivative factorises,
$\partial_\phi\tilde V=\phi\,F(\phi,T)$, with
\begin{equation}
F(\phi,T)=-m_3^2+\frac{\lambda_3}{T}\,\phi^2-\frac{g_3^3}{2\pi\sqrt{T}}\,\phi
-\frac{h_3}{8\pi}\sqrt{m_D^2+\frac{h_3\,\phi^2}{2T}}\,,
\label{eq:F}
\end{equation}
so that the non-trivial extrema are the positive roots of $F(\phi,T)=0$. When two positive
roots exist, the larger one is the broken minimum $\phi_{\rm true}$ and the smaller one is
the top of the barrier, $\phi_{\rm bar}$.

\paragraph{Critical temperature.}
The critical temperature $T_c$ is defined by degenerate minima, i.e.\ by the simultaneous
solution for $(\phi_{\rm true},T_c)$ of
\begin{equation}
\tilde V(\phi_{\rm true},T_c)=0\,,\qquad F(\phi_{\rm true},T_c)=0\,,\qquad \phi_{\rm true}>0\,.
\label{eq:Tc}
\end{equation}

\paragraph{Spinodal temperatures.}
The symmetric phase ceases to be metastable when $\partial^2_\phi\tilde V(0,T)=-F(0,T)=0$,
which is equivalent to
\begin{equation}
\bar m_3^2(T_{\rm sp}^{\rm low})=0\,,\qquad
\bar m_3^2\equiv m_3^2+\frac{h_3\,m_D}{8\pi}\,,
\label{eq:Tsplow}
\end{equation}
with $\bar m_3^2$ as in Eq.~\eqref{eq:softstep}. The upper spinodal, at which the broken
minimum first appears, is where $\phi_{\rm true}$ and $\phi_{\rm bar}$ merge, i.e.\
$\partial_\phi\tilde V=\partial_\phi^2\tilde V=0$ at some $\phi_{\rm sp}>0$, which is equivalent to
\begin{equation}
F(\phi_{\rm sp},T_{\rm sp}^{\rm up})=0\,,\qquad
\partial_\phi F(\phi_{\rm sp},T_{\rm sp}^{\rm up})=0\,,
\label{eq:Tspup}
\end{equation}
with
\begin{equation}
\partial_\phi F=\frac{2\lambda_3}{T}\,\phi-\frac{g_3^3}{2\pi\sqrt{T}}
-\frac{h_3^2\,\phi}{16\pi T\sqrt{m_D^2+h_3\phi^2/(2T)}}\,.
\end{equation}
These temperatures satisfy $T_{\rm sp}^{\rm low}<T_c<T_{\rm sp}^{\rm up}$. Expanding the square
root in Eq.~\eqref{eq:F} to first order in $h_3\phi^2/(2Tm_D^2)$ reproduces the polynomial
potential $A\phi^2+B\phi^3+C\phi^4$, for which the conditions reduce to the closed forms
$A=B^2/(4C)$, $\phi_{\rm true}=-B/(2C)$ and $9B^2=32AC$. These are reliable only
for $\phi^2\ll T^2/3$ and are not used here.

The quantities recorded at the transition are
\begin{equation}
\phi_{\rm false}=0\,,\qquad \phi_{\rm true}\,,\qquad \phi_{\rm bar}\,,\qquad
\Delta V=\tilde V(\phi_{\rm true},T)\,,
\end{equation}
together with $m_3^2,\lambda_3,g_3^2,m_D^2,h_3$ in place of $A(T),B(T),C(T)$.

\subsection{Scan strategy}
	
For efficiency, the parameter scan is organised in stages. A fast
analytic filter first checks that the effective potential permits
the required first-order transition structure. The nucleation/
percolation calculation is then carried out for points passing
this filter, and the successful points are mapped to the GW
spectrum using the LTE wall-velocity treatment. This is the
calculation used in the Bayesian scans shown in the main text.
	
The debugging-stage output table contains
\begin{equation}
\begin{split}
&g,\ \lambda,\ y,\ n_f,\ q_f,\ m^2,\ v,\ T,\  \phi_{\rm false},\ \phi_{\rm true},\ \Delta V,\\
&A(T),\ B(T),\ C(T),\ T_c,\ T_{\rm spinodal},\ v_w,\ L_w,\ v_w^{\rm LTE},
\end{split}
\end{equation}
together with the collision-cache directory and solver status. This
provides a compact record of the thermodynamic, nucleation, and
wall-dynamics information before connecting the calculation to the full
FOPT and GW analysis pipeline.

To examine
the validity of the LTE wall treatment, a small separate set of
successful transition points is passed to the full
collision-based \texttt{WallGo} solver. For these validation
points we additionally retain $v_w^{\rm full}$ when a stationary
solution exists, $v_w^{\rm LTE}$, the native solver classification,
the wall width when returned, the collision-cache path, and the
momentum grid size. The full collision-based solver is not
used at each likelihood evaluation of the production scan.

\subsection{Wall velocity calculation}
	
The wall velocity in the production likelihood is obtained within
the local-thermal-equilibrium (LTE) treatment of \texttt{WallGo},
using the phase thermodynamics and hydrodynamic matching from the
NLO dimensionally reduced effective potential. This avoids
computing the off-equilibrium collision integrals at every sampled
point. An LTE limiting result $v_w^{\rm LTE}=1$ should be understood
as the outcome of this approximation, not as an independent full
collision-based demonstration of runaway propagation.

For the separate validation points, we solve the full wall
problem with off-equilibrium perturbations. Schematically, the
scalar equation of motion in the wall frame may be written as
\begin{equation}\label{wall-eom-appendix}
\frac{d^2\phi}{dz^2}
-\frac{\partial V_{\rm eff}}{\partial\phi}
-\mathcal{F}_{\rm fric}[\phi,v_w]=0,
\end{equation}
where $\mathcal{F}_{\rm fric}$ represents plasma friction arising
from deviations from LTE. In \texttt{WallGo} these perturbations
are obtained by solving linearised Boltzmann equations for the
species included in the collision model. The matrix elements are
generated using \texttt{WallGoMatrix}, with collision integrals
computed by \texttt{WallGoCollision} and cached for reuse. The
collision calculation uses leading-order thermal-mass inputs;
these should be distinguished from the NLO effective potential
used for the equilibrium phase thermodynamics.

Each validation point is evaluated at its archived transition
temperature $T_\ast$, without temperature-offset retries. We
check the native solver status rather than treating any numerical
return as a stationary solution. The stationary branches are
separated from cases classified RUNAWAY, for which the
code does not return a stationary wall velocity. Grid-size tests
are reported where available, including $N=5$ and $7$ for points
C and E and $N=5$, $7$ and $9$ for a separate stationary reference
point. The candidate at E is located at the Jouguet branch
boundary and is therefore not assigned a precision-converged
terminal velocity. These comparisons are used only as a limited
validation of the LTE treatment and do not replace the LTE
calculation in the nested-sampling likelihood.

\subsection{Numerical convergence of the Abelian nested-sampling analysis}
\label{sec:abelian_logz_convergence}
We assess the numerical stability of the $U(1)_X$ posterior using the evidence-consistency statistic $R_Z$ defined in \eq{eq:logz_convergence_statistic}. For the O4a dataset, comparisons are performed separately for the linear and logarithmic priors, holding fixed the likelihood, physical parameters and all prior bounds while varying the nested-sampling resolution. For each pair of independently initialised runs, we calculate $|\Delta\log Z|$, the combined uncertainty
$\sigma_{\rm comb}=\sqrt{\sigma_a^2+\sigma_b^2}$, and the ratio
$R_Z=|\Delta\log Z|/\sigma_{\rm comb}$.

\begin{table}[tbp]
\centering
\setlength{\tabcolsep}{4pt}
\begin{tabular}{@{}lcccc@{}}
\hline
Prior & $\log Z_{200}$ & $\log Z_{20000}$ & $|\Delta\log Z|$ & $R_Z$ \\
\hline
Linear & $-7246.4423\pm 0.1523$ & $-7246.4859\pm 0.0237$ & $0.0435$ & $0.283$ \\
Logarithmic & $-7247.0031\pm 0.2647$ & $-7247.0795\pm 0.0428$ & $0.0765$ & $0.285$ \\
\hline
\end{tabular}
\caption{Resolution-stability check of the log Bayesian evidence for the
Abelian $U(1)_X$ NLO-3D LTE analysis with the combined O3+O4a dataset.
Within each row the model, likelihood and prior bounds are identical;
the initial number of live points changes from $200$ to $20\,000$.
The uncertainties are those reported by \texttt{dynesty}.}
\label{tab:u1_logz_convergence}
\end{table}

A value $R_Z\leq 1$ indicates consistency of the two evidence estimates within their quoted combined numerical uncertainty. We interpret this result together with the precision of the individual evidence estimates and the stability of the one-dimensional posterior distributions and credible regions. Thus, numerical convergence within a given prior is assessed independently of the changes induced by replacing the linear
prior with a logarithmic one; the latter are the subject of the
prior-robustness analysis in the following subsections.

The nominal evidence-consistency statistic $R_Z=|\log Z_{20000}-\log Z_{200}|/ \sqrt{\sigma_{20000}^2+\sigma_{200}^2}$ is $0.283$ for the linear prior and $0.285$ for the logarithmic prior, both below the adopted threshold $R_Z=1$. This indicates stability of the estimated evidence when the sampling resolution increases, for each prior separately. The saved
samples reveal that the $200$-live-point runs share their initial pseudo-random samples with the corresponding $20\,000$-live-point runs; thus this is a resolution-stability diagnostic rather than an independent-realisation test. Agreement of the evidences alone does not establish convergence of all marginal posterior distributions.

\bibliographystyle{apsrev4-2}
\bibliography{GW_ST_FOPT}

\end{document}